 \documentclass[final,5p,times,twocolumn,authoryear]{elsarticle}

\usepackage{amssymb}
\usepackage{lipsum}
\usepackage{graphicx}%
\usepackage{multirow}%
\usepackage{amsmath,amssymb,amsfonts}%
\usepackage{amsthm}%
\usepackage{mathrsfs}%
\usepackage[title]{appendix}%
\usepackage{xcolor}%
\usepackage{textcomp}%
\usepackage{manyfoot}%
\usepackage{booktabs}%
\usepackage{algorithm}%
\usepackage{algorithmicx}%
\usepackage{algpseudocode}%
\usepackage{listings}%
\usepackage[colorlinks=true, citecolor=blue]{hyperref}
\usepackage{natbib}%
\usepackage{subcaption}%
\usepackage{comment}%
\usepackage[normalem]{ulem}
\journal{High Energy Astrophysics}

\begin{document}

\begin{frontmatter}



\title{Time lags and their association with the Boundary Layer structure in a Z source GX 349+2}


\author[1]{Abhishek M.V.R.\corref{cor1}}
\affiliation[1]{organization={Department of Astronomy, Osmania University},%Department and Organization
            addressline={}, 
            city={Hyderabad},
            postcode={500049}, 
            state={},
            country={India}}
\author[1]{Sriram K}

\author[1]{Gouse SD}
\cortext[cor1]{Corresponding author. 
mvrabhishek@gmail.com (M.V.R. Abhishek)}

\begin{abstract}
Studying the cross-correlation function between the soft and hard X-ray emission in Neutron Star Low Mass X-ray Binaries provides crucial insight into the structure and dynamics of the innermost accretion regions. In this work, we investigate the CCF of the Z-source GX 349+2 using an XMM-Newton observation. We noted that asymmetric CCFs with lags of a few hundred seconds between soft and hard band light curves in the horizontal branch, whereas CCFs remained symmetric in normal and flaring branches. We also performed a CCF study during the flux transition duration and observed lags of the order of a few tens to hundreds of seconds. Monte Carlo simulations were performed to assess the robustness of these CCFs, confirming their significance at a  95\% confidence level. 
%\uline{Spectral analysis during the flux transitions further suggests that the inner accretion disk extends close to the last stable orbit.} 
We propose that the observed hard lags arise from the readjustment of the boundary layer/coronal region located near the inner edge of the accretion disk. From the measured lags, we estimate the characteristic size of the boundary layer. We show that the observed lags could also be associated with the depletion timescale of the boundary layer with low viscosity.
\end{abstract}




\end{frontmatter}

%\tableofcontents



\section{Introduction}\label{sec1}
Neutron star low-mass X-ray binaries (NS-LMXBs) are particularly valuable laboratories for studying accretion physics under strong gravity, as the presence of a solid surface and often a Boundary Layer(BL) distinguishes them fundamentally from black hole systems. Despite decades of observational and theoretical work, several key aspects of the accretion process in NS-LMXBs remain poorly understood, including the geometry and evolution of the corona, the physical origin of spectral state transitions, and the coupling between the disk, boundary layer, and Comptonizing region.

NS-LMXBs are further classified based on the characteristic tracks they trace in Hardness–Intensity Diagrams (HIDs) into atoll and Z sources. Z sources exhibit three distinct branches in the HID: the Horizontal Branch (HB), Normal Branch (NB), and Flaring Branch (FB), which are generally associated with changes in mass accretion rate and spectral–timing properties. Only a small number of Z sources are known, and they are further subdivided according to their phenomenological behaviour into 1) Sco-like sources, e.g., Sco X-1, GX 17+2, and GX 349+2 (Sco X-2) and 2) Cyg-like sources, e.g., Cyg X-2, GX 5–1, and GX 340+0 (Hasinger \& van der Klis  \citeyear{1989ESASP.296..203V}).

GX 349+2 is a persistently bright NS-LMXB and a well-known member of the Sco-like subclass of Z sources (Hasinger \& van der Klis  \citeyear{1989ESASP.296..203V}). The source traces a characteristic Z-shaped pattern in Hardness–Intensity and Colour–Colour diagrams, primarily exhibiting the normal and flaring branches, while the horizontal branch is generally weak or absent, distinguishing it from Cyg-like Z sources (Kuulkers \& van der Klis  \citeyear{1995xrbi.nasa..252V}; O’Neill et al. \citeyear{o2002x}). Extensive timing studies using RXTE have revealed strong low-frequency noise components and peaked variability in the $\sim$ 3–7 Hz range along the normal and flaring branches, with the timing properties evolving systematically along the Z track (O’Neill et al. \citeyear{o2002x}; Agrawal \& Sreekumar \citeyear{10.1111/j.1365-2966.2003.07147.x}). Additionally, twin kilohertz quasi-periodic oscillations at frequencies of $\nu$ $\sim$ 712 and $\sim$978 Hz have been detected, indicating that the inner accretion flow extends close to the neutron star surface (Zhang et al. \citeyear{zhang1998discovery}). 
Cross-correlation Function (CCF) analysis between two energy band light curves provides us with timing information of which band leads/lags, giving us a crucial understanding of the hard and soft band photons emitting regions and their interaction, which could be manifesting in the form of the said lags/leads.
Cross-correlation studies of this source using RXTE observations have revealed both soft and hard X-ray lags, ranging from a few seconds to several hundred seconds (Ding et al.  \citeyear{ding2016cross}). Broadband spectral analyses suggest that the X-ray emission of Sco X-2 is dominated by a combination of thermal emission from the accretion disk and BL, along with a Comptonized component originating from a hot corona, reflecting the complex accretion geometry in this system (Agrawal \& Sreekumar \citeyear{10.1111/j.1365-2966.2003.07147.x}; Iaria et al. \citeyear{iaria2009ionized}; Lin et al. \citeyear{lin2009spectral}; Ng et al. \citeyear{refId0}).

In this work, we use the XMM-Newton observations to perform a CCF analysis of the Z-source GX 349+2 (Sco X-2), with the aim of constraining the physical properties of the system through a combined temporal and spectral investigation.

\section{Observations}

The XMM-Newton observatory carries the European Photon Imaging Camera (EPIC), which consists of two MOS CCD detectors and one pn CCD detector. The EPIC instruments provide a field of view of approximately 30 arcmin and are sensitive over the energy range, 0.15–15.0 keV. Owing to the high brightness of GX 349+2, the MOS1 and MOS2 cameras were not operated during this observation to avoid photon pile-up effects. Consequently, only the EPIC-pn camera was used, operated in timing mode with the medium optical blocking filter. The source was observed by XMM-Newton for a total exposure of 22.5 ks on 2008 March 19, between 16:42:41 UT and 22:58:55 UT (ObsID: 0506110101), during satellite revolution 1516. (Jansen et al. \citeyear{jansen2001xmm}) \\

\begin{figure*}[hbt!]
    \centering
    % Left Image
    \begin{minipage}{0.48\linewidth}
        \centering
        \includegraphics[width=\linewidth]{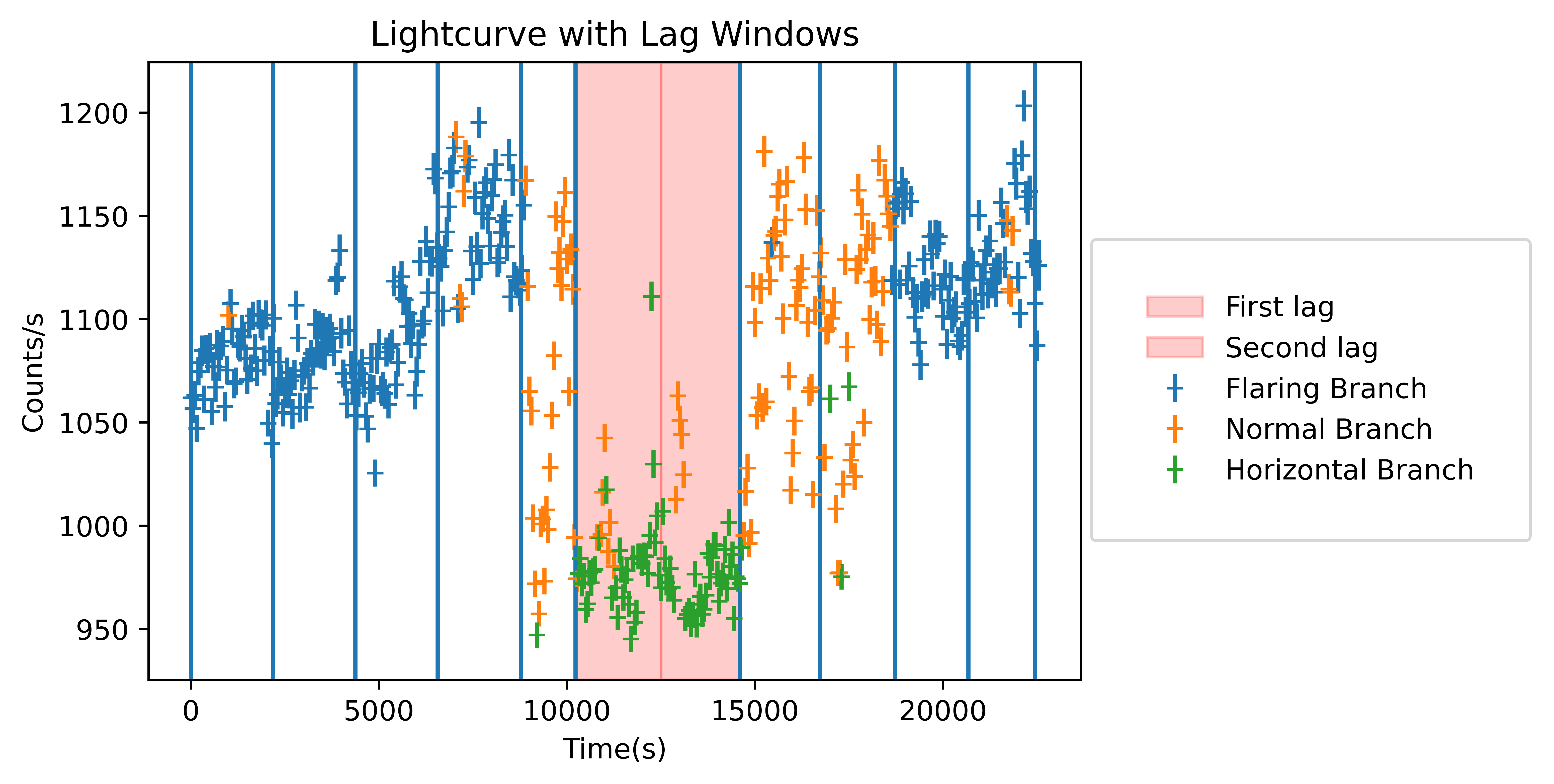}
    \end{minipage}
    \hfill % This pushes the two minipages to the far left and right
    % Right Image
    \begin{minipage}{0.48\linewidth}
        \centering
        \includegraphics[width=\linewidth]{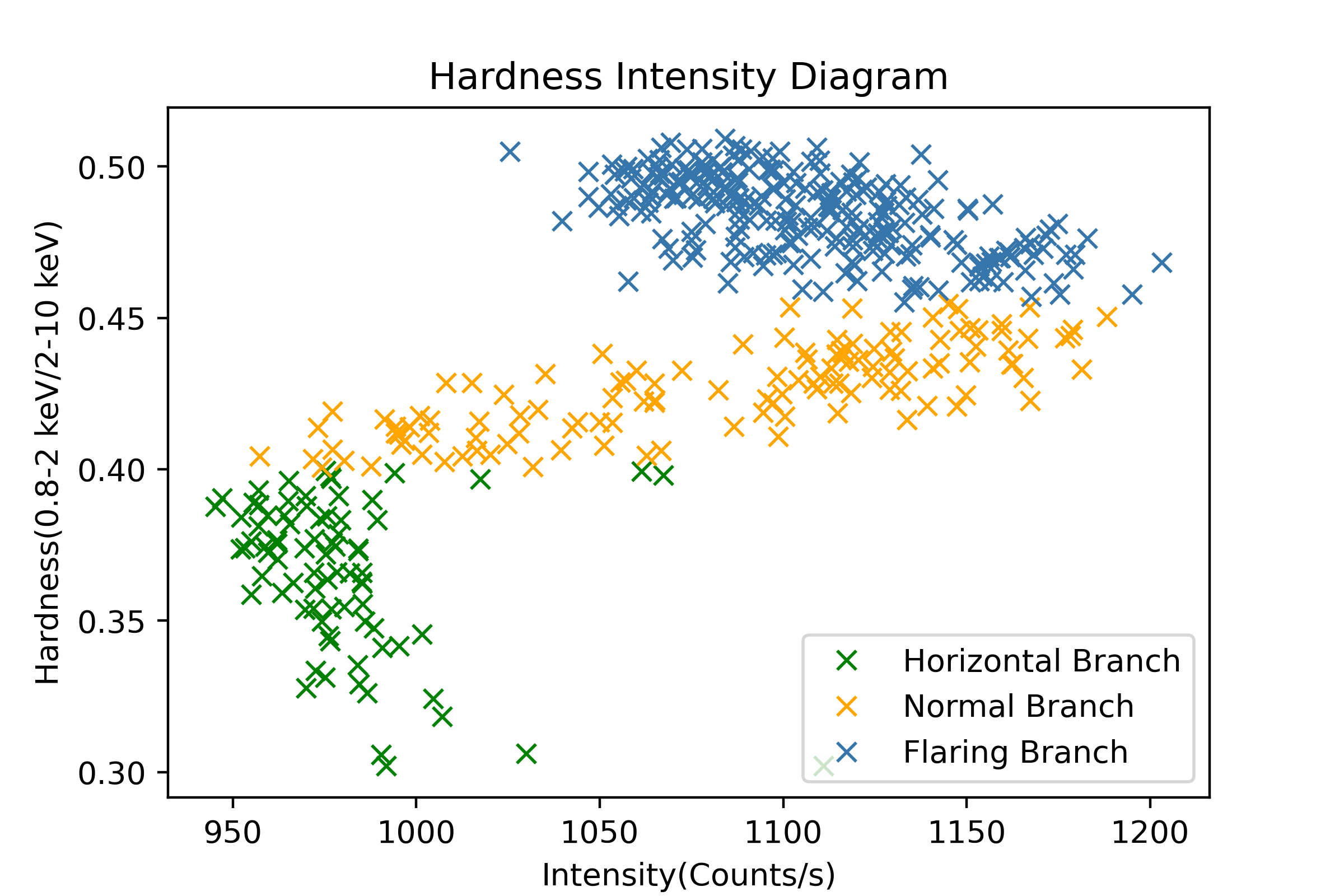}
    \end{minipage}

    \caption{The left panel is a plot of the light curve in the 0.8-10 keV band, with the sections where lags are found (horizontal branch) highlighted; the right panel is a plot of the Hardness Intensity Diagram of the observation.}
    \label{fig:my_combined_figure}
\end{figure*}

%\begin{figure}[hbt!]
 %   \centering
  %  \includegraphics[width=1\linewidth]{Total_LC.png}
   % \caption{Lightcurve of the entire 22.5 ks observation.}
    %\label{fig:Fig-1}
%\end{figure}

\section{Data Reduction and Analysis}

X-ray light curves for GX 349+2 were extracted using the \textit{evselect} task within the XMM-Newton Science Analysis Software (SAS). For the EPIC-pn camera operating in timing mode, we applied the standard event selection criteria (PATTERN $\le$ 4) and $XMMEA\_EP$. During the observation (ObsID: 0506110101), the EPIC-pn instrument experienced full scientific buffer conditions throughout, resulting in a low and highly variable effective live-time fraction per frame. To account for this effect, we applied the \textit{epiclccor} task to correct the extracted light curves. This procedure applies the appropriate Good Time Intervals (GTIs), corresponding to periods when reliable flux measurements are obtained, and corrects for instrumental effects such as dead time and exposure variations, see (Iaria et al. \citeyear{iaria2009ionized}).

Fig. 1 (left panel) shows the background-subtracted light curve(LC) of the entire observation in the 0.8–10.0 keV energy band, and the shaded region depicts the HB, and different colours display the different branches (blue: FB, green: HB, orange: NB). To mitigate potential photon pile-up effects, we excluded the central bright RAW column (RAWX = 37) along with one adjacent column on either side (RAWX = 36 \& 38). As evident from Fig. 1 (left panel), the source exhibits strong variability throughout the observation (see Iaria et al. \citeyear{iaria2009ionized}). The right panel of Fig. 1 displays the colour-coded hardness–intensity diagram (HID), where the hardness ratio (HR) is defined as the ratio of count rates in the 2.0–10.0 keV and 0.8–2.0 keV energy bands. The classification of the source HID into three branches, viz., horizontal branch (HB; blue points in Fig. 1), normal branch (NB; orange), and flaring branch (FB; green), was done by identifying the apexes in the HID track. HR $\ge 0.45$  was used to define the FB, while HR $\le 0.4$ was defined as HB, and $0.40 < $ HR $< 0.45$ as the NB.

%To guide the temporal analysis, we first constructed a hardness–intensity diagram (HID). The HID represents the source intensity (count rate) on the x-axis and the hardness ratio on the y-axis, defined here as the ratio of count rates in the 2.0–10.0 keV and 0.8–2.0 keV energy bands. The resulting HID (Figure 10) clearly confirms the classification of GX 349+2 as a Z source.

For the detailed analysis, the total exposure of 22.5 ks was divided into 11 segments of approximately 2000 s each. We performed a CCF analysis between the soft (0.8–2.0 keV) and hard (2.0–10.0 keV) X-ray light curves using the \textit{crosscorr} task from the XANADU data analysis package. The cross-correlations were computed with a 50s bin and applied to all the segments of the observation. Choosing an appropriate time bin size is crucial for reliable CCF analysis. In this work, we detect a characteristic lag of a few hundred seconds and therefore adopt a time bin of 50 s to adequately resolve this delay. This choice ensures that the lag is sampled over multiple bins (i.e., 
$\sim$2-4 bins across the lag timescale), thereby minimizing any smearing of the correlation signal that would occur for larger bin sizes. At the same time, using significantly smaller bins would reduce the signal-to-noise ratio per bin and introduce larger statistical uncertainties in the CCF. Thus, a bin size of 50 s represents an optimal bin time between temporal resolution and statistical robustness. Asymmetric CCFs exhibiting non-zero lags of $\sim$ 200 s, accompanied by relatively low cross-correlation coefficients, were detected in the segments associated with the horizontal branch of the HID (Fig. 1, shaded region). In each row, the left panel shows the soft band light curve, the middle panel displays the hard band light curve, and the right panel presents the corresponding CCF for each of the sections (see Fig. 2). These segments are hereafter referred to as Sections 1 \& 2. In contrast to Sections 1 and 2, shown in Fig. 2, the CCFs shown in Fig. 3 are highly symmetric and peak at zero lag, indicating a near simultaneous variation of the soft and hard X-ray emissions. These segments are predominantly associated with the normal branch (NB) and flaring branch (FB) of the HID. In a separate division of data covering the transition windows of approximately 2200 s each, see Fig. 4 (top). The highlighted windows are where the non-zero lags with asymmetric CCFs were found, and these windows are referred to as sections A, B and C, hereafter, to perform spectral studies. 

\subsection{Validation of Lags}
To validate the lags found from the original observations CCF analysis, in Sections 1 and 2 as well as Sections A, B and C, we performed simulations by generating 10,000 pairs of synthetic soft and hard light curves, which were produced by introducing random variations to each data point within the standard deviation of the corresponding observed light curve, thereby preserving the intrinsic variability characteristics of the source (Gouse et al. \citeyear{gouse2025asymmetric}, \citeyear{gouse2025association}).\\
CCF functions were then computed for each simulated light curve pair. Each resulting CCF was fitted with a composite model consisting of a linear baseline plus a Gaussian component (line + Gaussian + line) to determine the peak cross-correlation coefficient and estimate the associated lag. The histogram of these estimated lags was fitted with a Gaussian to validate the observed lag; plots are shown in Fig. 5 \& 6 (left panels of each row).

\subsection{Cross Correlation Confidence Interval}
Soft and hard light curves were simulated using the method proposed by Timmer \& Konig (\citeyear{timmer1995generating}), which is widely used to produce stochastic variability in time series characterized by power-law noise. Using the observed power spectral properties of the source, $\Gamma_{PSD} = 2.1$, and the mean same as the original light curve, we generated 10,000 simulated light curves independently for the soft (0.8–2.0 keV) and hard (2.0–10.0 keV) energy bands (Gouse et al.  \citeyear{gouse2025asymmetric},\citeyear{gouse2025association}). The CCF was then computed for each pair of simulated light curves in order to construct the statistical distribution of CCFs and derive the corresponding 95\% confidence intervals (Fig. 5 \& 6, right panels). The simulated light curves were generated using the AstroML package(VanderPlas et al. \citeyear{vanderplas2012introduction}) time-series module. \\
In Fig. 5 \& 6 (right panels), the CCFs corresponding to the 10,000 simulated light curves are shown in blue, while the observed CCF is overlaid in red. The dotted black curves represent the 95\% confidence limits derived from the simulations. This procedure is essential to demonstrate that the asymmetric CCFs and associated time lags detected in the data are not the result of random fluctuations, but instead arise from intrinsic temporal incoherence between the soft and hard X-ray emitting regions.

\begin{figure*}[hbt!]
    \centering

    % Group 1: The 6 images belonging to Subfigure 2(a)
    \begin{subfigure}[b]{\textwidth}
        \centering
        % First row of Subfigure (a)
        \includegraphics[width=2in, height=2in]{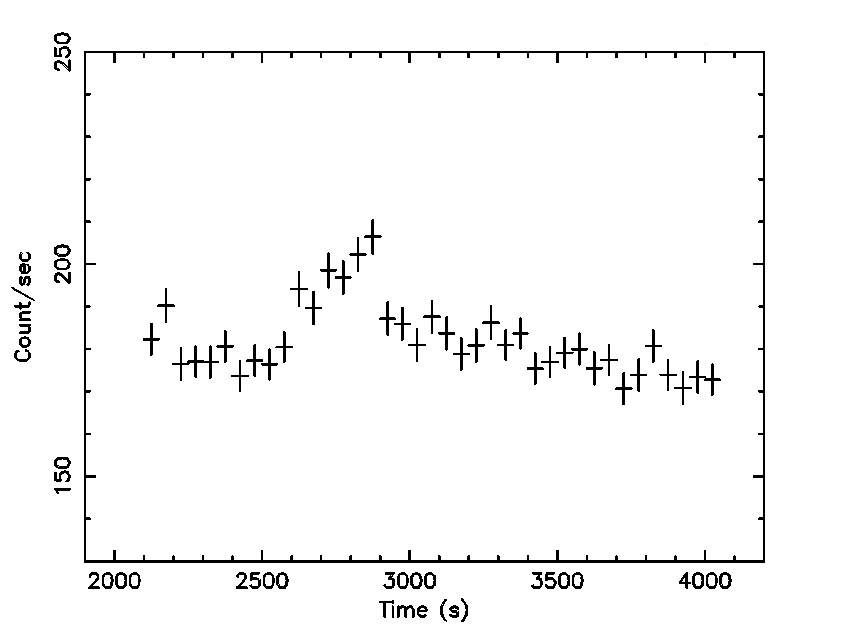}
        \includegraphics[width=2in, height=2in]{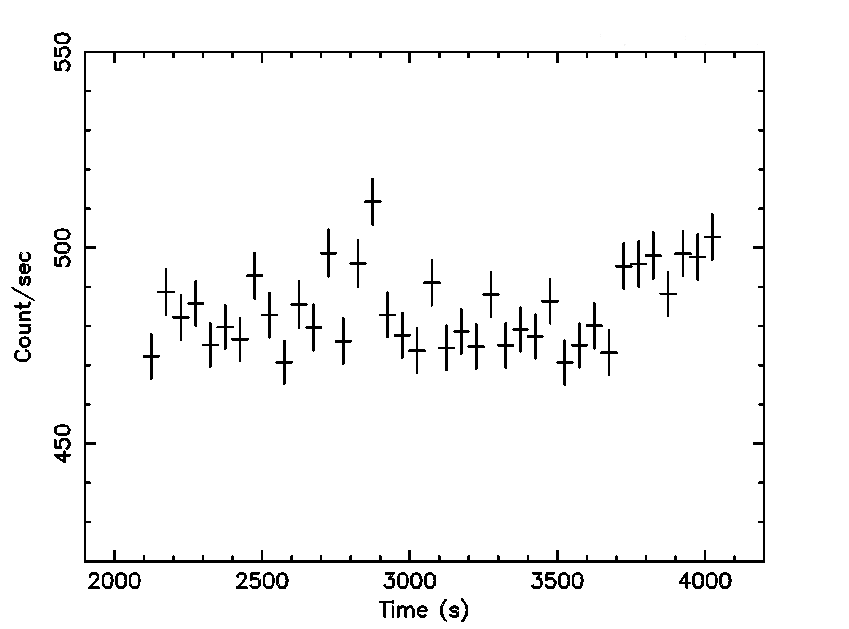}
        \includegraphics[width=2in, height=2in]{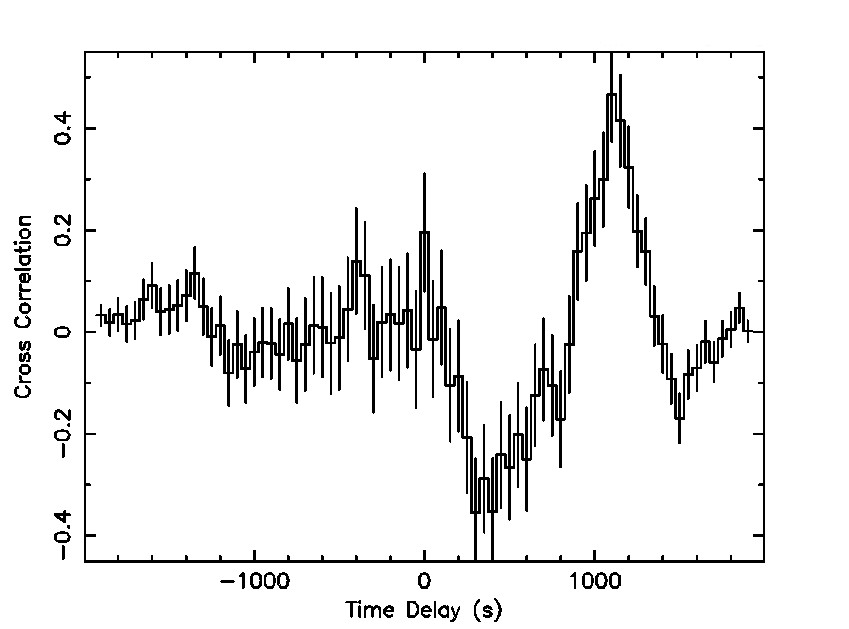} \\ \vspace{0.1cm}
        
        % Second row of Subfigure (a)
        \includegraphics[width=2in, height=2in]{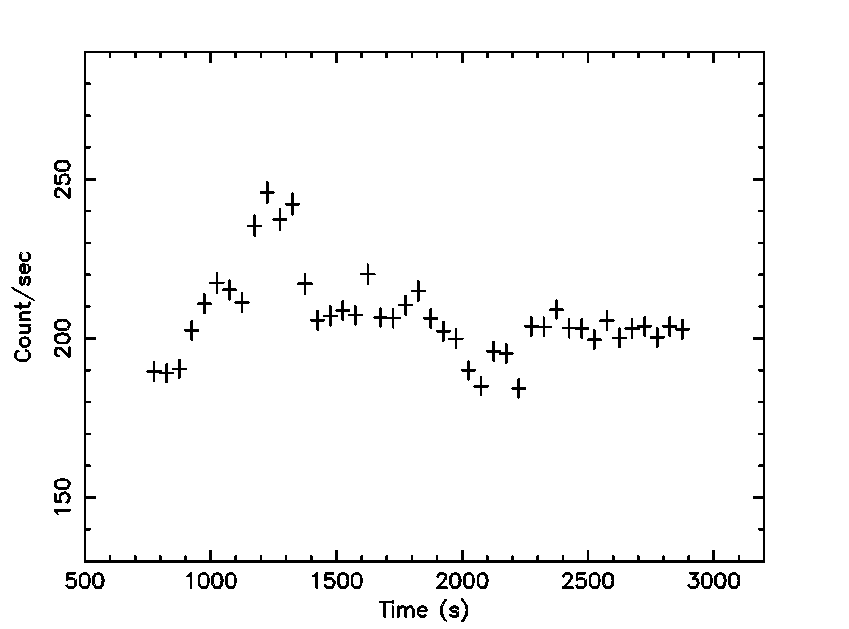}
        \includegraphics[width=2in, height=2in]{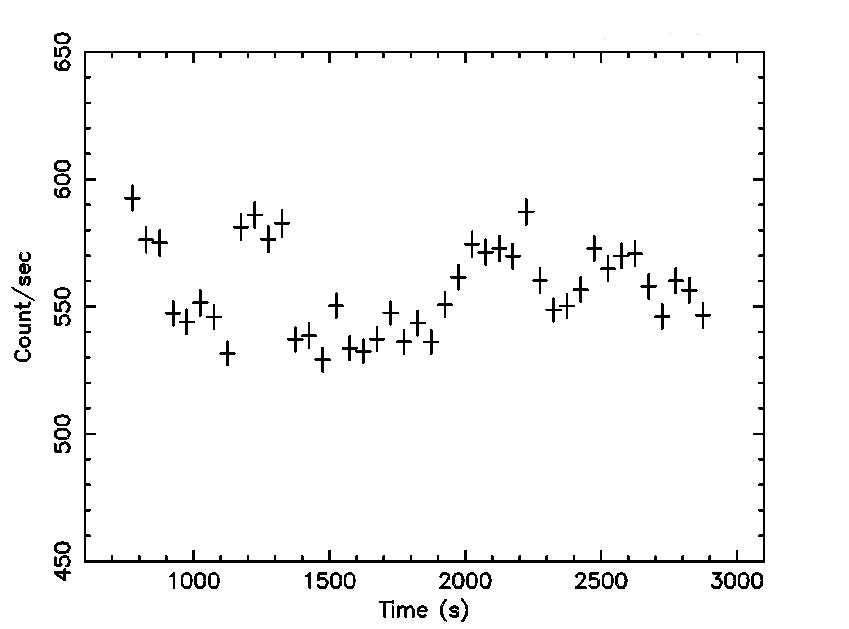}
        \includegraphics[width=2in, height=2in]{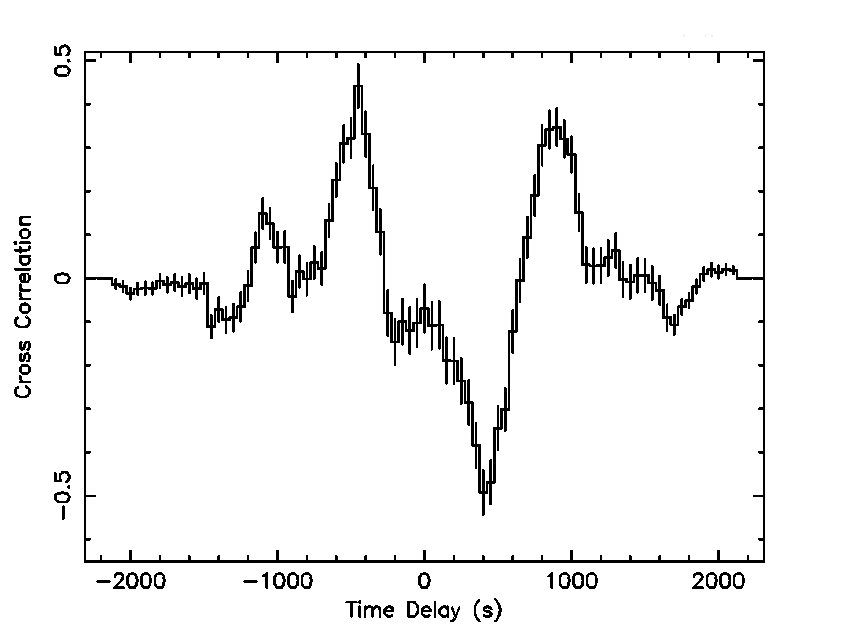}
        
        \caption{Soft and hard LCs, and CCF plots for HB1 and HB2.}
        \label{fig:lightcurves_ccf}
    \end{subfigure}

    \vspace{0.6cm} % Distinct separation between 2(a) and 2(b)

    % Group 2: The 2 images belonging to Subfigure 2(b)
    \begin{subfigure}[b]{\textwidth}
        \centering
        \includegraphics[width=3in, height=2in]{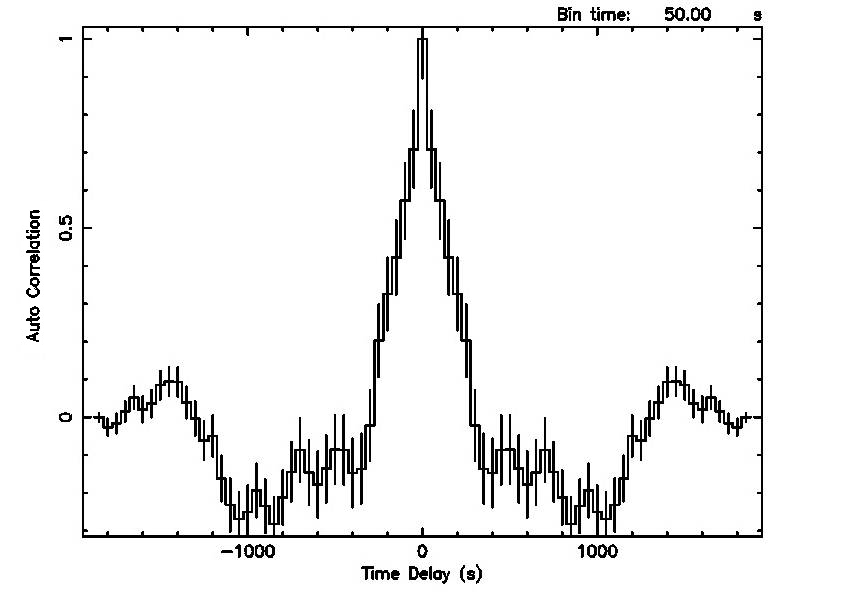}
        \includegraphics[width=3in, height=2in]{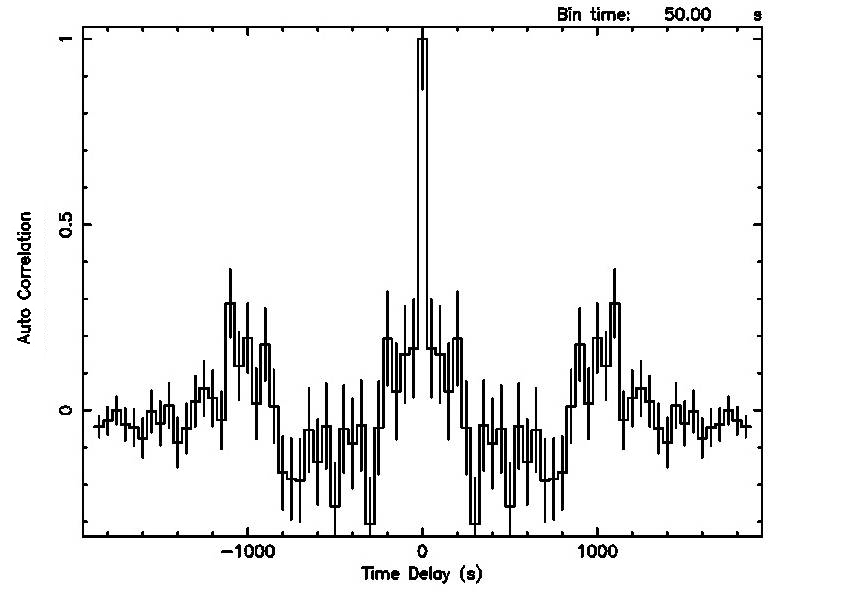}
        \includegraphics[width=3in, height=2in]{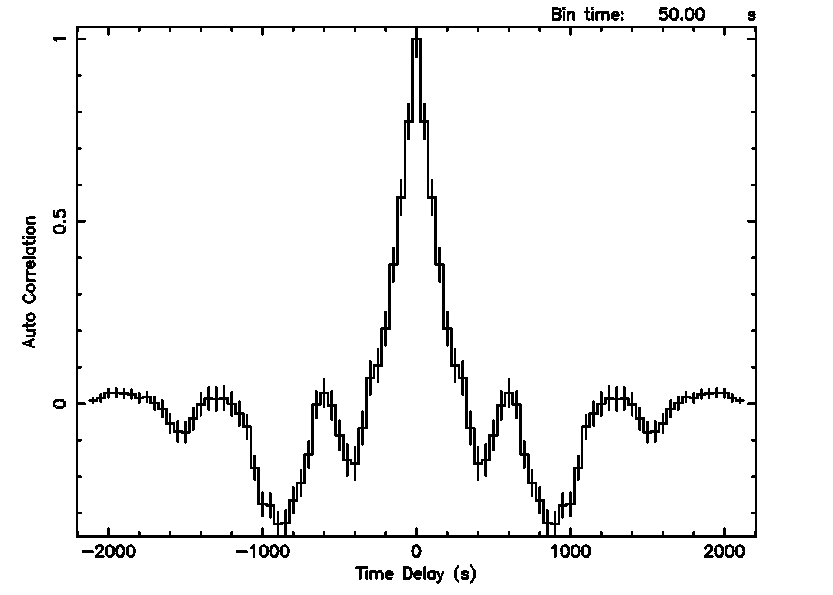}
        \includegraphics[width=3in, height=2in]{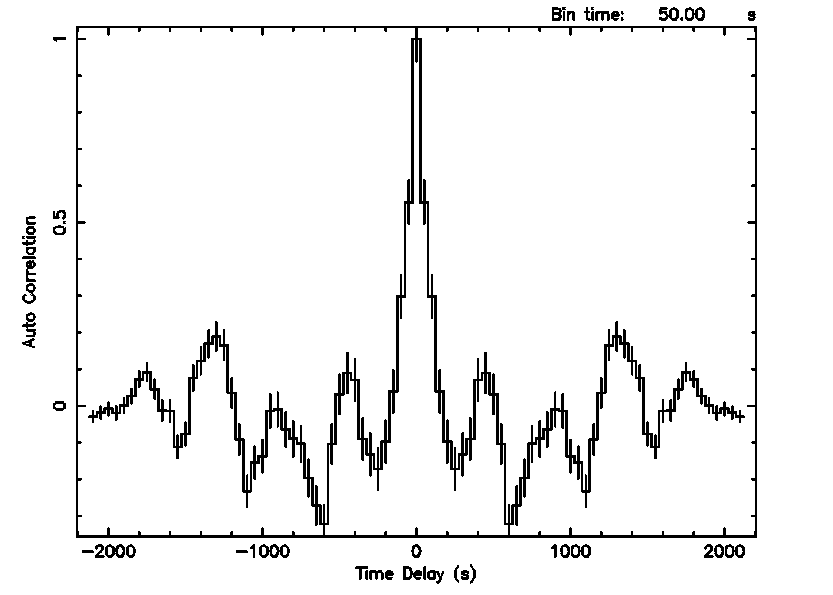}
        
        \caption{ACF Plots for Soft band(left panel) and Hard band(left panel); Row 1: HB1 and Row 2: HB2.}
        \label{fig:next_plots}
    \end{subfigure}

    % Main Figure Caption covering the whole set
    \caption{Comprehensive overview of the light curves and correlation functions. First two sub-plots in each row of (a) are the soft and hard LCs, respectively, and the third is the CCF plot.
    \textbf{The 2nd sub-plot (b) shows ACF plots of HB1 (row 1) and HB2 (row 2)}.}
    \label{fig:main_figure_2}
\end{figure*}

\begin{figure*}
    \centering
  %\begin{minipage}[b]{0.8\linewidth}
    \includegraphics[width=2in, height=2in]{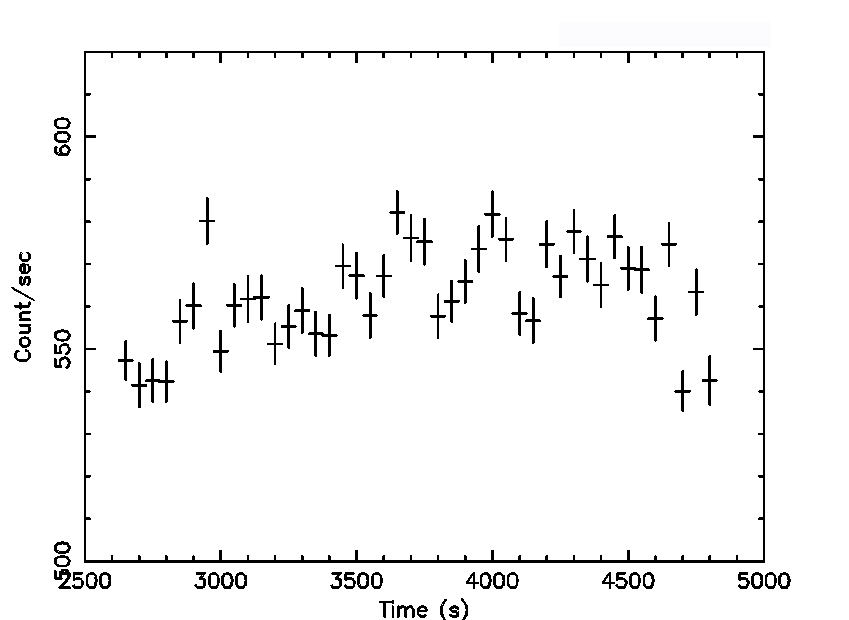}
    \includegraphics[width=2in, height=2in]{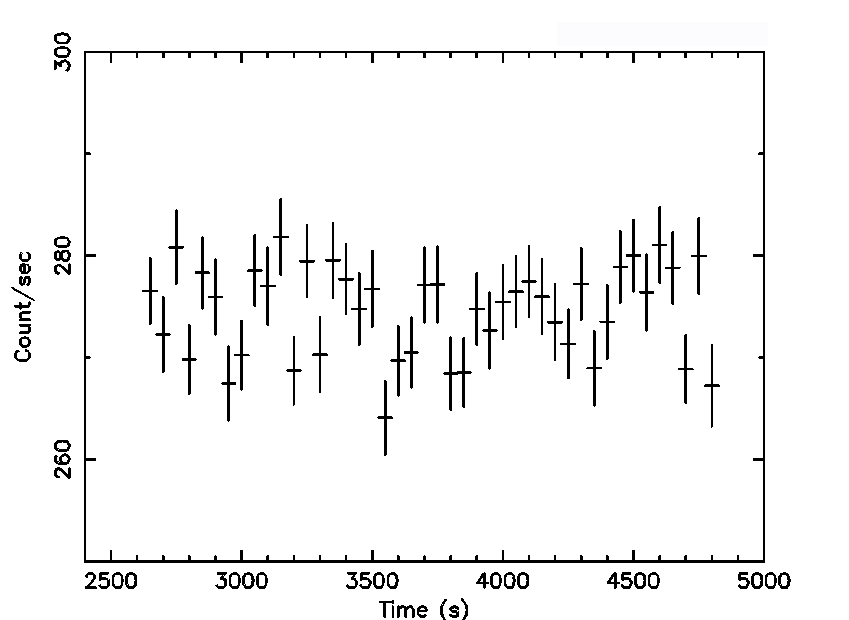}
    \includegraphics[width=2in, height=2in]{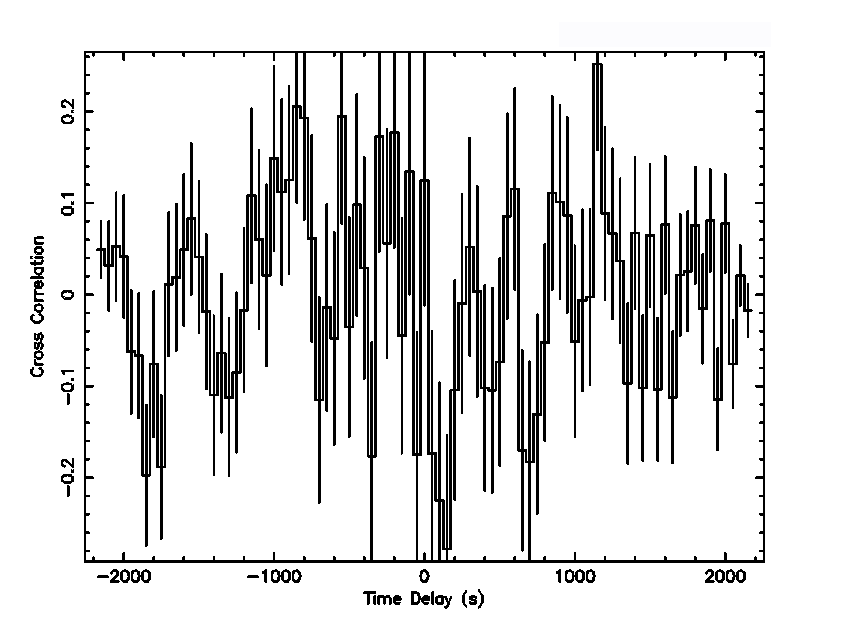}\\
    \includegraphics[width=2in, height=2in]{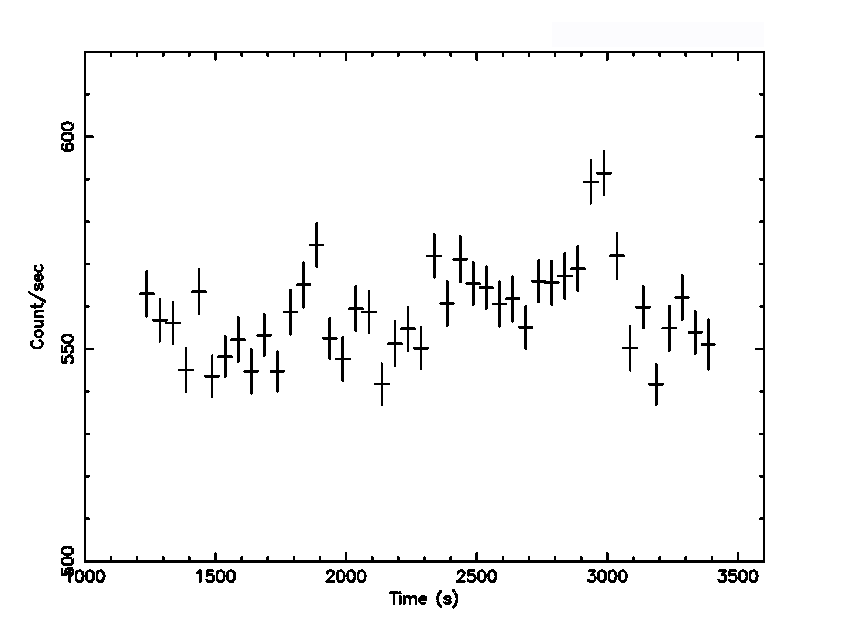}
    \includegraphics[width=2in, height=2in]{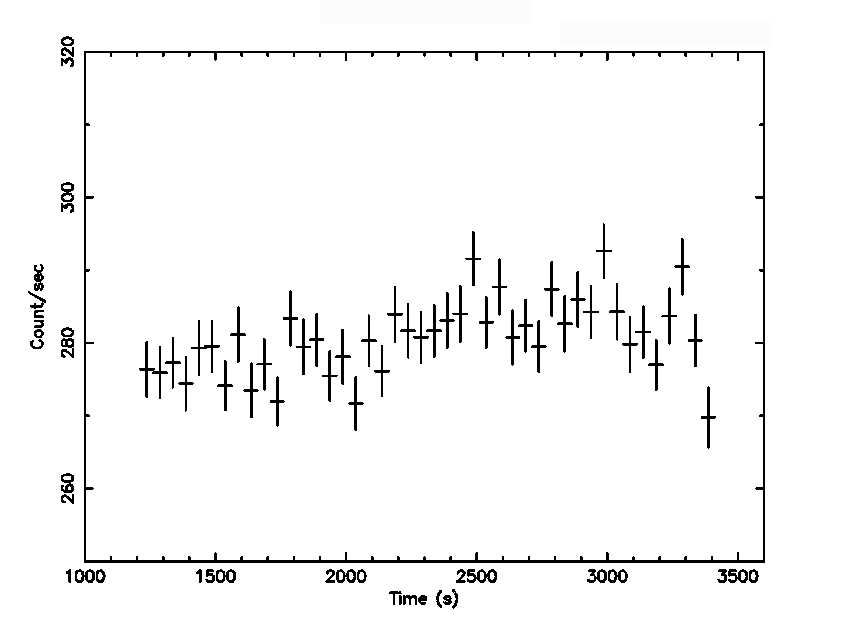}
    \includegraphics[width=2in, height=2in]{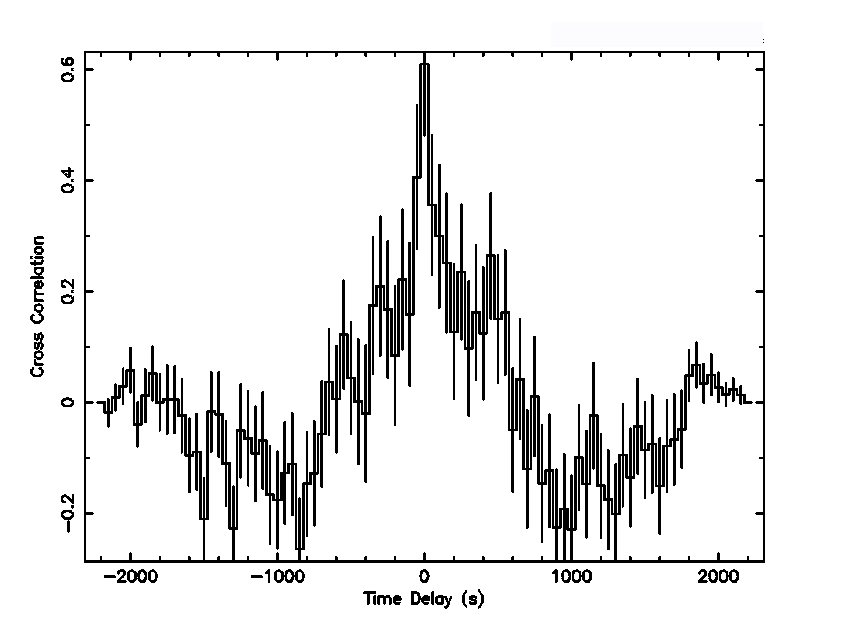}
    \includegraphics[width=2in, height=2in]{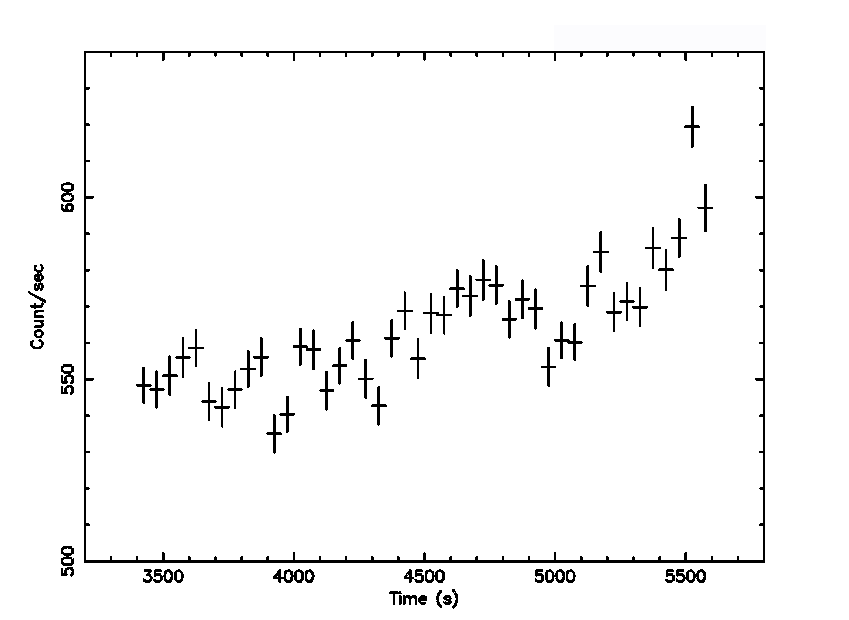}
    \includegraphics[width=2in, height=2in]{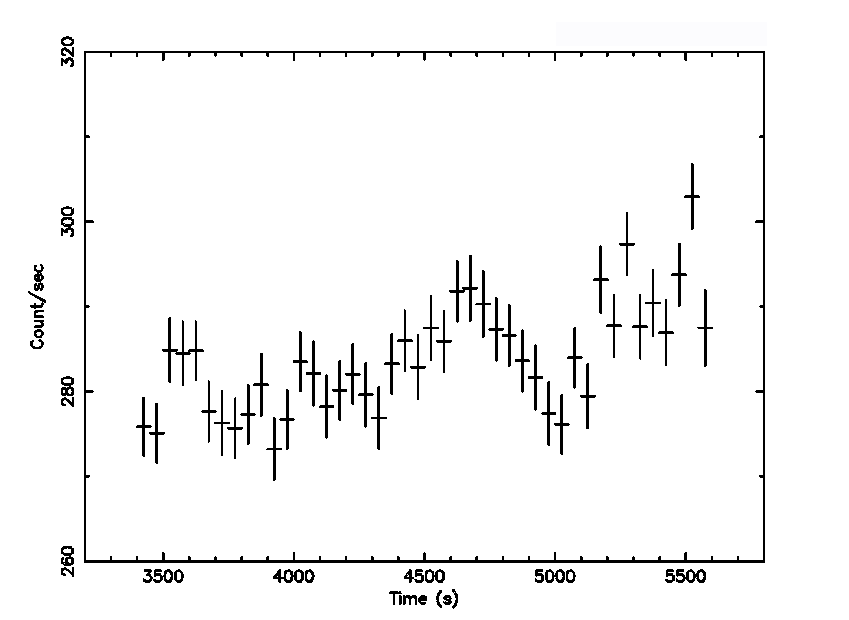}
    \includegraphics[width=2in, height=2in]{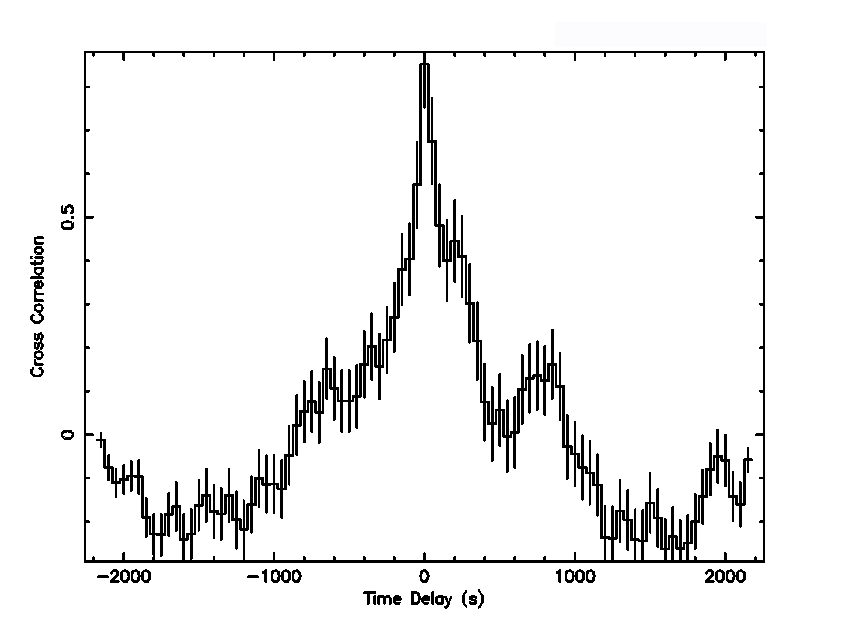}\\
    \includegraphics[width=2in, height=2in]{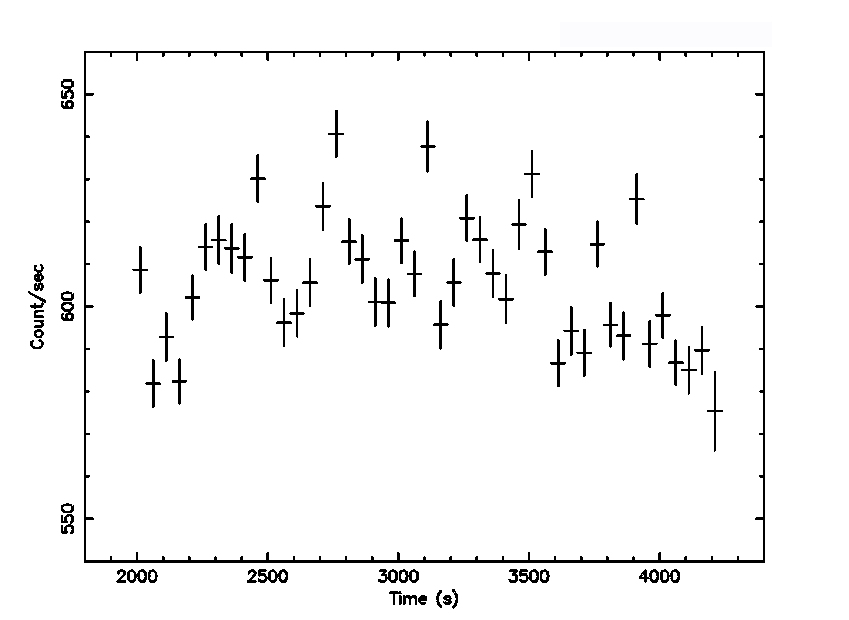}
    \includegraphics[width=2in, height=2in]{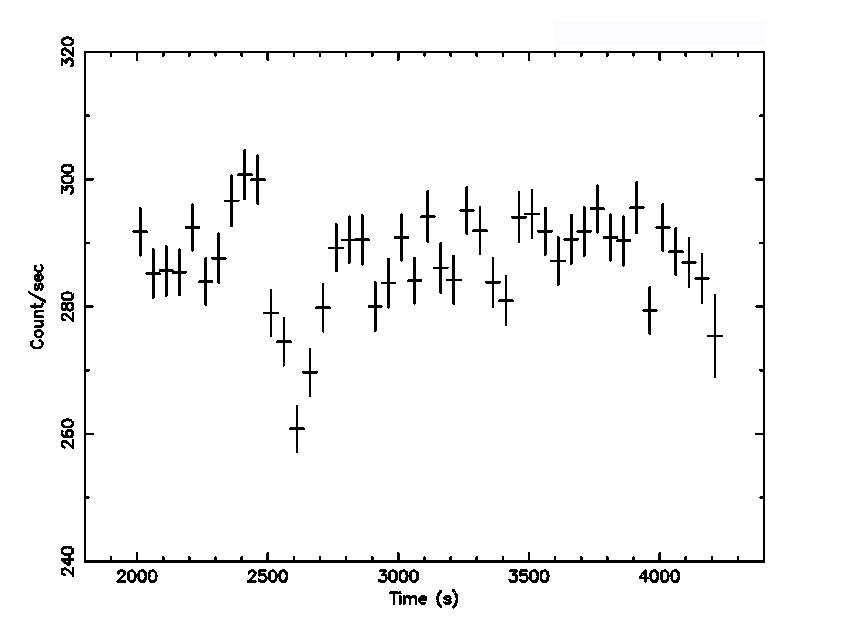}
    \includegraphics[width=2in, height=2in]{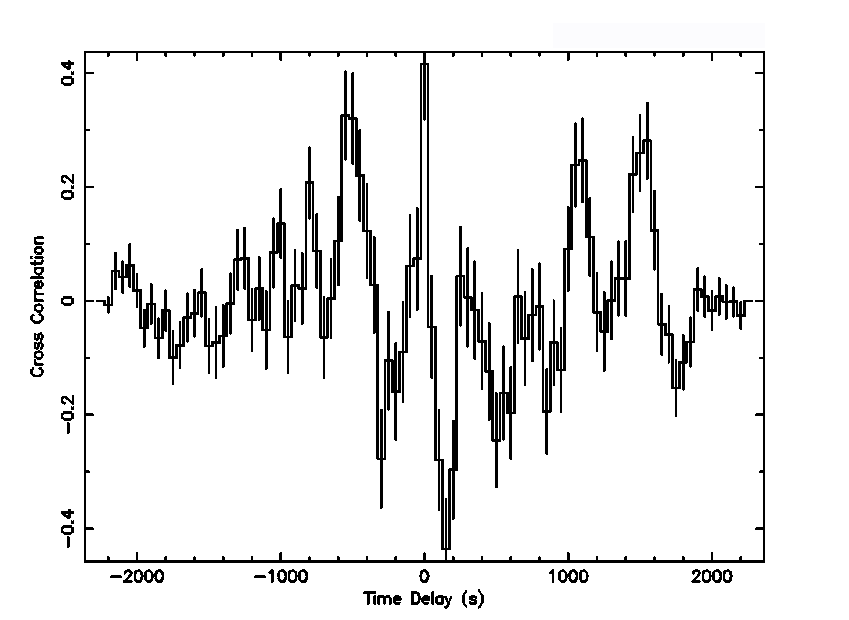}
 \caption{The first two plots in each row are the soft and hard LCs, respectively, and the third is the CCF. All rows belong to the FB.}
 % \end{minipage}
   
\end{figure*}

\begin{figure*}
    \ContinuedFloat
    \centering
  %\begin{minipage}[b]{0.8\linewidth}
    \includegraphics[width=2in, height=2in]{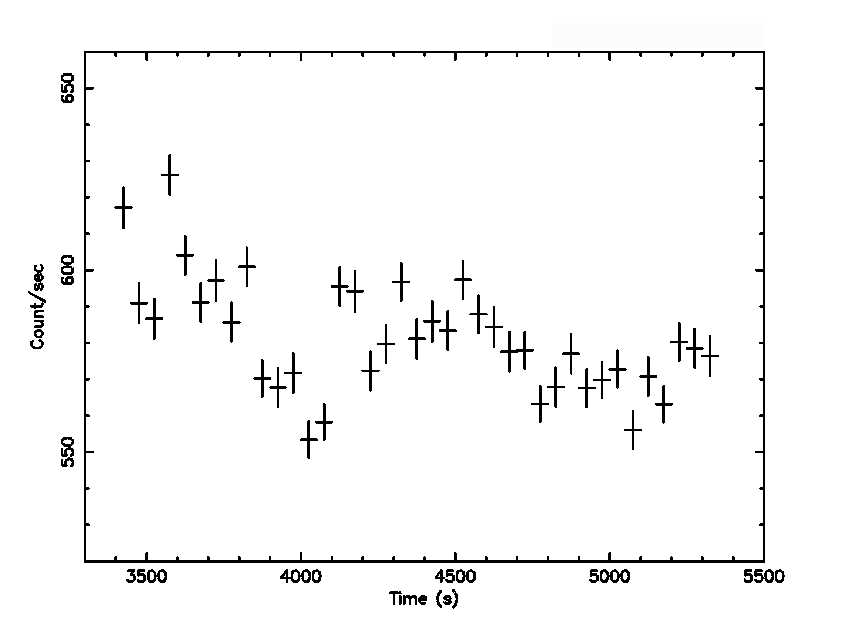}
    \includegraphics[width=2in, height=2in]{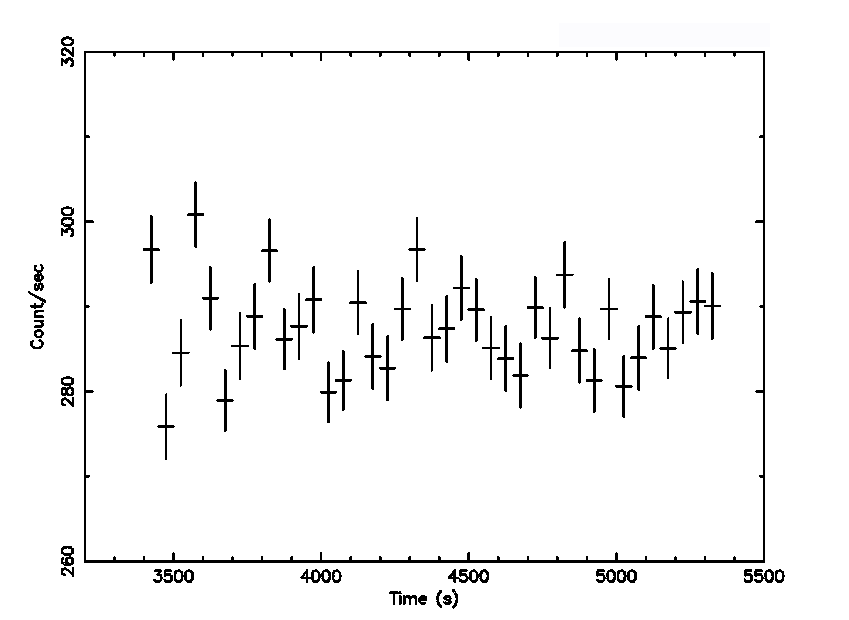}
    \includegraphics[width=2in, height=2in]{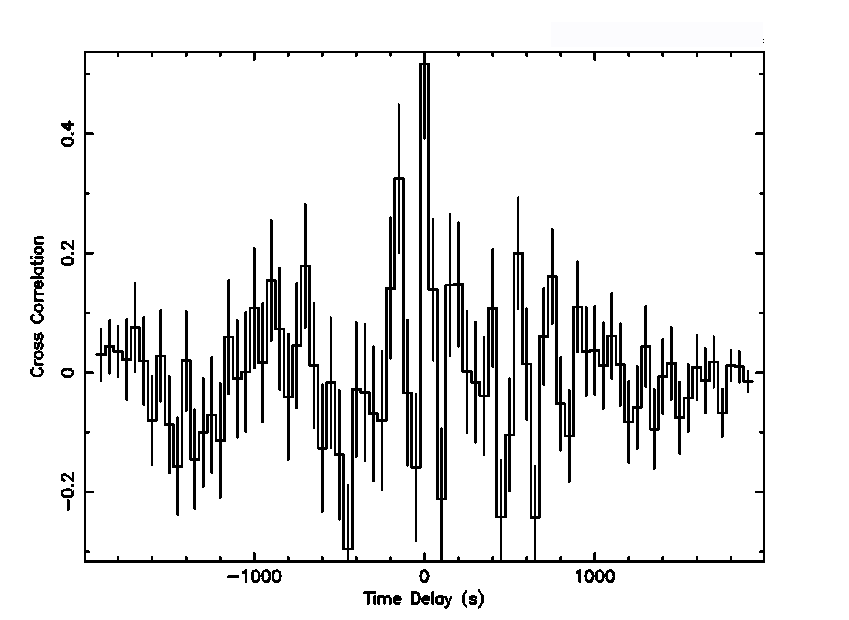}\\
    \includegraphics[width=2in, height=2in]{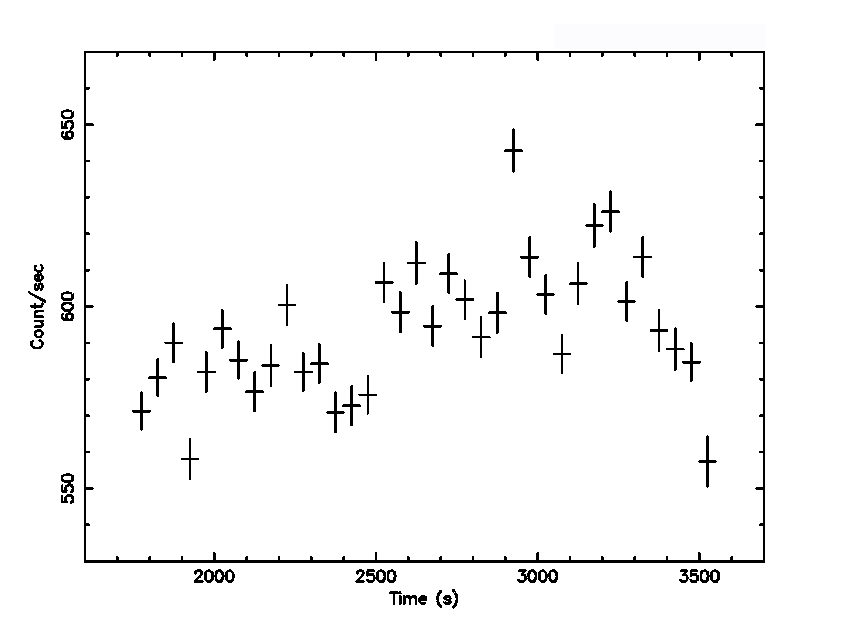}
    \includegraphics[width=2in, height=2in]{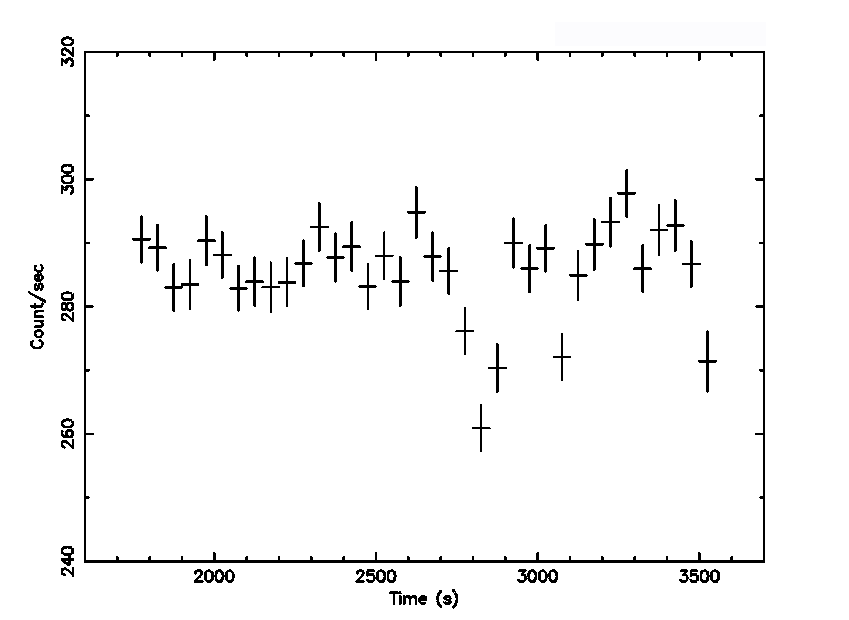}
    \includegraphics[width=2in, height=2in]{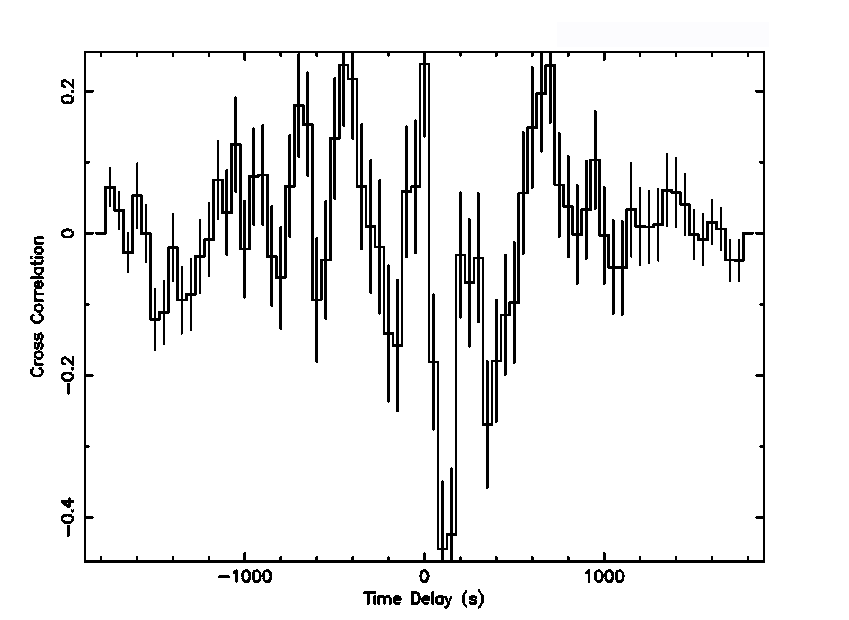}\\
    \includegraphics[width=2in, height=2in]{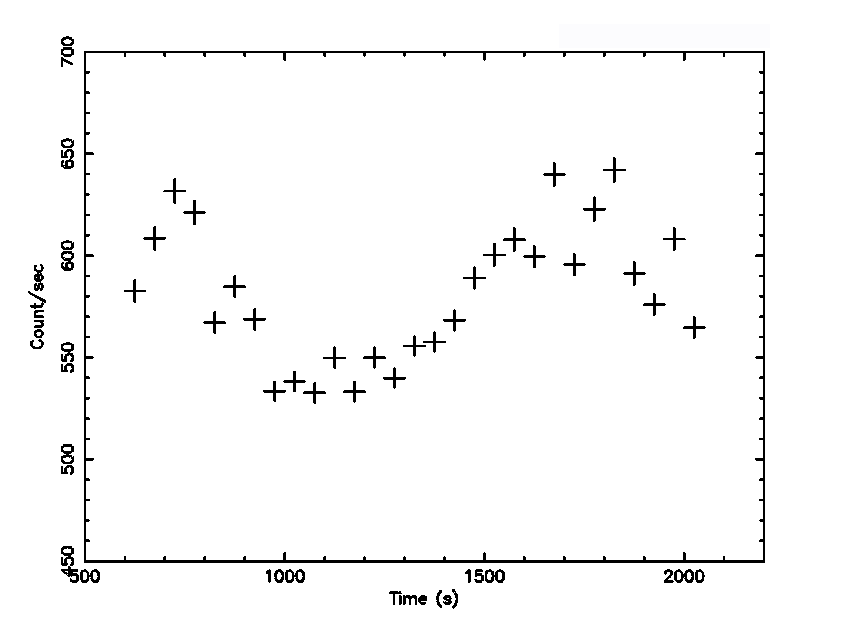}
    \includegraphics[width=2in, height=2in]{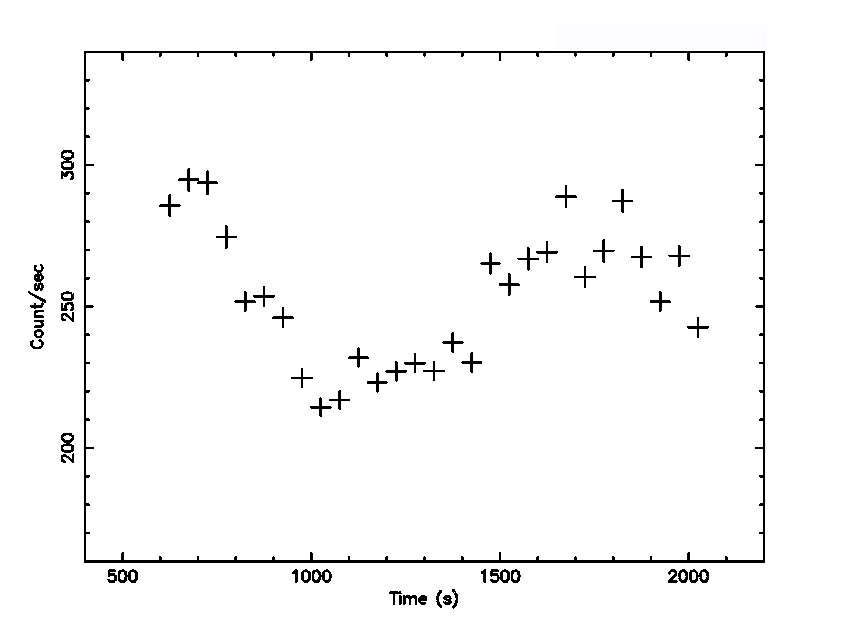}
    \includegraphics[width=2in, height=2in]{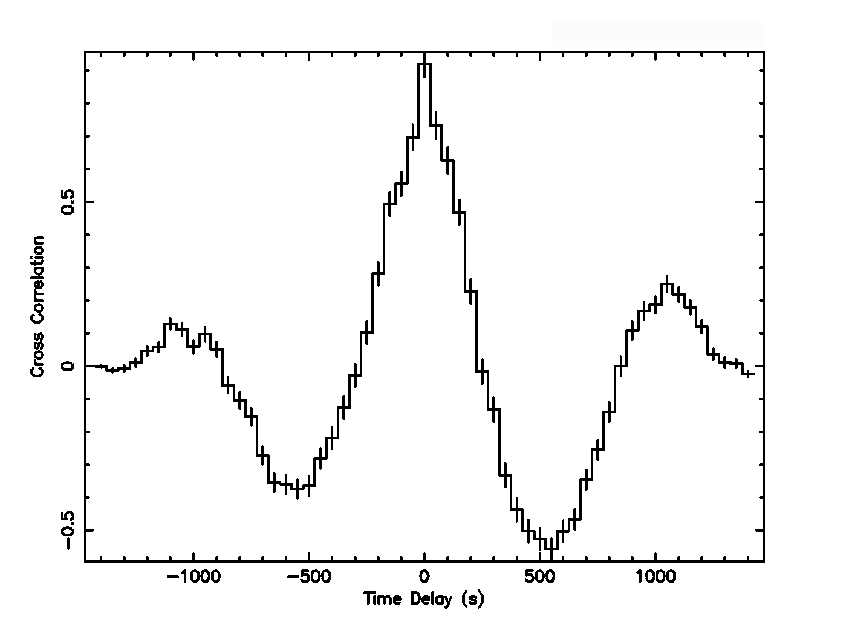}
    \includegraphics[width=2in, height=2in]{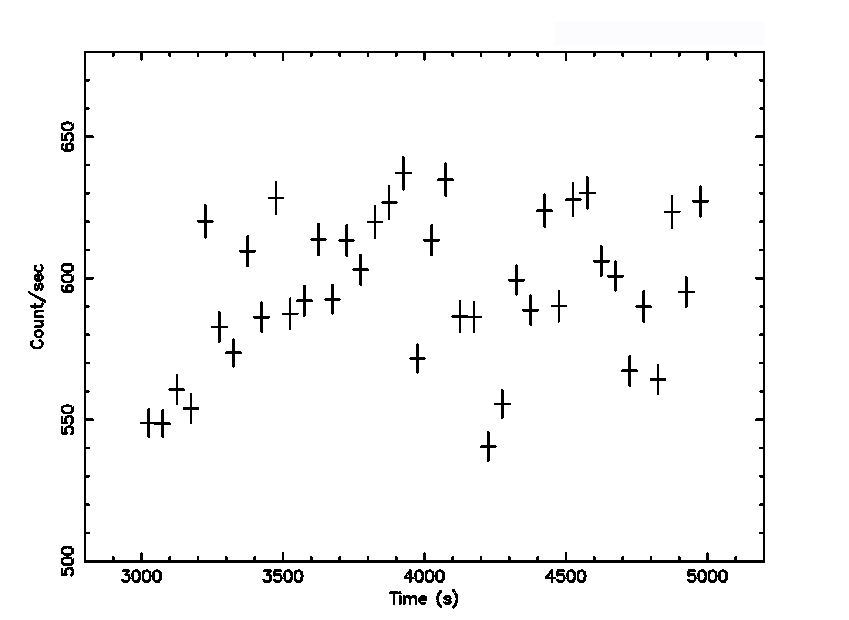}
    \includegraphics[width=2in, height=2in]{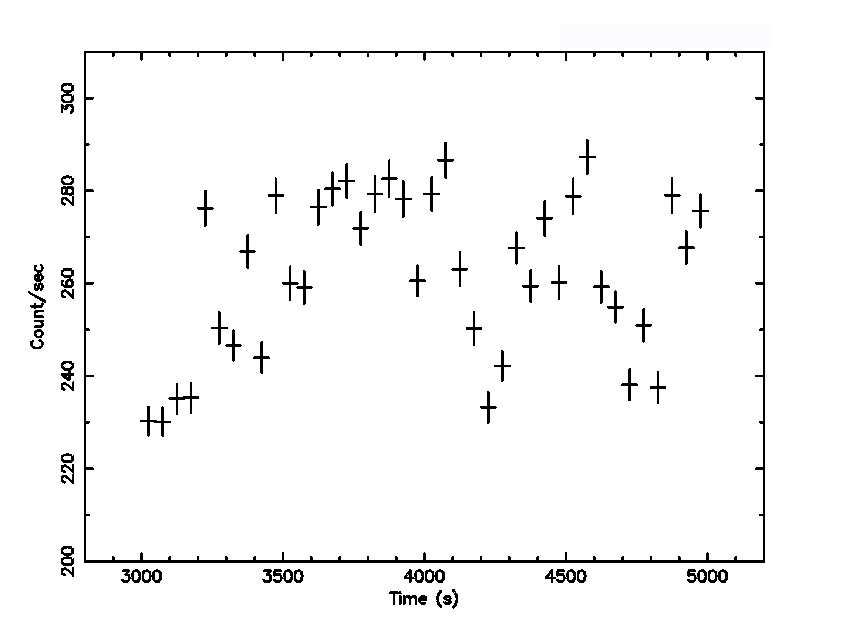}
    \includegraphics[width=2in, height=2in]{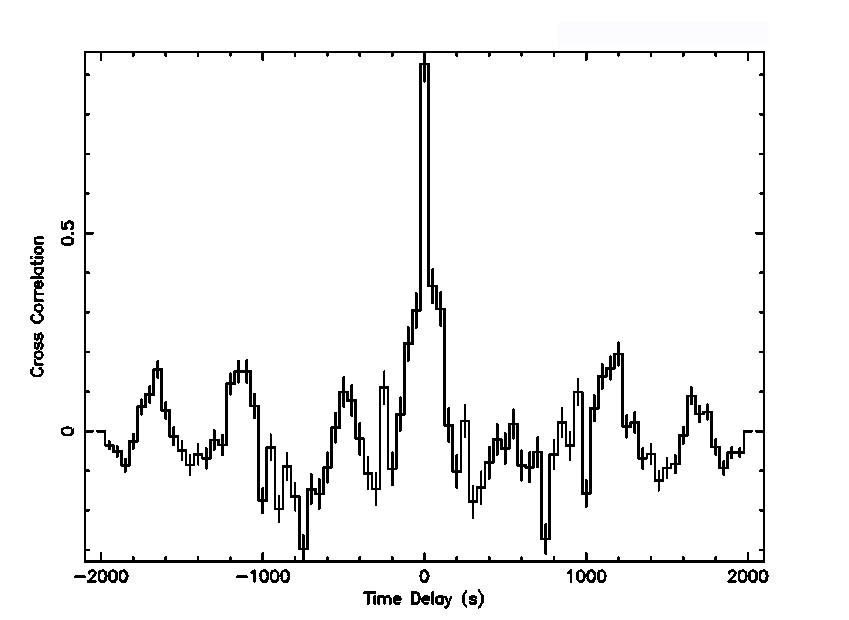}\\
\caption{Continued.. The top 2 rows display the LCs and CCFs of the FB, and the bottom 2 panels show the NB section.}
 % \end{minipage}
   
\end{figure*}

\begin{figure*}
    \ContinuedFloat
    \centering
  %\begin{minipage}[b]{0.8\linewidth}
    \includegraphics[width=2in, height=2in]{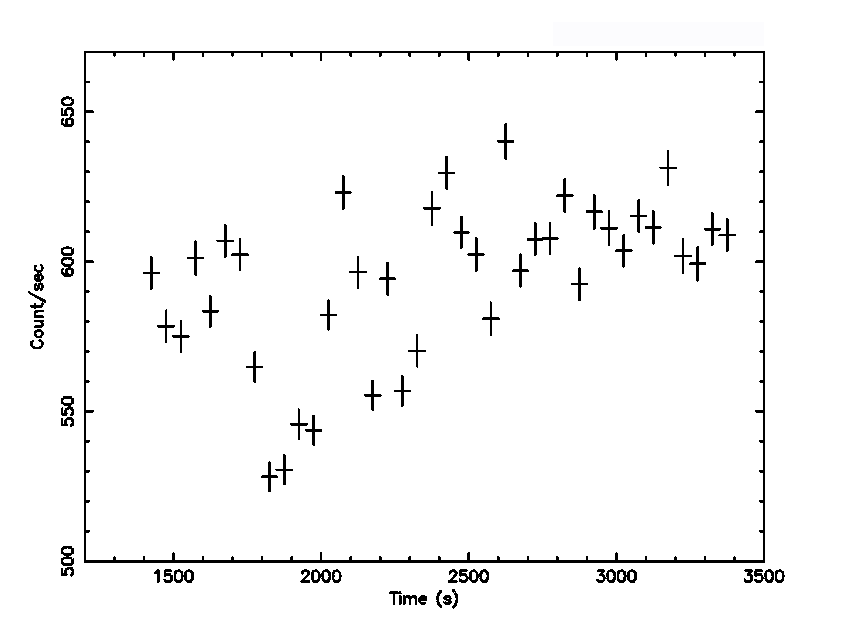}
    \includegraphics[width=2in, height=2in]{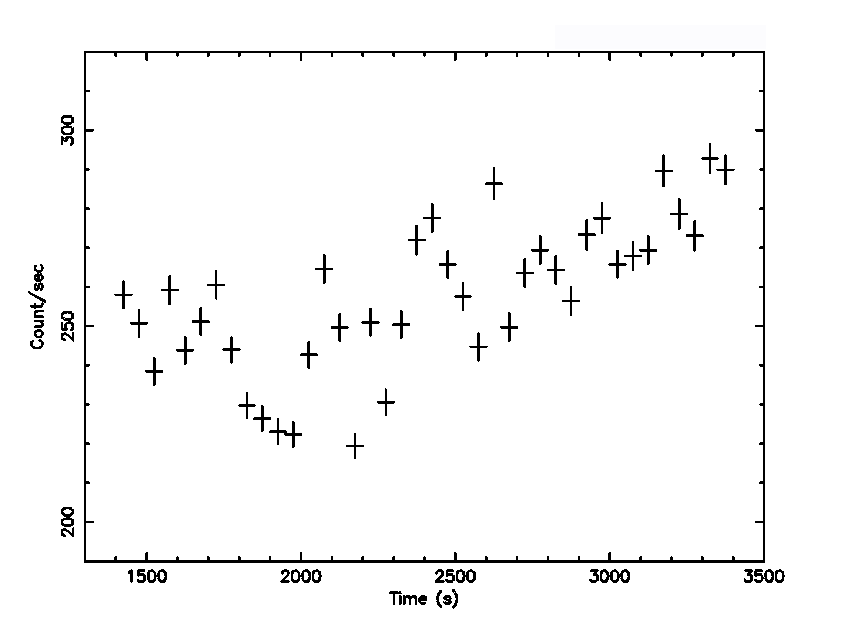}
    \includegraphics[width=2in, height=2in]{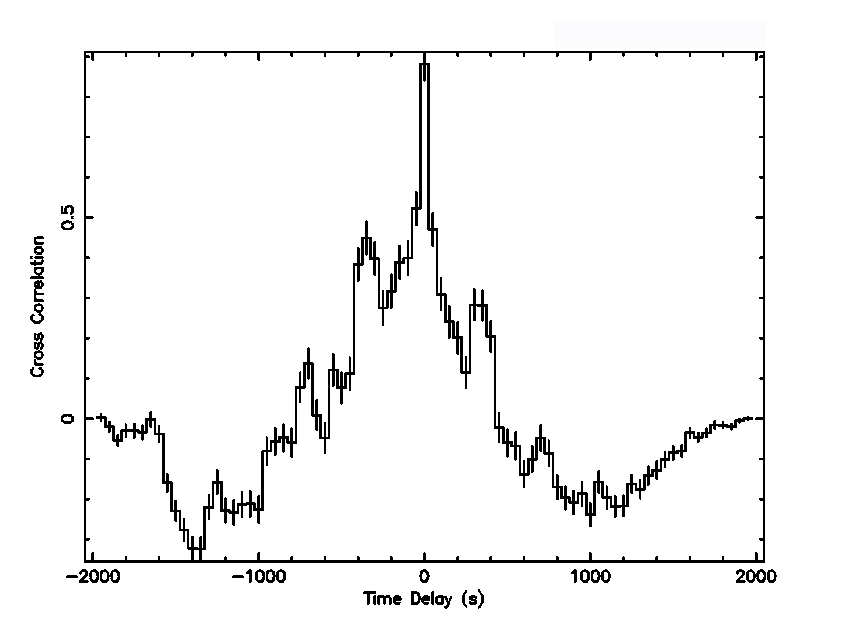}
 \caption{continued.. NB section LCs and CCF.}
 % \end{minipage}
   
\end{figure*}

\begin{comment}
\begin{figure*}[hbt!]
    \centering
    \includegraphics[width=6in, height=3in]{Lag_windows_old.png}
  %\begin{minipage}[b]{0.8\linewidth}

    \includegraphics[width=2in, height=2in]{soft3.png}
    \includegraphics[width=2in, height=2in]{hard3.png}
    \includegraphics[width=2in, height=2in]{ccf3.png}\\
    \includegraphics[width=2in, height=2in]{soft5.png}
    \includegraphics[width=2in, height=2in]{hard5.png}
    \includegraphics[width=2in, height=2in]{ccf5.png}
    \includegraphics[width=2in, height=2in]{soft8.png}
    \includegraphics[width=2in, height=2in]{hard8.png}
    \includegraphics[width=2in, height=2in]{ccf8.png}\\
 \caption{This first plot is the entire observation's LC with Sections A, B, and C highlighted. In the following plots, the first two images in each row are the soft and hard LCs, respectively, and the third is the CCF plot. Row 2, 3 and 4 belong to Sections A, B and C, respectively.}

 % \end{minipage}
  
\end{figure*}
\end{comment}

% --- PAGE 1: Subfigure 4(a) ---
\begin{figure*}[hbt!]
    \centering

    % Group 1: All 10 plots belonging to Subfigure 4(a)
    \begin{subfigure}[b]{\textwidth}
        \centering
        % Top large observation plot
        \includegraphics[width=6in, height=3in]{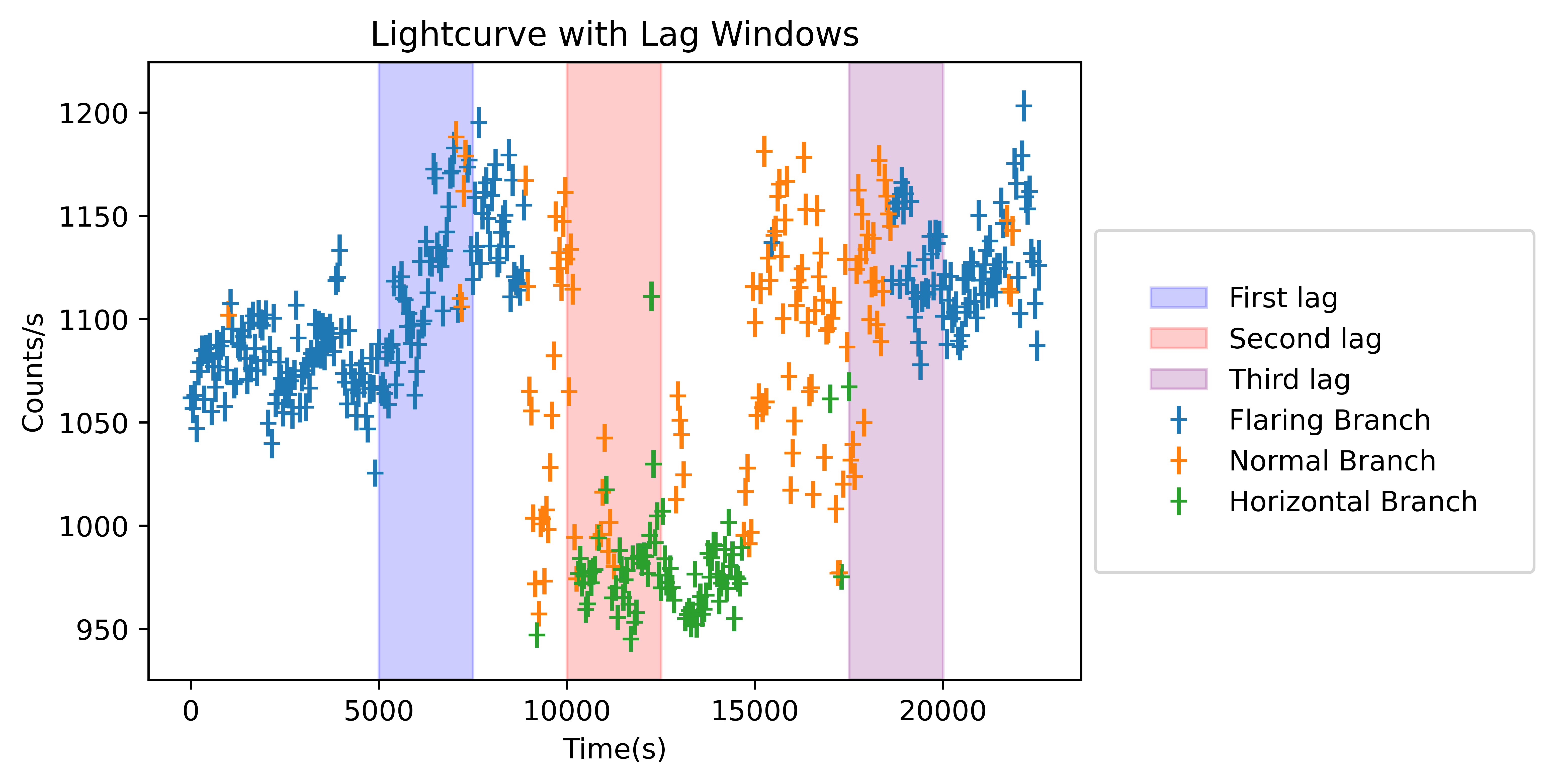} \\ \vspace{0.3cm}

        % Section A plots
        \includegraphics[width=2in, height=2in]{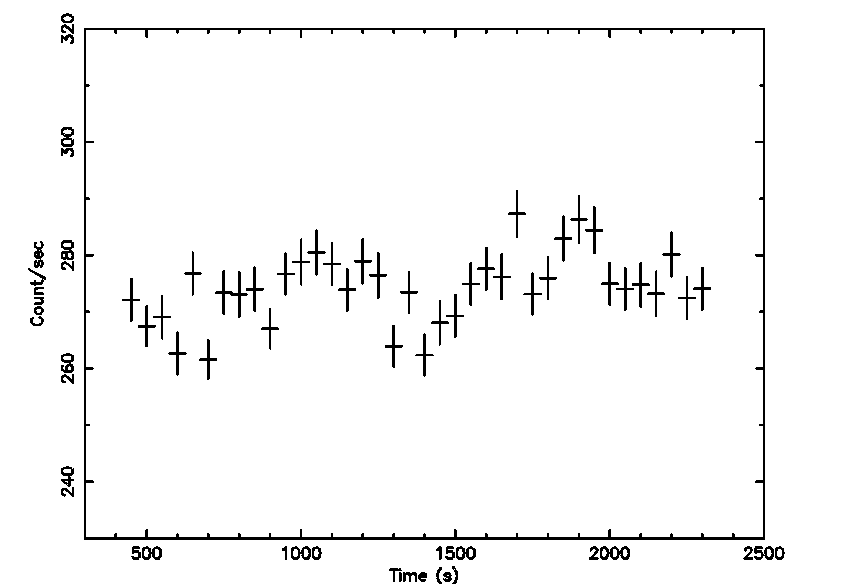}
        \includegraphics[width=2in, height=2in]{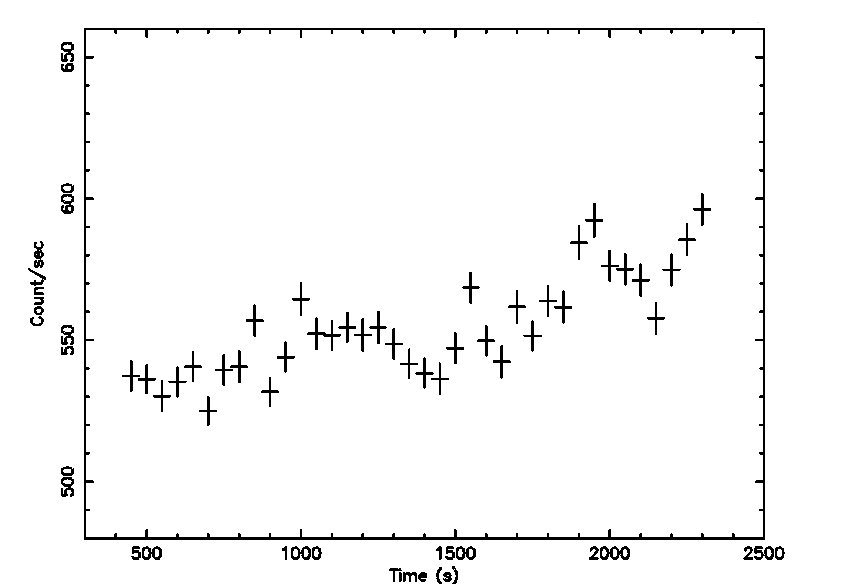}
        \includegraphics[width=2in, height=2in]{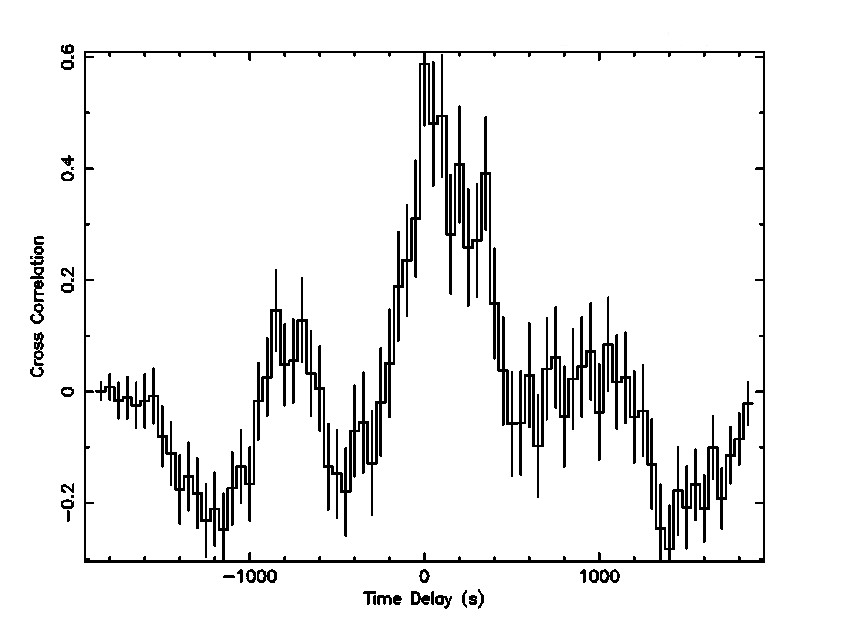} \\ \vspace{0.1cm}
        
        % Section B plots
        \includegraphics[width=2in, height=2in]{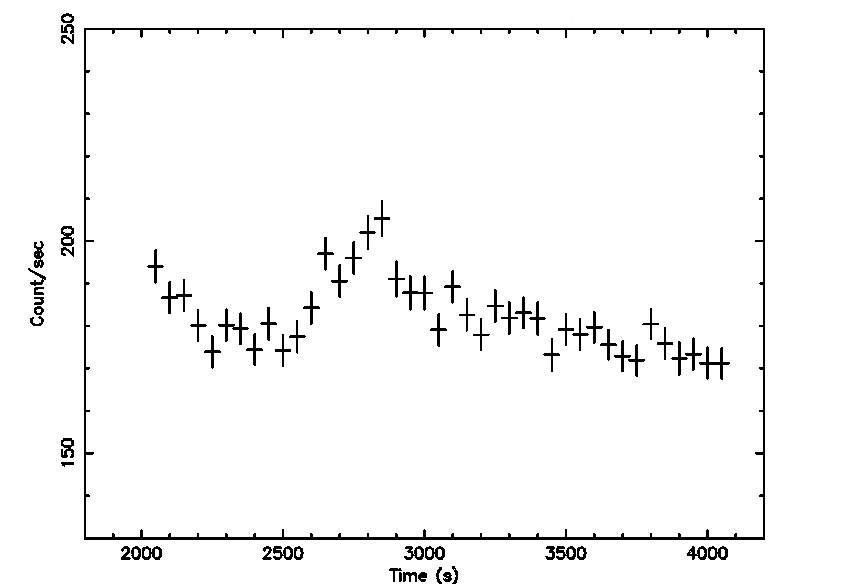}
        \includegraphics[width=2in, height=2in]{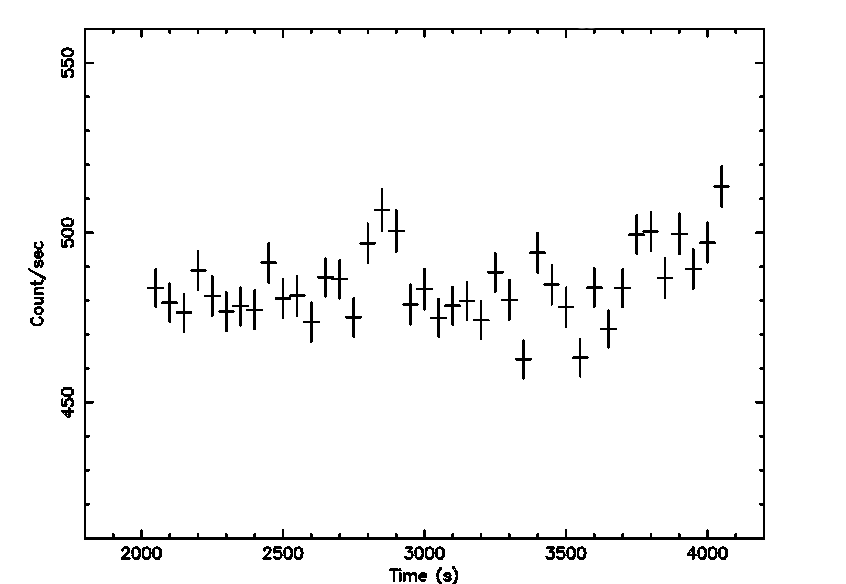}
        \includegraphics[width=2in, height=2in]{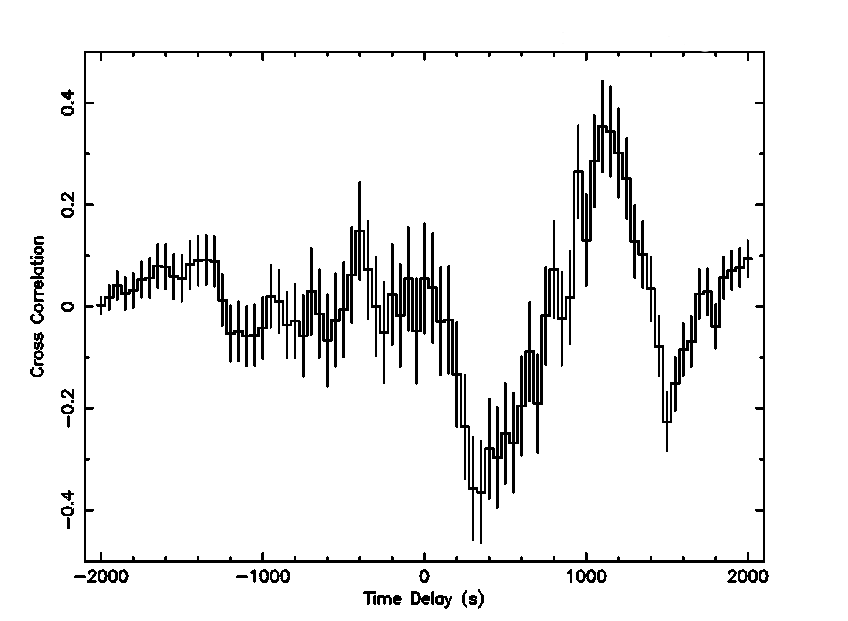} \\ \vspace{0.1cm}
        
        % Section C plots
        \includegraphics[width=2in, height=2in]{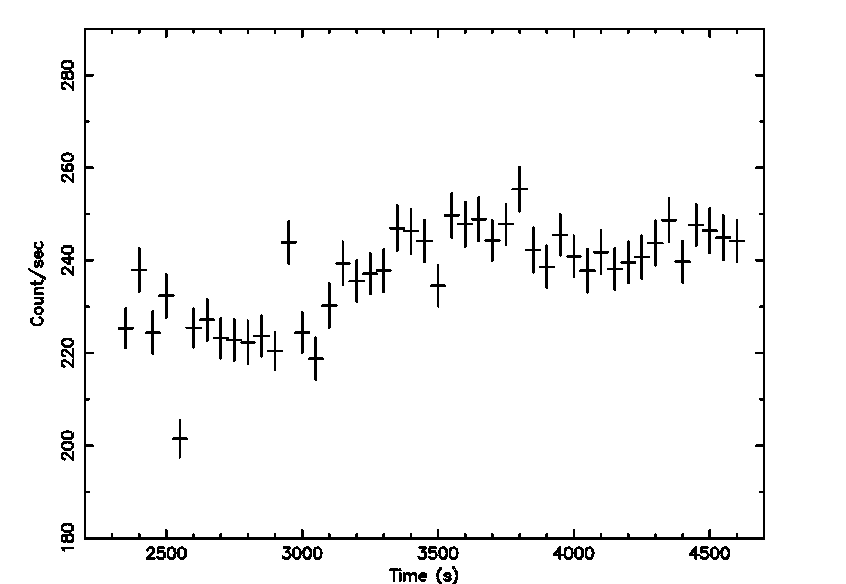}
        \includegraphics[width=2in, height=2in]{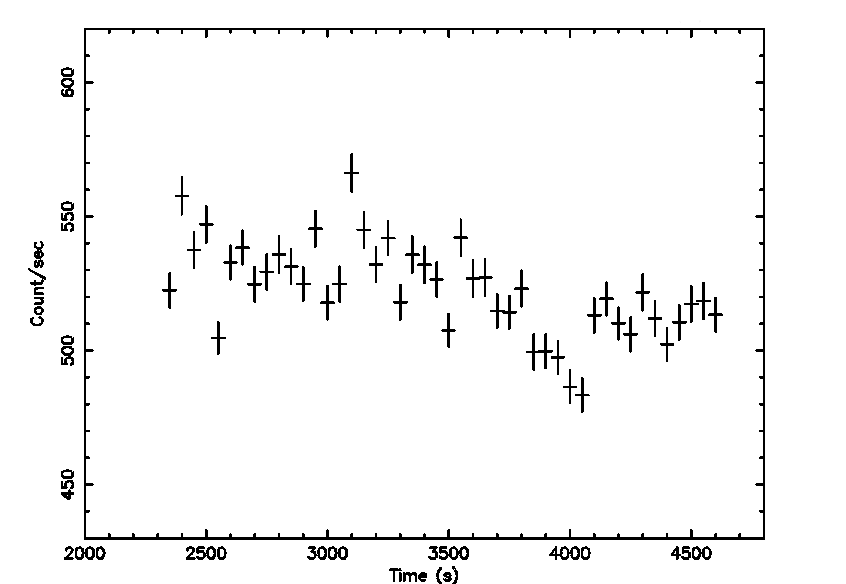}
        \includegraphics[width=2in, height=2in]{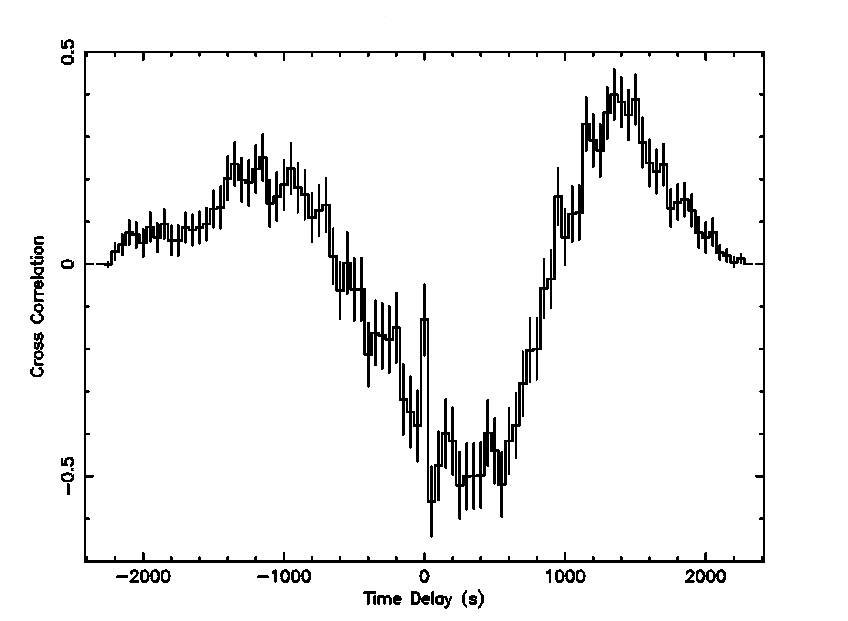}
        
        \subcaption{Entire observation LC (top) and individual section breakdowns for Sections A, B, and C.}
        \label{fig:obs_sections_a}
    \end{subfigure}

    % Main caption for the first part
    
\end{figure*}

% --- PAGE 2: Subfigure 4(b) ---
\begin{figure*}[hbt!]
    \ContinuedFloat % <--- CRITICAL: Tells LaTeX this is still Figure 4!
    \centering

    % Group 2: The next two images belonging to Subfigure 4(b)
    \begin{subfigure}[b]{\textwidth}
        \centering
        \includegraphics[width=3in, height=2in]{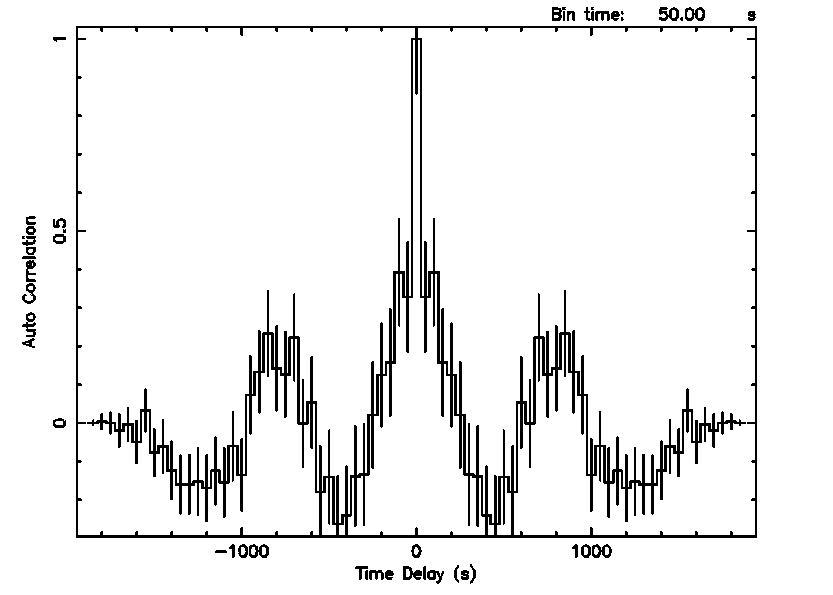}
        \includegraphics[width=3in, height=2in]{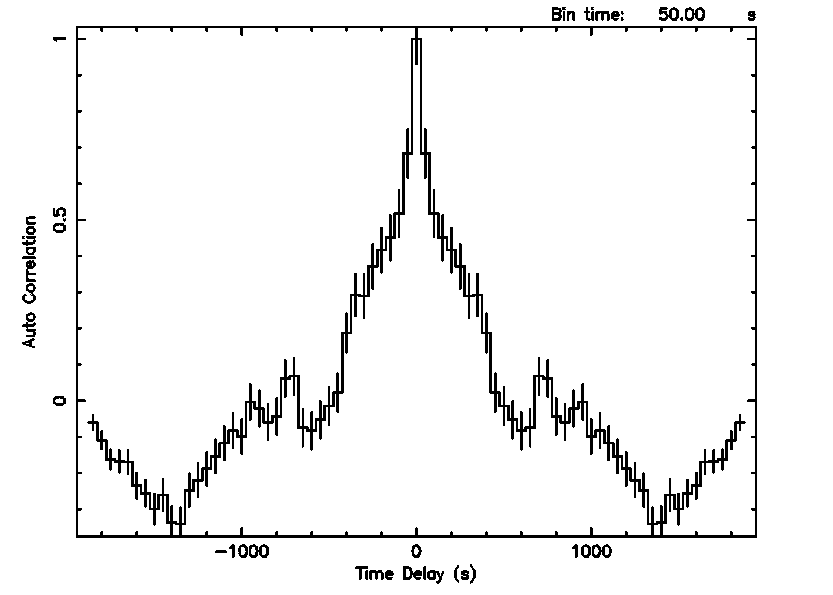}
        \includegraphics[width=3in, height=2in]{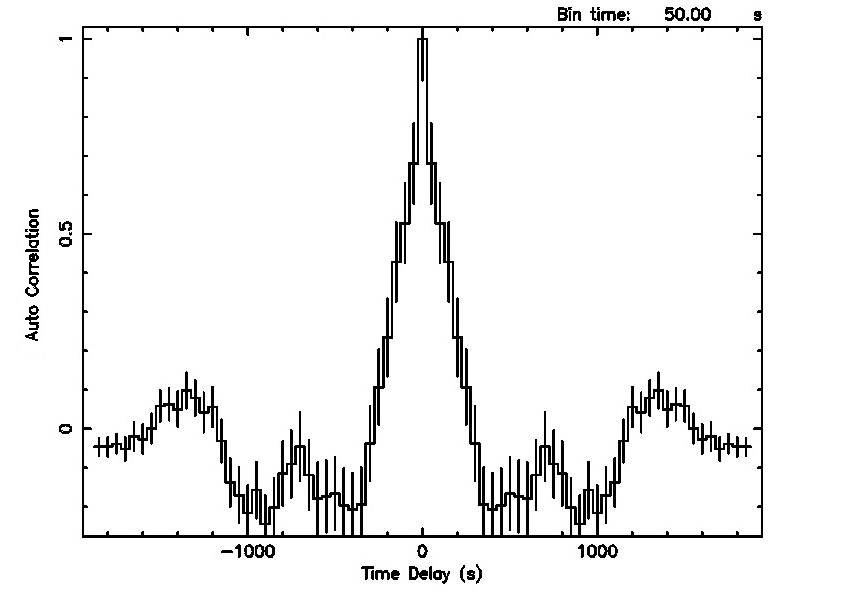}
        \includegraphics[width=3in, height=2in]{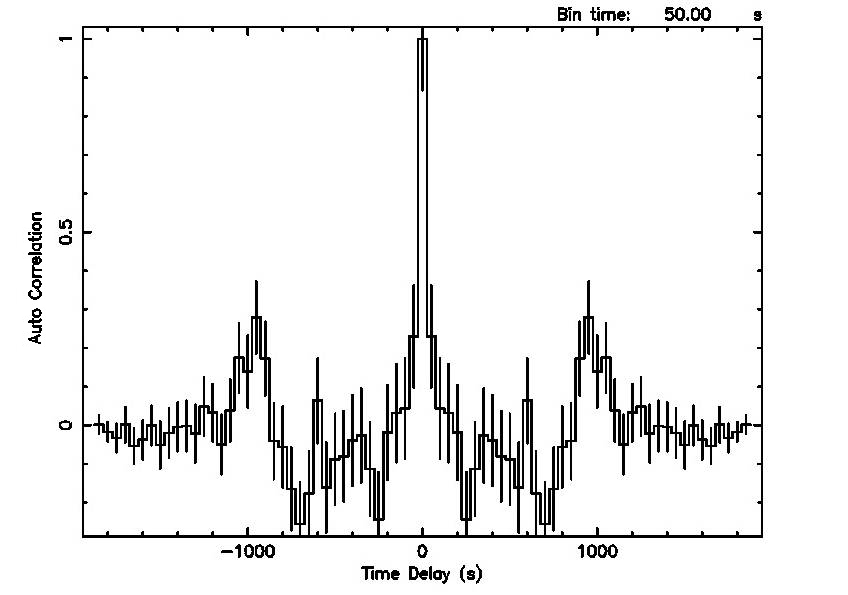}
        \includegraphics[width=3in, height=2in]{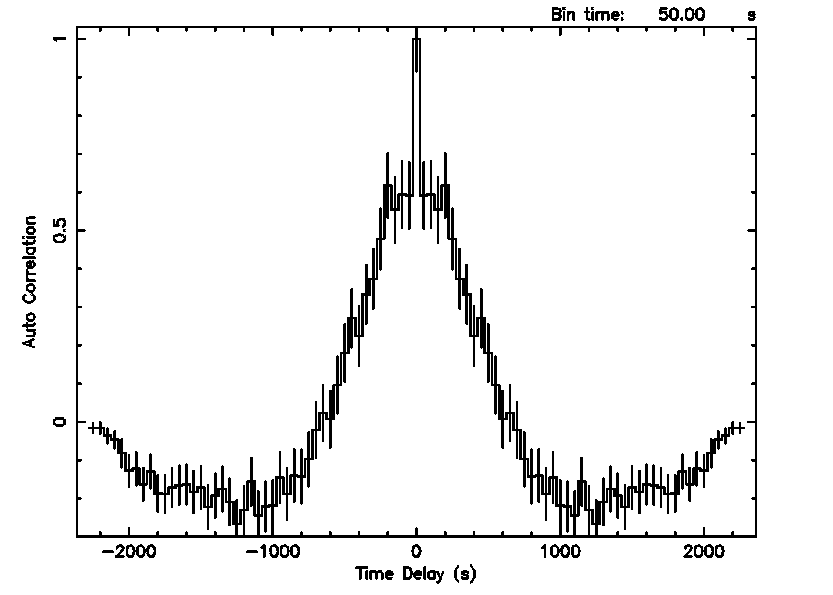}
        \includegraphics[width=3in, height=2in]{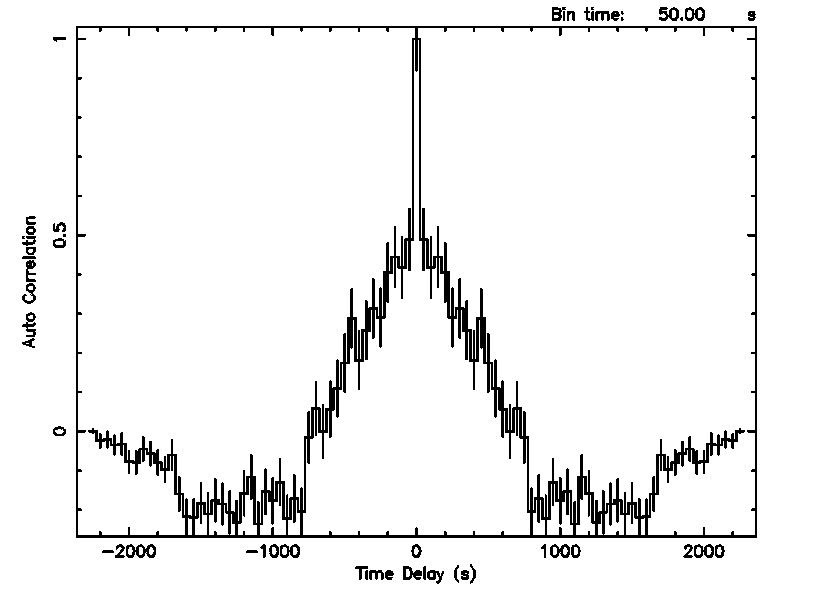}
        
        \subcaption{Soft and Hard LC Auto-Correlation Functions (ACF) for Section A, B and C.}
        \label{fig:obs_sections_b}
    \end{subfigure}

    % Main unified caption repeated/continued
    \caption{Overview of observations and lag window correlations. For (a): The first plot is the entire observation's LC with Sections A, B, and C highlighted. In the following grid plots, the first two images in each row are the soft and hard LCs, respectively, and the third is the CCF plot. Rows 2, 3, and 4 belong to Sections A, B, and C, respectively. (b) \textbf{ACF plots for Section A, B and C, soft band (left) and hard band (right)}}

\end{figure*}

\begin{figure*}[hbt!]
    \centering
    \includegraphics[width=3in, height=2in]{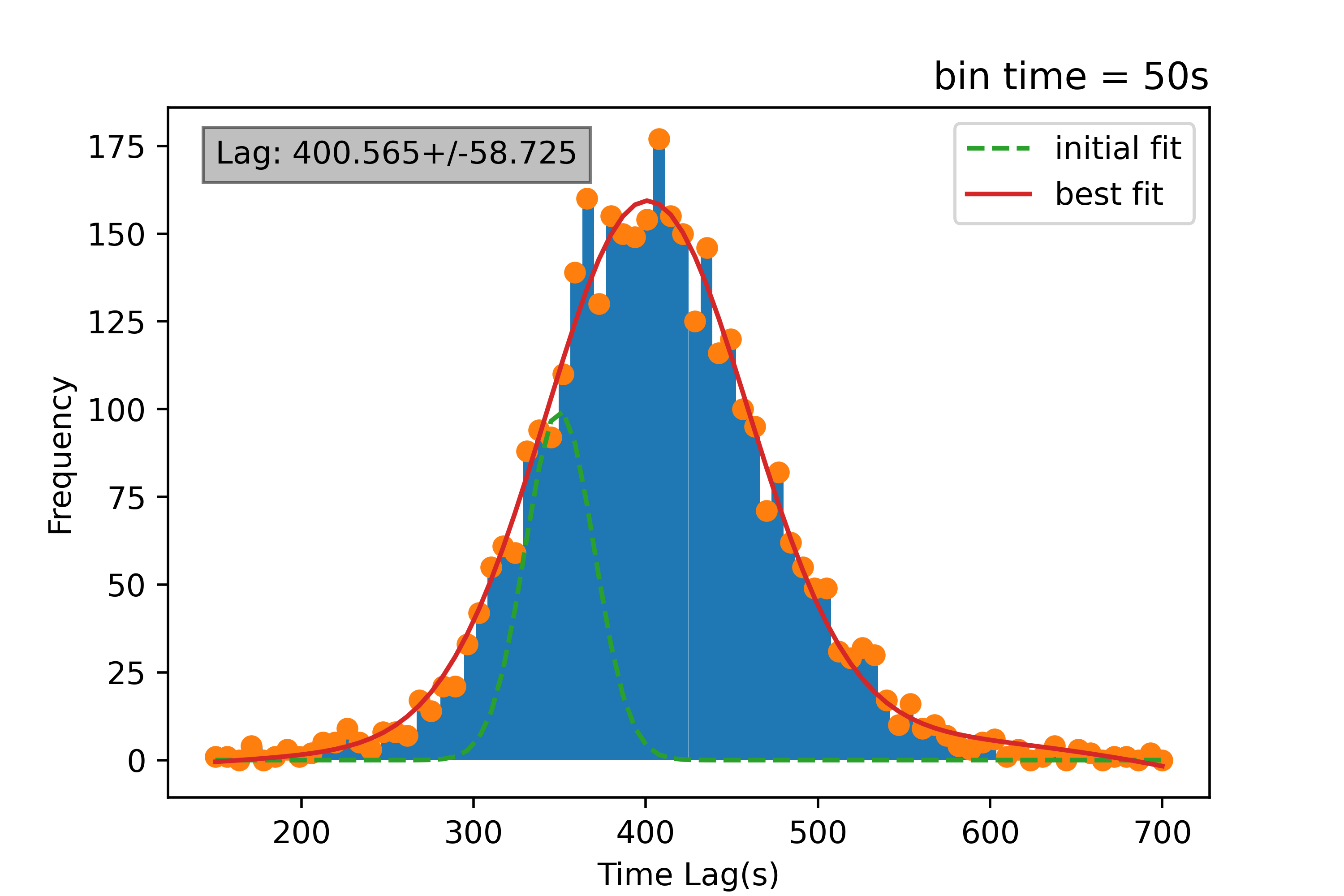}
    \label{fig:CC3}
    \includegraphics[width=3in, height=2in]{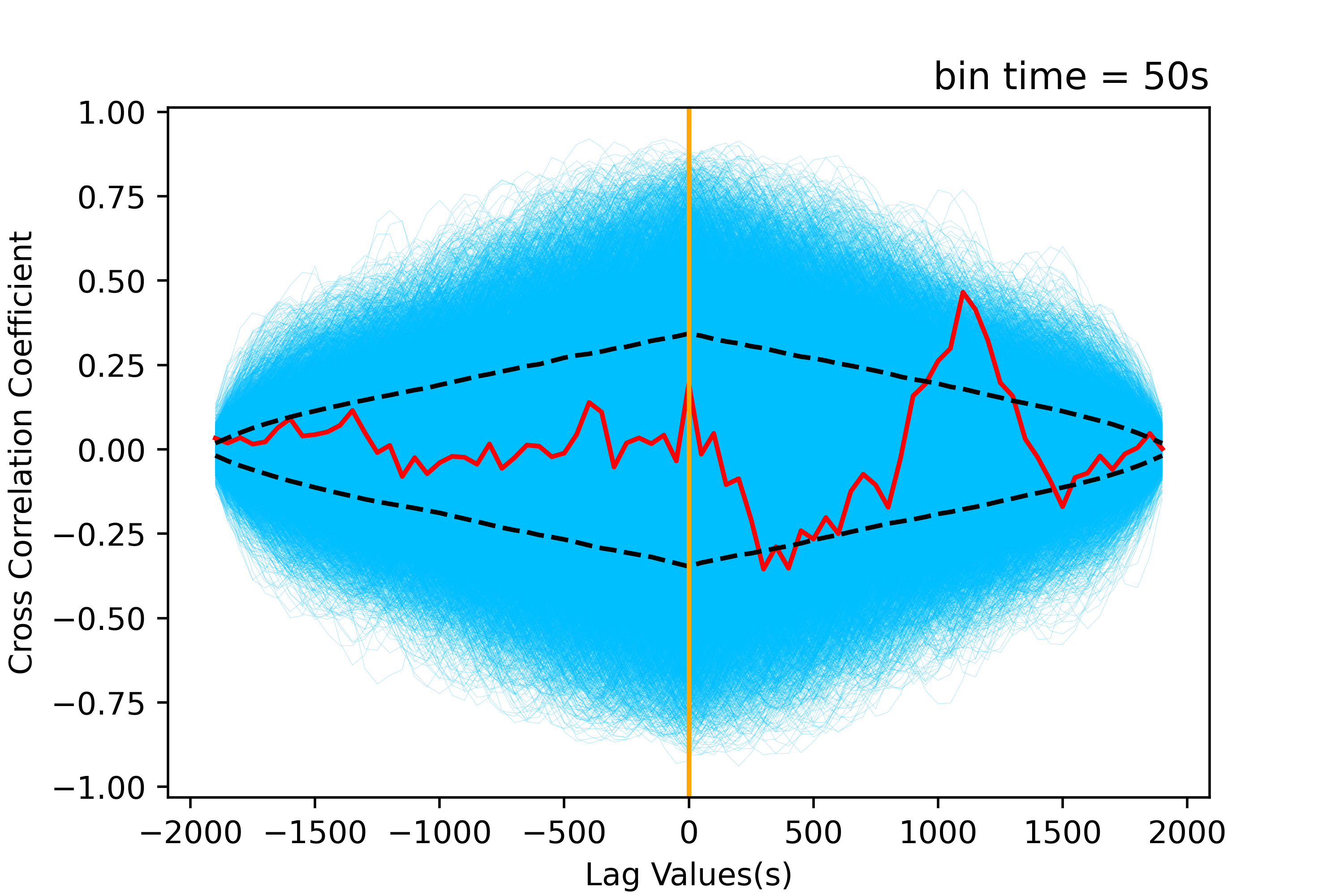}
    \includegraphics[width=3in, height=2in]{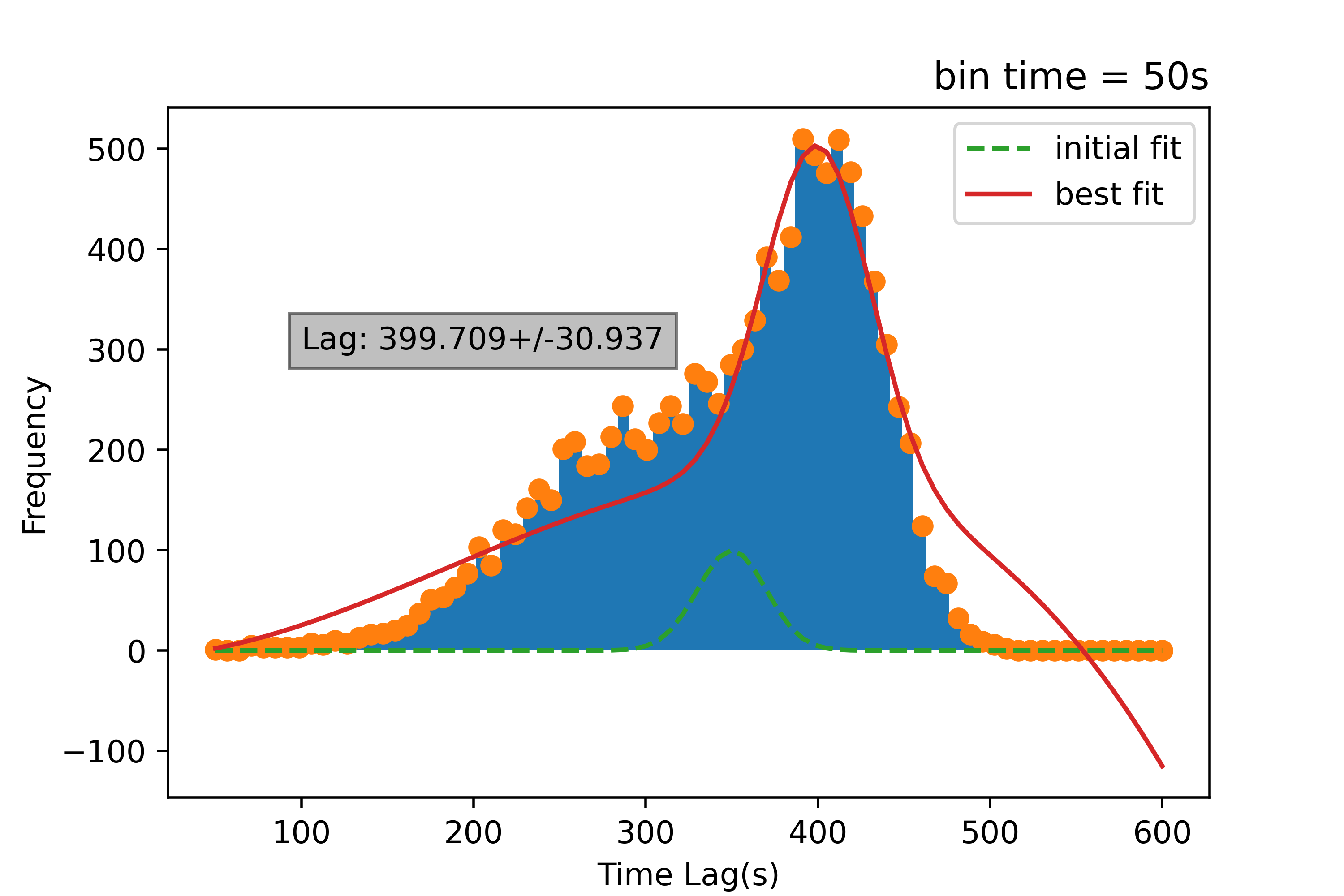}
    \label{fig:CC3}
    \includegraphics[width=3in, height=2in]{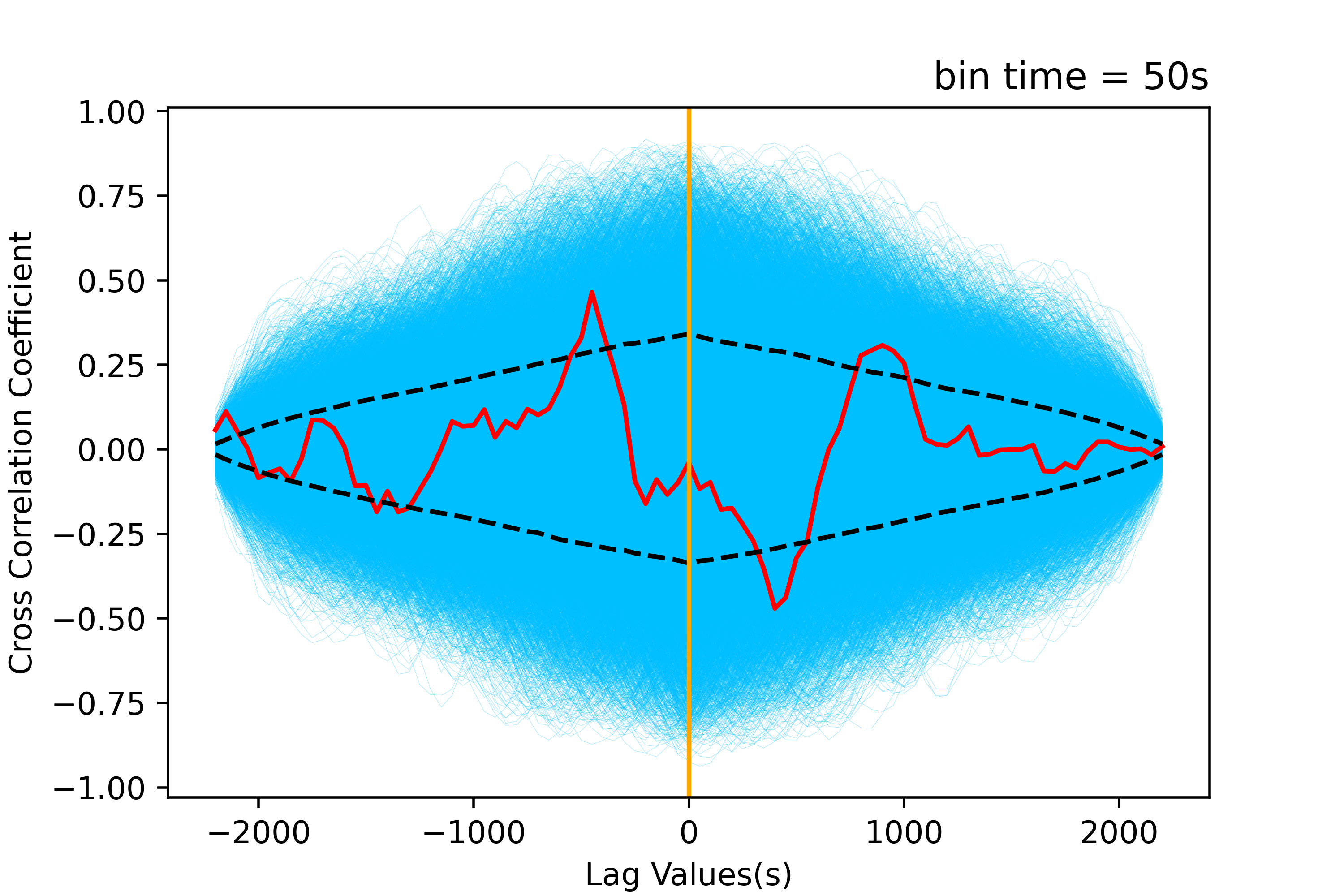}
    \caption{In each row, the left plot is of the histogram of the simulated CCFs and the right plot is of the associated Confidence Interval for that section using LCs simulated with the Timmer-Konig method. \citeyear{timmer1995generating} method, where the $95\%$ confidence windows are shown with the black dotted line and the simulated CCF plots are shown in a blue shaded area. The red colour plot overlaid is of the observed CCF.}
    \label{fig:CC3}
\end{figure*}

\begin{figure*}[hbt!]
    \centering
    \includegraphics[width=3in, height=2in]{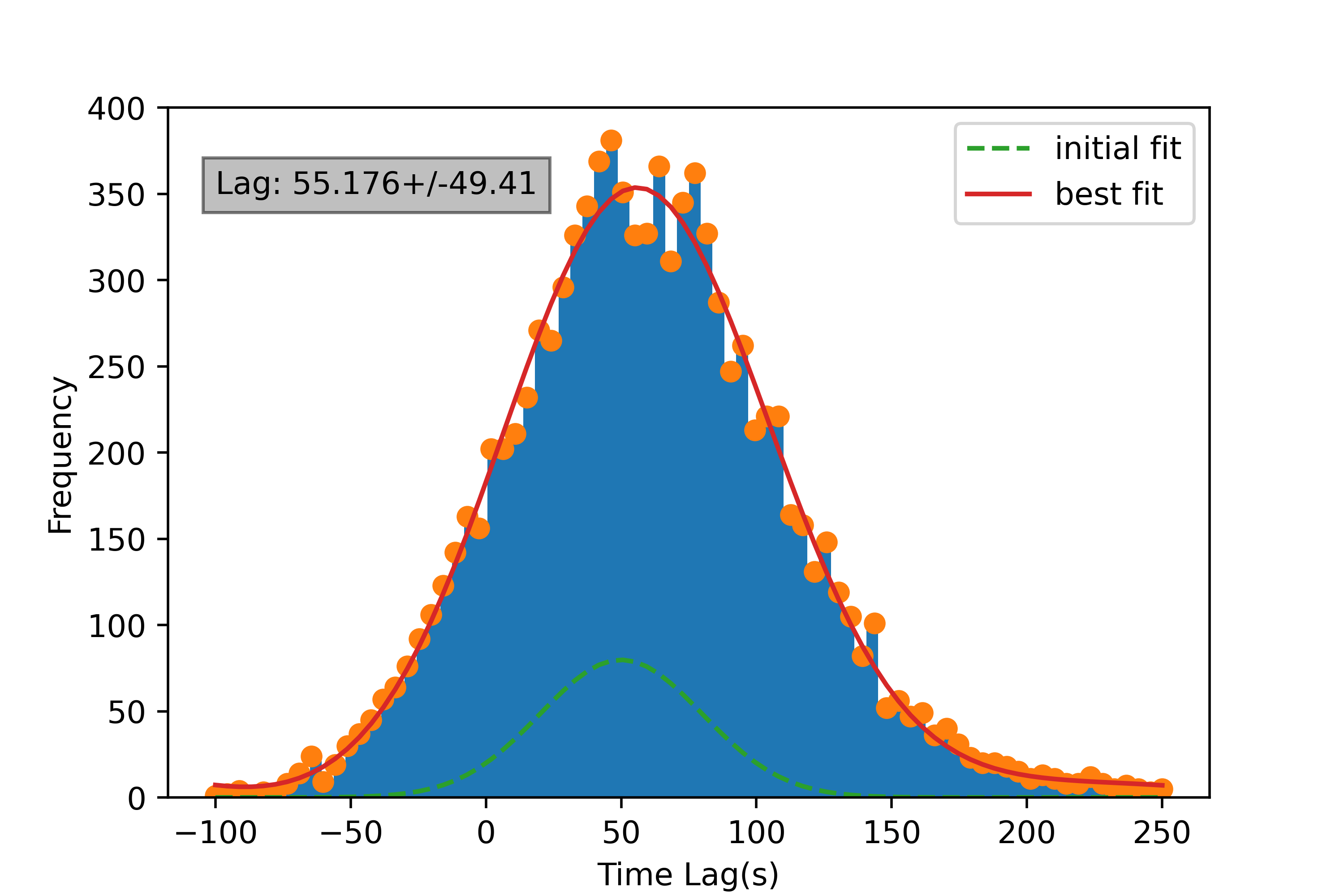}
    \includegraphics[width=3in, height=2in]{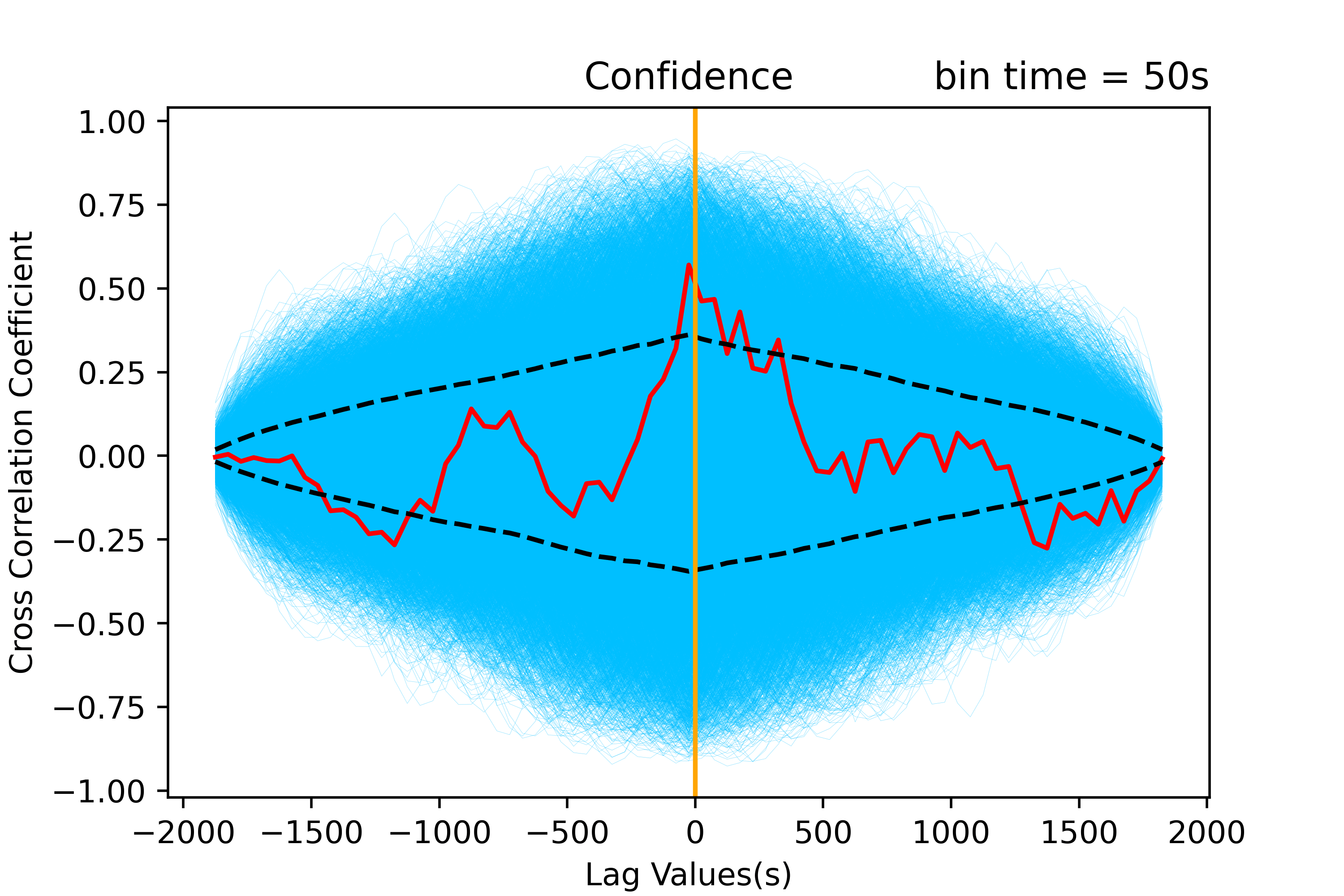}
    \includegraphics[width=3in, height=2in]{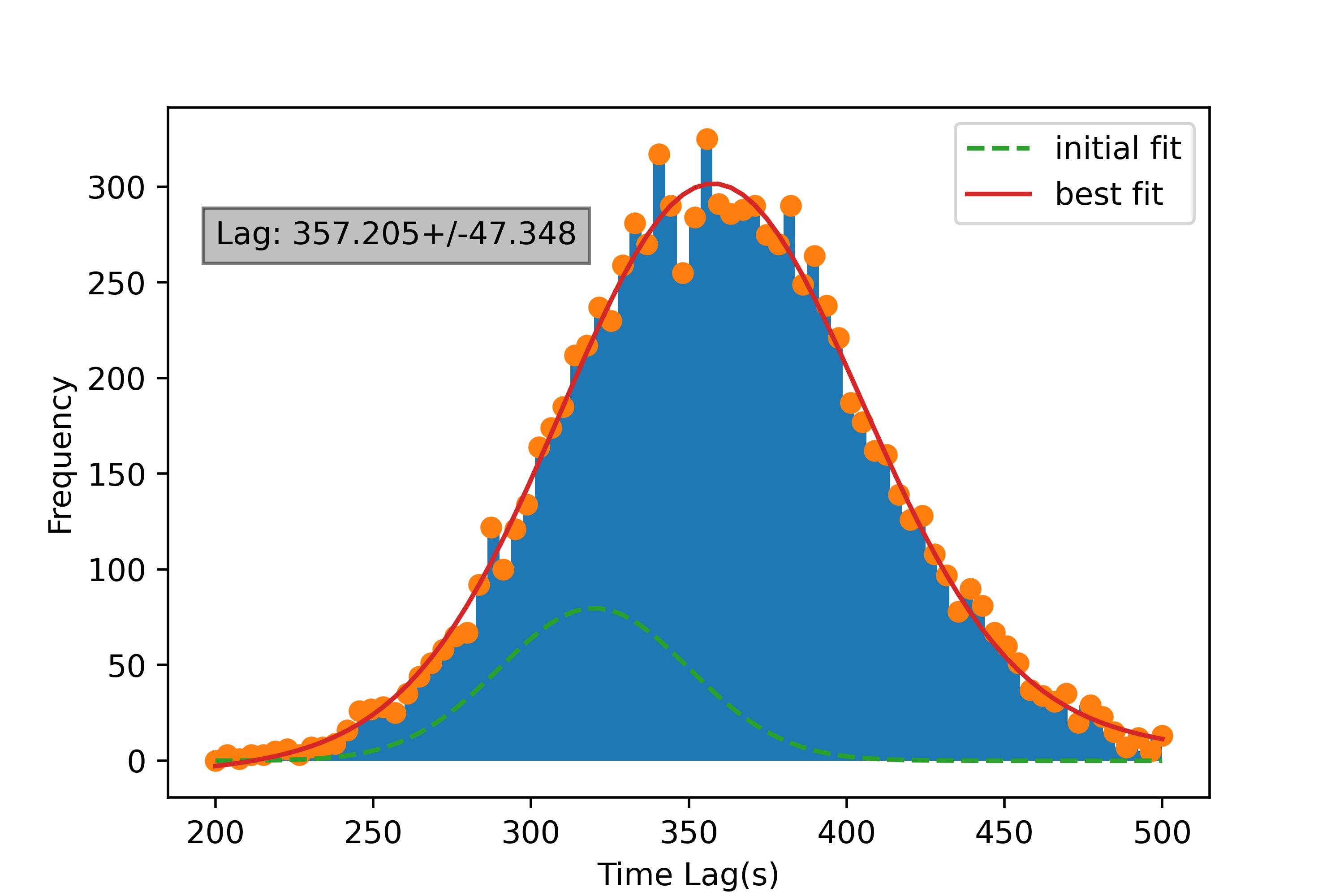}
    \includegraphics[width=3in, height=2in]{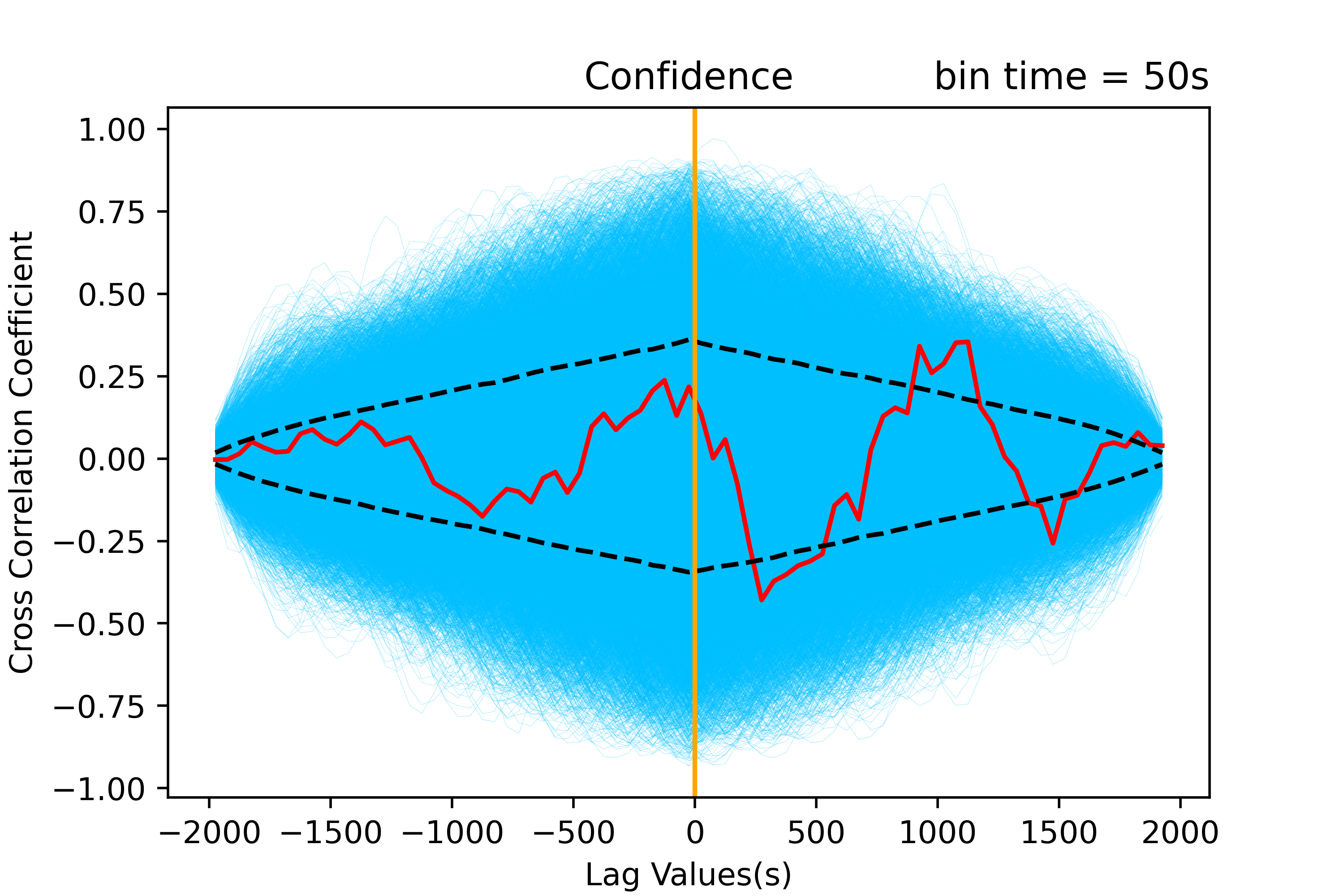}
    \includegraphics[width=3in, height=2in]{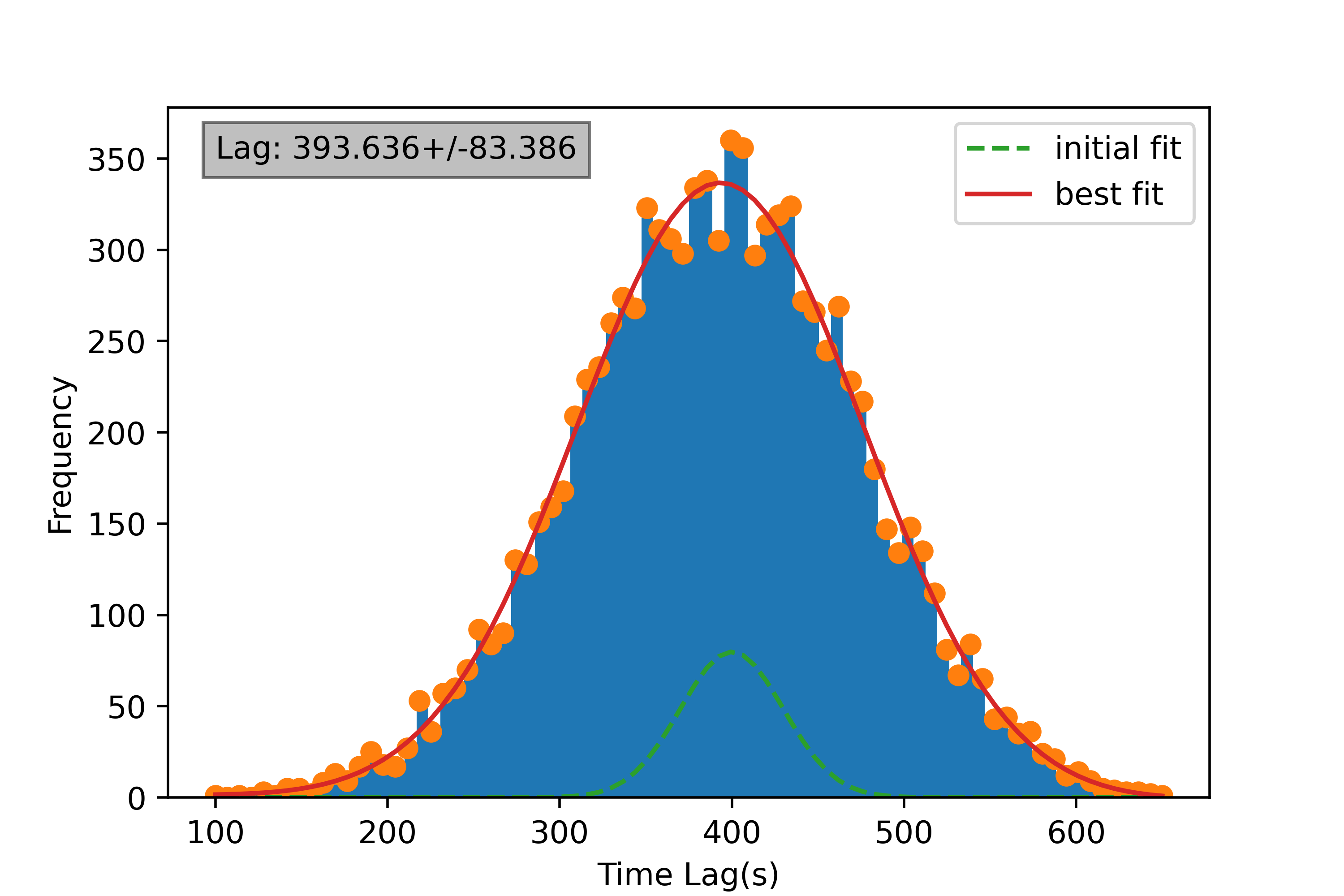}
    \includegraphics[width=3in, height=2in]{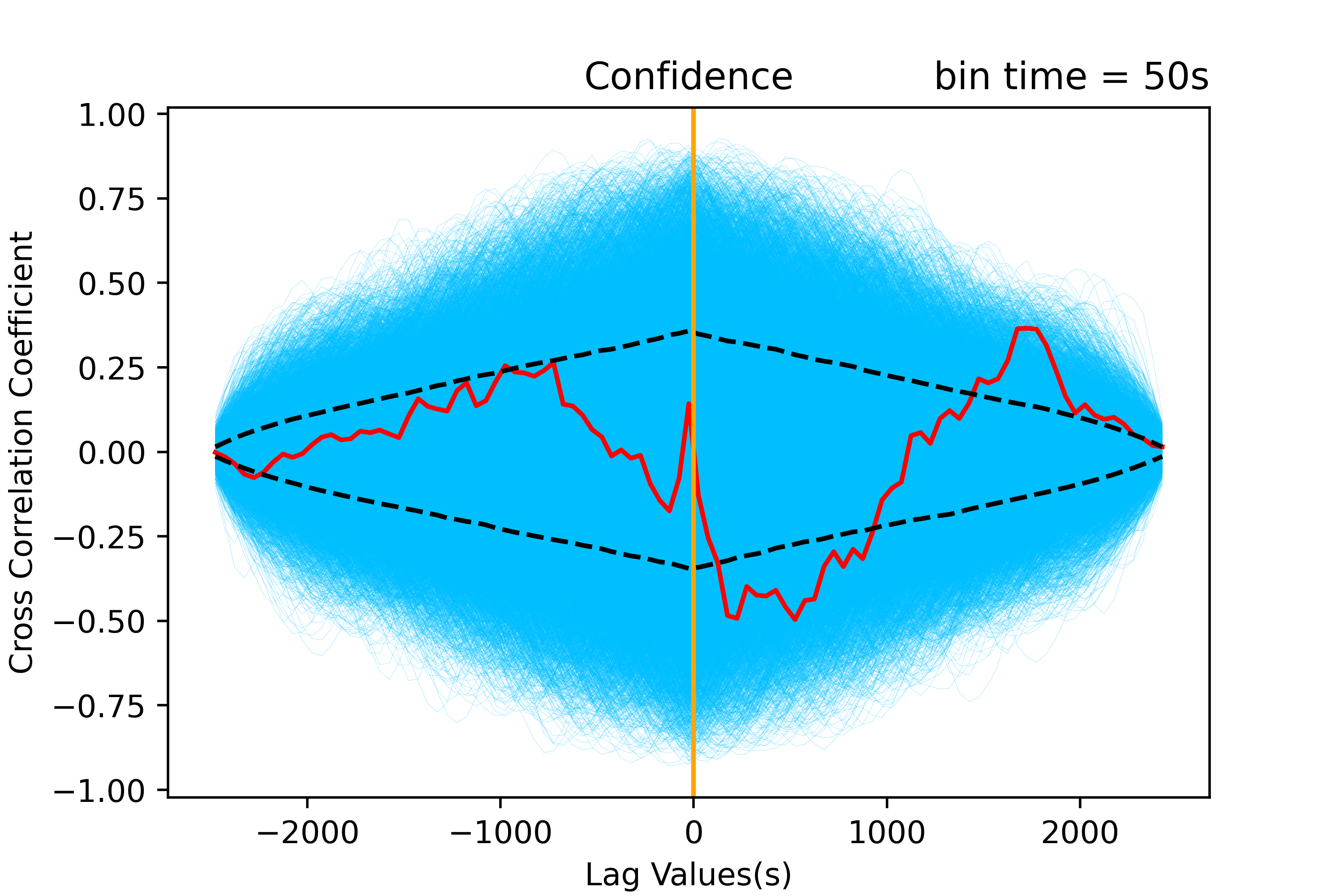}
    \caption{Simulated CCFs histogram (left) and associated Confidence Interval plots (right) using Timmer-Konig method (\citeyear{timmer1995generating}), the $95\%$ confidence windows shown with the black dotted line and the simulated CCF plots shown in a blue shaded area. The red color is the observed CCFs for sections A and B in rows 1 and 2, respectively.}
    \label{fig:CC3}
\end{figure*}

\section{Spectrum Analysis}
The spectra were extracted in the energy range 0.8–10 keV during the intervals corresponding to the detected lags for sections A, B, and C. For the source extraction, we used RAWX columns 29–45, excluding the central bright columns 37 and 38 to mitigate the effects of pile-up (Iaria et al. \citeyear{iaria2009ionized}). The background spectra were extracted from RAWX columns 2–18. In accordance with the XMM-Newton User Handbook (May \citeyear{xmm_uhb_2016}), we applied the selection criterion {\it (FLAG=0)} to ensure robust spectral analysis. The EPIC-pn spectrum shows calibration issues between the energy band 2 - 2.5 keV, due to the presence of the Au edge at $\sim2.3$ keV, we excluded the 2.1 - 2.5 keV band when extracting the spectrum, Iaria et al. (\citeyear{iaria2009ionized}). The spectra were grouped to a minimum of 25 counts per bin.

The spectral modelling of this source was performed using XSPEC (Arnaud et al. \citeyear{arnaud2003x}) version 12.12.1. Based on the study of Iaria et al. (\citeyear{iaria2009ionized}), the hydrogen column density was fixed at $N_{\rm H} = 0.734 \times 10^{22}$ cm$^{-2}$. Moreover, Iaria et al. (\citeyear{iaria2009ionized}) interpreted that the observed Fe--K$\alpha$ emission feature was a blend of the Fe\,{\sc xxv} (6.7 keV) and Fe\,{\sc xxvi} (6.97 keV) lines, which was modeled using a \textit{diskline} component. In the present work, we instead modeled the Fe--K$\alpha$ emission using a simple Gaussian profile with a width of $\sim$ 0.5 keV, allowing the line energy (\textit{$E_{line}$}) to vary freely within the range 6.4-7.0 keV.

Initially, the spectra were modeled with the {\it Tbabs $\times$ (bbodyrad + powerlaw + Gaussian)} (Model 1). The \textit{bbodyrad} component represents blackbody emission, with its normalization proportional to the emitting surface area. The normalization provides an estimate of the radius of the blackbody emitting surface according to $N_{bbr} = \frac{R_{km}^2}{D_{10}^2}$. D$_{10}$ is the source distance in units of 10 kpc. We adopt a source distance of 9.2 kpc (Coughenour et al. \citeyear{coughenour2018nustar}). The best-fit spectral parameters for the first and last 500 s of sections A, B, and C are listed in Table 1, and Fig. 7 (left and right panels) shows the spectral fits for section A's first and last 500 s. 

\begin{figure*}[hbt!]
    \centering
    \includegraphics[width=3in, height=2in]{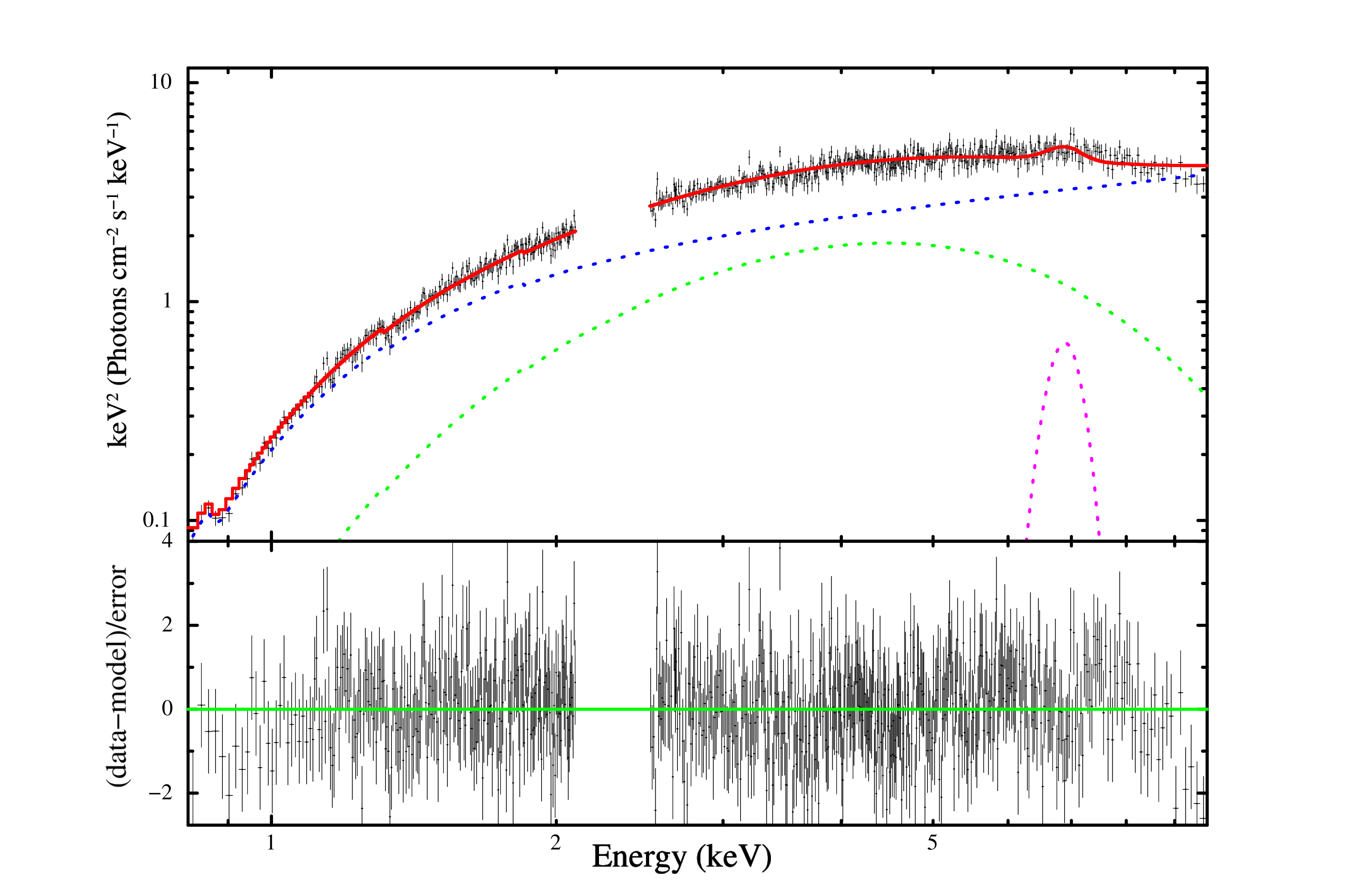}
    \includegraphics[width=3in, height=2in]{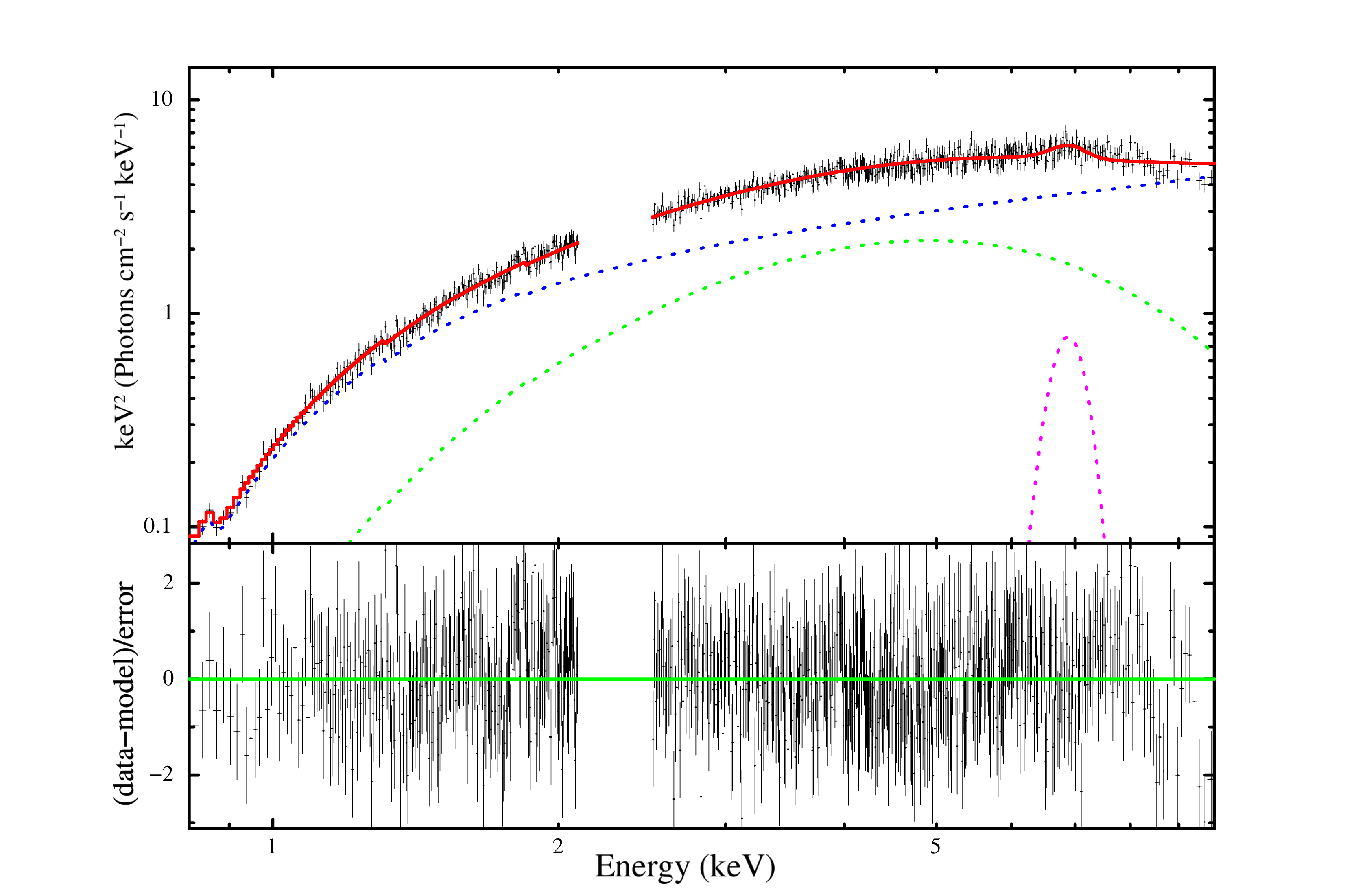}
    \caption{Spectrum of Section A, fitted with model 1; first 500 s (left), last 500 s (right).}
    \label{fig:CC3}
\end{figure*}

%The spectrum was generated between the energy values of 0.8-10 keV for a duration where the lags were found during the transitions, i. e. sections A, B and C. For the source regions, we used the RAWX columns 29-45 and removed the central bright columns 37 and 38, to avoid any pile-up (Iaria et al. 2009). For the background region, we used the RAWX columns 2-18. We used the '(FLAG==0)' condition as recommended in the handbook (User Handbook for XMM Newton May 2016) to be set when performing serious spectral analysis. For analysing the spectra we grouped the data with a bin size of 25 counts/bin. 
%The spectrum was analysed using the Xspec 12.12.1 package, we fit the spectrum using the model,$\boldsymbol{tbabs*(bbodyrad+powerlaw+gaussian)}$ 'bbodyrad' is just the black body spectrum with normalization proportional to the surface area. Where the norm component of this model tells us about the Radius of the black body(the Neutron star in our case) from which we receive the radiation, from the following relation.\\ $Norm = \frac{R_{km}^2}{D_{10}^2}$ \\ Where $D_{10}$ is the source distance (9.2 kpc) in units of 10 kpc. Table 1 gives us the spectral fit values for the first and last 500 s of sections A, B and C:

The spectra were also fitted with another model $\it {Tbabs*(bbodyrad+diskbb+gaussian)}$ (Model 2), as done by Coughenour et al. (\citeyear{coughenour2018nustar}), where they used this as the initial model to fit the data from NuSTAR for this source, and tried to model it separately during each branch of its HID diagram. We applied this model to sections A, B and C, and reported the best-fit parameters for the same in Table 2.

\begin{table*}[h]
\centering
\caption{Spectral Fitting Parameters across Sections A, B, and C, using the model $Tbabs*(bbodyrad+powerlaw+gaussian).$}
\label{tab:spectral_params}
\small 
\begin{tabular}{l cc cc cc}
\toprule
& \multicolumn{2}{c}{\textbf{Section A}} & \multicolumn{2}{c}{\textbf{Section B}} & \multicolumn{2}{c}{\textbf{Section C}} \\
\cmidrule(lr){2-3} \cmidrule(lr){4-5} \cmidrule(lr){6-7}
\textbf{Parameter} & \textbf{First 500s} & \textbf{Last 500s} & \textbf{First 500s} & \textbf{Last 500s} & \textbf{First 500s} & \textbf{Last 500s} \\
\midrule
$N_H$ ($10^{22}$ cm$^{-2}$) & 0.734 (f) & 0.734 (f) & 0.734 (f) & 0.734 (f) & 0.734 (f) & 0.734 (f) \\
\addlinespace
$kT_{bbr}$ (keV) & $1.10_{-0.03}^{+0.03}$& $1.21_{-0.03}^{+0.03}$& $1.33_{-0.04}^{+0.04}$& $1.32_{-0.04}^{+0.04}$& $1.28_{-0.04}^{+0.04}$& $1.15_{-0.03}^{+0.03}$\\
\addlinespace
$N_{bbr}$ & $272_{-22}^{+25}$& $214_{-17}^{+19}$& $140_{-11}^{+12}$& $144_{-12}^{+13}$& $180_{-13}^{+15}$& $253_{-19}^{+21}$\\
\addlinespace
$\Gamma_{pl}$ & $1.55_{-0.03}^{+0.03}$& $1.49_{-0.03}^{+0.03}$& $1.30_{-0.03}^{+0.03}$& $1.25_{-0.03}^{+0.04}$& $1.42_{-0.03}^{+0.03}$& $1.56_{-0.03}^{+0.03}$\\
\addlinespace
$N_{pl}$ & $1.33_{-0.03}^{+0.03}$& $1.32_{-0.03}^{+0.03}$& $0.92_{-0.02}^{+0.02}$& $0.86_{-0.02}^{+0.02}$& $1.13_{-0.03}^{+0.03}$& $1.37_{-0.03}^{+0.03}$\\
\addlinespace
$E_{line}$ (keV) & $6.93_{-0.16}^{+0.16}$& $7.01_{-0.15}^{+0.15}$& $6.88_{-0.15}^{+0.14}$& $6.88_{-0.15}^{+0.16}$& $6.78_{-0.17}^{+0.17}$& $6.86_{-0.12}^{+0.12}$\\
\addlinespace
$\sigma$ (keV) & 0.50 (f)& 0.50 (f)& 0.50 (f)& 0.50 (f)& 0.50 (f)& 0.50 (f)\\
\addlinespace
%$N_{gauss}$ & $1.825 \times 10^{-2}$ & $2.170 \times 10^{-2}$ & $1.692 \times 10^{-2}$ & $1.915 \times 10^{-2}$ & $1.866 \times 10^{-2}$ & $2.156 \times 10^{-2}$ \\
\midrule
$Flux_{bbr}$ & 4.14 & 4.26 & 4.93 & 4.67 & 4.40 & 4.67 \\
$Flux_{pl}$ & 9.11 & 9.80  & 9.16 & 9.22 & 8.75 & 9.22 \\
%$Flux_{gauss}$ & 0.24 & 0.20 & 0.20 & 0.24 & 0.20 & 0.24 \\
$Flux_{total}$ & 13.45 & 14.31 & 13.33 & 13.43 & 14.29 & 14.12 \\
\midrule
$\chi^2 / \text{dof}$ & 643 / 624 & 641 / 648 & 713 / 665 & 734 / 671 & 679 / 654 & 703 / 636 \\
\bottomrule
\end{tabular}
\vspace{1mm}

\footnotesize{The $N_H$ value adopted from Iaria et al. (\citeyear{iaria2009ionized}); (f): fixed parameter; Flux values are in units of $10^{-9}$ $ergcm^{-2}s^{-1}$; all errors are reported at 90\% confidence.}
\end{table*}

\begin{table*}[h]
\centering
\caption{Spectral Fitting Parameters across Sections A, B, and C, using the model 2 $Tbabs*(bbodyrad+diskbb+gaussian).$}
\label{tab:spectral_params}
\small 
\begin{tabular}{l cc cc cc}
\toprule
& \multicolumn{2}{c}{\textbf{Section A}} & \multicolumn{2}{c}{\textbf{Section B}} & \multicolumn{2}{c}{\textbf{Section C}} \\
\cmidrule(lr){2-3} \cmidrule(lr){4-5} \cmidrule(lr){6-7}
\textbf{Parameter} & \textbf{First 500s} & \textbf{Last 500s} & \textbf{First 500s} & \textbf{Last 500s} & \textbf{First 500s} & \textbf{Last 500s} \\
\midrule
$N_H$ ($10^{22}$ cm$^{-2}$) & $0.734$ (f)& $0.734$ (f)& $0.734$ (f)& $0.734$ (f)& $0.734$ (f)& $0.734$ (f)\\
\addlinespace
$kT_{bbr}$ (keV) & $1.81_{-0.07}^{+0.09}$& $1.85_{-0.06}^{+0.08}$& $1.99_{-0.08}^{+0.12}$& $2.13_{-0.04}^{+0.04}$& $1.84_{-0.06}^{+0.07}$& $1.73_{-0.05}^{+0.06}$\\
\addlinespace
$N_{bbr}$ & $79_{-17}^{+16}$& $90_{-17}^{+16}$& $68_{-18}^{+14}$& $49_{-13}^{+11}$& $95_{-16}^{+15}$& $107_{-17}^{+17}$\\
\addlinespace
$kT_{diskbb}(keV)$& $1.07_{-0.07}^{+0.08}$& $1.07_{-0.07}^{+0.09}$& $1.20_{-0.10}^{+0.08}$& $1.44_{-0.05}^{+0.04}$& $1.06_{-0.08}^{+0.10}$& $0.99_{-0.06}^{+0.07}$\\
\addlinespace
$N_{diskbb}$& $255_{-51}^{+56}$& $246_{-53}^{+59}$& $122_{-40}^{+42}$& $69_{-14}^{+16}$ & $215_{-51}^{+56}$& $322_{-63}^{+71}$\\
\addlinespace
$E_{line}$ (keV) & $6.76$ (f)& $6.76$ (f)& $6.76$ (f)& $6.76$ (f)& $6.76$ (f)& $6.76$ (f)\\
\addlinespace
$\sigma$ (keV) & $0.50$ (f)& $0.50$ (f)& $0.50$ (f)& $0.50$ (f)& $0.50$ (f)& $0.50$ (f)\\
\addlinespace
%$N_{gauss}$ & $6.03 \times 10^{-3}$& $8.23 \times 10^{-3}$& $7.08 \times 10^{-3}$& $1.01 \times 10^{-2}$& $7.05 \times 10^{-3}$& $8.73 \times 10^{-3}$\\
\midrule
$Flux_{bbr}$ & 7.50& 9.05& 8.59& 7.79& 9.39& 8.61\\
$Flux_{diskbb}$& 5.77& 5.60& 4.60& 5.48& 4.74& 5.29\\
%$Flux_{gauss}$ & 0.06& 0.09& 0.08& 0.01& 0.08& 0.09\\
$Flux_{total}$ & 13.33& 14.74& 13.26& 13.37& 14.21& 14.00\\
\midrule
$\chi^2 / \text{dof}$ & $627 / 625$& $653 / 649$& $683 / 666$& $696 / 673$& $661 / 655$& $703 / 637$\\
\bottomrule
\end{tabular}
\vspace{1mm}

\footnotesize{The $N_H$ value adopted from Iaria et al. (\citeyear{iaria2009ionized}); (f): fixed parameter; Flux values are in units of $10^{-9}$ $ergcm^{-2}s^{-1}$; all errors are reported at 90\% confidence.}
\end{table*}

\begin{comment}
\begin{table*}[hbt!]
\centering
\caption{Best Fit parameters using the model $Tbabs*(bbodyrad+diskbb+gaussian).$ The $Fe_{K\alpha}$ line is fixed at 6.76keV and the $N_H$ value fixed at $0.8\times 10^{22}$ atoms cm$^{-2}$.}
\begin{tabular}{l c l c l c l}
\toprule
Parameter & \multicolumn{2}{c}{Section A } & \multicolumn{2}{c}{Section B} & \multicolumn{2}{c}{Section C} \\
\midrule
 & First 500s & Last 500s & First 500s & Last 500s & First 500s & Last 500s \\
 \midrule
$kT_{bbr} (keV)$ & $1.87^{+0.10}_{-0.08}$& $1.89^{+0.08}_{-0.07}$& $2.04^{+0.14}_{-0.09}$& $2.11^{+0.06}_{-0.04}$& $1.88^{+0.08}_{-0.06}$& $1.75^{+0.06}_{-0.05}$\\
$N_{bbr}$ & $68^{+15}_{-15}$& $81^{+15}_{-15}$& $60^{+15}_{-17}$& $51^{+1}_{-1}$& $86^{+14}_{-16}$& $100^{+15}_{-16}$\\
kT$_{in}$ (keV) & $1.15^{+0.09}_{-0.08}$& $1.14^{+0.1}_{-0.08}$& $1.31^{+0
22}_{-0.14}$& $1.46^{+0.07}_{-0.06}$& $1.14^{+0.12}_{-0.09}$& $1.04^{+0.07}_{-0.06}$\\
$N_{dbb}$& $195.80^{+44.94}_{-39.95}$& $192.98^{+46.03}_{-41.25}$& $89.97^{+35.61}_{-31.61}$& $63.03^{+8.10}_{-7.84}$& $164.82^{+44.67}_{-40.77}$& $265.72^{+55.45}_{-49.71}$\\
$Flux_{bbr}$ & 7.04& 8.83& 8.33& 8.01& 9.17& 8.41\\
$Flux_{dbb}$& 6.19& 5.79& 4.84& 5.39& 4.92& 5.46\\
$Flux_{Total}$ & 13.2& 14.6& 13.2& 13.4& 14.1& 13.9\\
 Luminosity& 1.34& 1.48& 1.33& 1.36& 1.43&1.41\\
$\chi^{2}/dof$ & 621/630& 648/660& 699/671& 722/677& 661/660& 699/642\\
\bottomrule
\end{tabular}

\footnotesize{bbr: bbodyrad, dbb: diskbb, Flux is the units of $10^{-9} erg cm^{-2} s^{-1}$, Luminosity has the units of $10^{38} erg/s.$}
\end{table*}
\end{comment}

Markov Chain Monte Carlo (MCMC) simulations were performed to investigate the posterior probability distributions of the model parameters. The simulations were carried out using the Goodman–Weare algorithm, with a total chain length of 200,000 steps and a burn-in phase of 50,000 steps. Fig. 8 presents the joint confidence (contour) plots of the key parameters, illustrating their variations across sections A, B, and C, where significant lags were detected.

\begin{table*}[hbt!]
\centering
\caption{Table for $t_{depl}$ (s) values at various radius for Model 1 (\textit{tbabs*(bbodyrad+powerlaw+gaus)}).}
\begin{tabular}{l r l l l l }
\toprule
Section & $R_{eff}$(km) & $\dot{m}$(g/s) & $\Omega_{k}$ (Hz) & $t_{depl,29km}$& $t_{depl,31km}$\\
\midrule
A First& 25.80& 1.18& 523.65& 2135.74& 1132.21\\
A Last & 22.92& 0.99& 625.16& 876.49& 565.03\\
B First& 18.51& 0.79& 861.78& 315.41& 224.99\\
B Last & 18.77& 0.81& 843.94& 333.90& 237.34\\
C First& 20.97& 0.96& 714.62& 545.75& 372.98\\
C Last & 24.91& 1.13& 551.86& 1554.51& 897.70\\
\bottomrule
\end{tabular}

\footnotesize{ units of $t_{depl}$ are in seconds;  $R_{eff}$ is the color corrected effective radius derived from $N_{bbr}$; $\dot{m}$ is in the units of $10^{18}$ }

\end{table*}

\section{Discussion}
For the first time, we performed CCF analysis on the soft (0.8-2.0 keV) and hard Lightcurves (2-10 keV) using EPIC-pn data of the XMM-Newton observatory for the source GX 349+2. The CCFs revealed time lags of a few hundred seconds during the horizontal branch (HB) as seen in sections 1 \& 2. We find that the CCFs associated with the HB are highly asymmetric and exhibit significant lags. In contrast, those corresponding to the normal and flaring branches (NB and FB) are symmetric and show no statistically measurable lags. Ding et al. (\citeyear{ding2016cross}) investigated soft–hard X-ray time lags in the Z-sources GX 349+2 using RXTE observations. They reported significant lags, typically ranging from tens to a few hundred seconds, predominantly in the HB, while little or no lag was detected in the normal and flaring branches, and they interpreted lags as the viscous timescale in the inner accretion flow, possibly linked to variations in the corona. EPIC-pn data allows us to use the soft energy band 0.8-2.0 keV, which was not possible with PCA/RXTE.
Recently, Gouse et al. (\citeyear{gouse2025asymmetric}, \citeyear{gouse2025association}) demonstrated that RXTE observations of Sco X-1 exhibit a clear dichotomy in soft–hard X-ray CCFs, which is closely associated with distinct modes of radio jet ejection. They found that asymmetric CCFs, characterized by weak correlations and time lags of a few hundred seconds, occur during ballistic (lobe) jet episodes and are accompanied by flat-topped noise and the absence of coherent oscillations in the power density spectrum (PDS). In contrast, symmetric CCFs occur during normal and flaring branch states, and coincide with ultra-relativistic flow (URF) events, and exhibit persistent oscillations (NBO/NBO+HBO/FBO). These results suggest two states of the inner accretion flow: ballistic jet launching disrupts the BL and/or corona, producing delayed CCFs and suppressing oscillations within a region of size $\sim$ 10--40 km, whereas URFs are associated with a stable, steady accretion configuration. Thus, CCF asymmetry provides a robust diagnostic of jet–disk coupling in Sco X-1. We also noted a very similar scenario in the case of GX 349+2, where asymmetric CCFs were associated with HB and relatively symmetric CCFs in the FB and NB. This analogy displays that the inner accretion disk was disrupted, probably due to the mechanism that launched the jet.
We also performed CCFs during the flux transition corresponding to branch variations in HID (Sections A, B, and C; See Fig. 5 \& 6). We again noted asymmetric CCFs with lags having more than 95\% significance during these phases. \\

The CCF is a convolution of the transfer function along with the continuum auto-correlation function (ACF), so the CCF depends on the continuum behaviour (Peterson \citeyear{PETERSON_2001}). As a direct consequence of this, any secondary peak (other than the zero peak) present in the ACF of any one of the time-series will imprint itself as a spurious peak in the CCF and should therefore not be considered as a physical lag/lead between the two time-series and should be considered as an artefact (Dean et al. \citeyear{dean2016dangers}). The hard band LC in both HB1 and Section B shows a secondary ACF peak (other than zero) at $\sim 1100$ s, which is precisely where the competing CCF peak appears. 
We also examined the ACF of the soft-band light curve in these sections and found no secondary peak near either $\sim 400s$ or $\sim 1100s$ (see figure 2(b) and 4(b)), confirming that the soft-band ACF does not introduce spurious features at either of the two candidate lag positions.
The $\sim 400$ s peak, by contrast, has no corresponding ACF feature and is therefore identified as the true physical lag and represents the temporal incoherence between the two energy bands. This interpretation is further supported by the 10000 CCF Monte-Carlo simulation lag histogram for this section; $\sim 62 \%$ of the realizations manifest at the $\sim 400$ s lag, but only $\sim 37\%$ of the realisations manifest at the $\sim 1100$ s lag. While we note that the Monte Carlo procedure preserves the temporal structure of the observed light curves and therefore cannot independently exclude ACF-induced features, and thus the combination of the ACF inspection along with the preferential robustness in the simulations provides strong grounds for identifying the $\sim 400s$ feature as the most probable physical lag candidate.

In the spectrum analysis, the contour plots from MCMC simulations (see Fig. 8) indicate that the \textit{Norm} of \textit{bbodyrad} showed a statistically significant variation in the sections A and C, suggesting that the BL size has varied during the said transition sections (since \textit{Norm} of \textit{bbodyrad} gives the direct measure of radius of the emitting surface, see Tables 3). Whereas the photon index ($\Gamma_{pl}$) showed an appreciable change only in section C, suggesting changes in the radiative and/or geometric properties of the BL or the corona during this interval. Both lags and spectral variation during the transitions clearly suggest that the BL has changed its radiative and physical structures. We suggest that if such structural variation had not happened, the CCFs would have been highly symmetric. On the other hand, it is to be noted that, in section B, the BL geometry remains stable, as evidenced by the $N_{bbr}$ remaining constant ($N_{bbr} = 140-144$; Table 1). Further, even the power-law components remain constant, 
%this is due to the stable $\dot{M}$, which suggests that the accretion flow has remained quite stable, 
but interestingly, the $N_{diskbb}$ showed a drop in its value ($N_{diskbb} \sim 122 \rightarrow 69 $; Table 2) along with an increase in the inner disk temperature (kT$_{diskbb}$ = 1.20$^{+0.08}_{-0.10}$ keV to 1.44$^{+0.05}_{-0.04}$ keV) suggesting that the inner disk front moved closer to the NS and this restructuring manifests in the form of a lag in the CCF.%, since this readjustment happens at viscous timescales.}

\subsection{Relation between lags and BL}

The observed soft--hard X-ray lag of \textbf{$t \sim 380$ s} in GX~349+2 can be interpreted as the mechanical readjustment timescale of the BL and the inner disk. In a previous study of a different Z source, it was noted that such observed time lags were interpreted as the readjustment time scale of corona or BL (Sriram et al. \citeyear{sriram2013spectral}; Sriram et al. \citeyear{sriram2019constraining}; Malu et al. \citeyear{malu2021exploring}; Chiranjeevi \& Sriram \citeyear{sriram2022anticorrelated}; Chiranjeevi et al. \citeyear{chiranjeevi2023detection}).  Assuming that the lag arises due to viscous redistribution of mass and angular momentum over a small radial extent $\Delta R \ll R$, the effective viscous timescale is related to the observed lag via
\begin{equation}
\Delta t \sim \frac{3}{2} \frac{\Delta R}{R} \, t_{\rm visc},
\end{equation}
where the viscous timescale $t_{\rm visc}$ at radius $R$ is given by the standard $\alpha$-disk prescription
\begin{equation}
t_{\rm visc} \sim \frac{1}{\alpha} \left(\frac{R}{H}\right)^2 \frac{1}{\Omega_K}, 
\quad \Omega_K = \sqrt{\frac{GM}{R^3}}.
\end{equation}
Here, $H$ is the scale height of the BL, $\Omega_K$ the Keplerian angular frequency, and $\alpha$ the dimensionless viscosity parameter. Solving for $\alpha$ in terms of the observed lag yields
\begin{equation}
\alpha \sim \frac{3}{2} \frac{\Delta R}{R} \frac{1}{\Delta t} \left(\frac{R}{H}\right)^2 \frac{1}{\Omega_K}.
\end{equation}

Using values for a neutron star with $R \sim 10$~km, $\Delta R \sim 3$ km, $H/R \sim 0.1$, and $\Omega_K \sim 1.36\times10^4$~s$^{-1}$, we obtain
\begin{equation}
\alpha \sim 8.7\times10^{-6}.
\end{equation}
This extremely low effective viscosity indicates that the mechanical adjustment of the BL proceeds much more slowly than standard turbulent viscosity ($\alpha \sim 0.01$--0.1) (Babkovskaia, Brandenburg, \&  Poutanen \citeyear{babkovskaia2008boundary}). 

Physically, this does not imply that the microscopic viscosity is unrealistically small; rather, it reflects the slower angular momentum transport in the BL, potentially mediated by radiative viscosity, magnetic stresses, or global coupling with the inner disk and corona. Such slow readjustment naturally explains the observed soft--hard lag on timescales of hundreds of seconds.
We point out that the $\alpha$ parameter inferred from the observed time lag is not intended to provide a direct quantitative constraint on the stress parameter adopted in the boundary-layer dynamical model. Instead, it should be regarded as an independent, order-of-magnitude estimate of the characteristic transport (or readjustment) viscosity based on the standard thin disk model. The stress parameter used in the boundary-layer model, defined through 
W$_{r \phi}$ = $\alpha_{BL}$ P, where W$_{r \phi}$ is the tangential stress and P is the pressure at the bottom of the layer; Abolmasov \&  Poutanen  (\citeyear{abolmasov2021mechanical}) and $\alpha_{BL}$ represents a distinct physical quantity associated with the angular momentum coupling at the disk -- neutron star interface.

In the context of a neutron star accretion system, the emitting region corresponds to the neutron star surface together with the boundary layer. $R_{bbr} = \sqrt{N_{bbr}}D_{10}$ - (found from the norm of bbodyrad) needs to be colour corrected such that $R_{eff} = f_c \times R_{bbr}$, adopting $f_c = 1.7$ from Shimura \& Takahara (\citeyear{1995ApJ...445..780S}). The apparent radius inferred from $N_{bbr}$ may therefore be expressed as an effective radius $R_{eff} = R_{NS} + \Delta R_{BL}$, we note that this quantity reflects the apparent thermal emitting area rather than a direct measurement of the physical radial width of the boundary layer, and should be treated as a qualitative indicator of changes in the emitting region. %re-written as, $\Delta R_2 = 10^{5.02 + 0.245[\log(\frac{\dot{M}}{10^{-9.85}M_{\odot}yr^{-1}})]^{2.19}}.$}
We calculate the $\Omega_k$ (Keplerian rotation frequency) at $R_{eff}$ inferred from $N_{bbr}$, $\Omega_{NS} = 266Hz$ (neutron star spin frequency) adopted from Zhang et al. (\citeyear{zhang1998discovery}) and $\Omega$ (at various distances), which will be used to find the t$_{depl}$ following the relation, 
$t_{depl}=\frac{(\Omega - \Omega_{NS})}{\alpha (\Omega_{k}^2-\Omega^2)}$ (Abolmasov \& Poutanen (\citeyear{abolmasov2021mechanical})). %Table 3 \& 4 shows us the $t_{depl}$ values for $\alpha$ value of $10^{-7}$ . 
The calculated time scales of $t_{depl}$ are in agreement with those found in the CCF of the source (see Table 3), the subscript under $t_{depl}$, 29 km \& 31 km represent the distances from the center at which the $\Omega$ were calculated and consecutively used to find $t_{depl}$ at those distances. Abolmasov \&  Poutanen (\citeyear{abolmasov2021mechanical}) showed that for $\alpha_{BL}$ values of $10^{-7}$, the $t_{depl}$ value turns out to be 740 s and lower for $\alpha_{BL}=10^{-6}$, these lags are well in agreement with the scales of lags observed in the source GX 349+2, hence the BL mechanical adjustments can be associated to the lags observed in this source. Such time scales were seen in an atoll source 4U 1728-34, the depletion associated with low viscosity $\alpha_{BL}$ $\le 10^{ -7}$ (Chiranjeevi et al. \citeyear{chiranjeevi2023detection}).

 As discussed above, the lags could be due to physical readjustment of the BL. We estimated the size of the BL using the following relation (Popham \& Sunyaev \citeyear{popham2001accretion}).

\begin{multline}
    \log(R_{BL}-R_{NS})
    \sim 5.02 + 0.245[\log(\frac{\dot{M}}{10^{-9.85}M_{\odot}yr^{-1}})]^{2.19}
\end{multline} 

where $\dot{M}$ is found using the equation $L=\frac{GM_{NS}\dot{M}}{R}$, where $M_{NS}$ is the mass of the neutron star ($M_{NS}=1.4 M_{\odot}$) and Luminosity $L$ is calculated using ($L = 4\pi D^2 F$), where $F$ is the total flux obtained for Model 1, between the energy range 0.8--10.0 keV, D is 9.2 kpc, and R is the effective radius obtained from the Norm of \textit{bbodyrad} from the spectral fit (see Table 1 and 2). 
We estimate the effective width of the boundary layer $\Delta R_{1} = R_{\rm eff} - R_{\rm NS}$ (we assume a $R_{NS} = 10$ km) and compare it with the effective boundary layer width $\Delta R_2$ derived from Eq. 5.
$\Delta R_1$ and $\Delta R_2$ values are reported in Table 4. The fall in the effective boundary layer width seen from both the approaches has been observed before; the reduction of the black body radius during the Horizontal branch has been reported before (Church et al. \citeyear{church2006explanation}).

\begin{comment}
\textbf{The corresponding values are presented in Table 4 for Model 1. We find that the inferred apparent boundary layer width to be around $\sim$ 3 km, which is broadly consistent with estimates reported for other Z-sources, e.g., GX 13+1, where the boundary layer was estimated to be $R_{BL} \sim 3$ km (Saavedra et al. \citeyear{saavedra2023relativistic})}. It can be seen that there is a difference in $\Delta R_{1}$ and $\Delta R_{2}$ in section B, which is due to the lower values of N$_{bbr}$ in HB, this reduction in the emitting surface area in the HB has been suggested in previous studies as well, where \textit{h} is the half-height of the black-body emitter (Church et al. \citeyear{church2006explanation}) was noted to be $\sim$ 1 km, similar to the reported in the present study (see Table 4). \\
\end{comment}

\begin{table}[hbt]
\centering
\caption{$\Delta R_1 $ and $\Delta R_2$ values for Sections A, B and C.}
\begin{tabular}{l l l }
\hline
 & \multicolumn{2}{c}{Model 1}  \\
\hline
Section & $\Delta R_1$ (km) & $\Delta R_2$ (km)\\
\hline
A First & 15.80& 19.69\\
A Last & 12.92& 15.63\\
B First & 8.51 & 11.90\\
B Last & 8.77& 12.21\\
C First & 10.97& 15.08\\
C Last & 14.91& 18.49\\
\hline
\vspace{1mm}

\end{tabular}

\footnotesize{ Model 1: tbabs*(bbodyrad+powerlaw+Gaussian); $\Delta R_{1} = R_{\rm eff} - R_{\rm NS}$; $\Delta R_2 = R_{BL}-R_{NS}$}%10^{5.02 + 0.245[\log(\frac{\dot{M}}{10^{-9.85}M_{\odot}yr^{-1}})]^{2.19}}.$}
\end{table}

Using equation 5, we obtained the effective size of BL ranging from 12-20 km (see Table 4 $\Delta R_2$),  using $\dot{M}$ values from Model 1 (see Table 3). It matches the BL radius estimated by Coughenour et al. (\citeyear{coughenour2018nustar}), where they found BL would extend from 19 to 40 km.

\subsection{Alternative scenario: Extended accretion disk corona}

Iaria et al. (\citeyear{iaria2009ionized}) showed that GX 349+2 hosts an extended accretion disk corona (ADC) spanning several tens to hundreds of kilometers, a result later supported by the systematic ADC framework presented by Church et al. (\citeyear{church2012spectral}). Recent X-ray polarization measurements of GX 349+2 revealed a polarization angle difference of 60$^{\circ}$ between the disk and Comptonized emission components, which can be explained by the presence of a boundary/spreading layer or an extended accretion disk corona (La Monaca et al. \citeyear{la2025x}). Furthermore, Ludlam et al. (\citeyear{ludlam2025structure}), using XRISM observations, independently reported evidence for a similarly extended coronal structure in the NS-LMXB, GX 340+0.

 Within the accretion disk corona (ADC) framework, the observed $\sim$ 200 s soft–hard lag cannot be attributed to light travel or Comptonization delays. Instead, it probably arises from the slow viscous readjustment of an extended corona in response to changes in the inner accretion flow. For a geometrically thick ADC extending over tens of kilometers, a low effective viscosity ($\alpha_{\rm eff}\sim10^{-3}$–$10^{-4}$) yields adjustment timescales of hundreds of seconds, consistent with the observed lags. For $\Delta R = \alpha (H/R)^2 v_{k} \Delta t$, where $\Delta t$ is observed lag, $v_{k}$ Keplerian velocity, assuming a H/R = 0.1 (eg. Smale, Church \& Balucinska-Church \citeyear{smale2001ephemeris}; Zhu \& Stone \citeyear{zhu2018global}), $\Delta t$ = 200 s, and $\alpha$ = 10$^{-3~ to~ -4}$, we get $\Delta$ R $\sim$ 30 km -- 300 km, this supports a picture in which the lag reflects the dynamical response of the disk-corona system rather than instantaneous radiative processes.

% Future X-ray spectral and timing studies, along with radio observations, would enable us to tightly constrain the underlying astrophysics of the observed lags and explore the possibility of such viscosity.  

\section{Conclusion}

This study presents a comprehensive timing and spectral analysis of the NS-LMXB GX 349+2 (Sco X-2), covering the full evolution of the source along the Z-track in the hardness–intensity diagram (HID), with the source spending most of the observation time in the flaring branch (FB). For the first time, Cross-correlation function analysis using EPIC-pn data between soft and hard X-ray LCs revealed both correlated and anti-correlated hard lags in several segments in this source. The main findings of this work are summarized below:

1. We noted a branch-dependent CCF behavior. Asymmetric CCFs exhibiting significant lags are observed exclusively during the horizontal branch (HB). In contrast, the normal branch (NB) and flaring branch (FB) are characterized by highly symmetric CCFs with no statistically significant delays between the soft and hard bands.

2. The symmetric CCFs seen in the NB and FB likely indicate a stable accretion geometry in the inner disk region, where soft and hard photon intensity variations maintain temporal coherence. The asymmetric CCFs with measurable lags in the HB suggest the presence of instabilities in the inner accretion flow, possibly linked to episodic jet activity similar to Sco X-1 (Motta \& Fender \citeyear{motta2019connection}). Such a scenario is consistent with the known association of the HB with radio jet emission, although simultaneous radio observations would be required to confirm this interpretation for GX 349+2.

3. We noted lags during flux transitions, as well. During transitions between different Z-track branches, asymmetric CCFs reappear with measurable time lags. This behavior points to a variable emission component in the innermost accretion region, most plausibly the BL and/or corona. The observed delays may represent the readjustment timescale of the varying physical structure in this region as the accretion flow reorganizes. This is further emphasized by the change in the \textit{Norm} parameter of \textit{bbodyrad} and \textit{Norm} parameter of \textit{diskbb}, observed in the spectral fitting and subsequently validated by the MCMC simulations.

4. We propose that the presence of these lags requires a relatively low effective viscosity in the inner accretion flow, allowing structural changes to propagate on observable timescales. This interpretation is consistent with the theoretical mechanical modeling of boundary layer presented by Abolmasov \& Poutanen (\citeyear{abolmasov2021mechanical}), where a low viscosity $\alpha_{BL}$ was observed, however our estimate of viscosity ($\alpha$) based on thin disk is similar but conceptually it is different from $\alpha_{BL}$.\\

%Point 5 not clear, in our study, we found lags in the NB, and we found NBO, which doesn't sit well with the Sco X-1 paper.
%5. Power density spectra (PDS) revealed the presence of normal branch oscillations (NBOs) during the NB. The detection of NBOs supports the picture of a stable inner accretion structure, consistent with the symmetric, zero-lag CCFs observed in this branch. In contrast, disturbances in the inner flow, such as those inferred in the HB, may suppress NBOs, coinciding with the emergence of asymmetric CCFs and measurable time lags.

A simultaneous study in the X-ray and radio band of this source would provide an important view on the relation between the inner-accretion region and the jet base formation, as studied by Gouse et al. (\citeyear{gouse2025asymmetric}, \citeyear{gouse2025association})  and Motta \& Fender (\citeyear{motta2019connection}). This would give a better understanding of the source's inner accretion disk geometry, structure and a coronal geometry, i. e., compact or extended corona scenario. 
 
\section{Acknowledgment}
We would like to thank the anonymous referee for his insightful comments and useful suggestions that improved the clarity and presentation of the manuscript.
 K. S. acknowledge the support from the ANRF CRG project, Government of India. This work is based on observations obtained with {\it XMM–Newton}, an ESA science mission with instruments and contributions directly funded by ESA Member States and NASA.

\bibliographystyle{elsarticle-harv} 
\bibliography{example}



%% For example:

%% \bibitem[Aladro et al.(2015)]{Aladro15} Aladro, R., Martín, S., Riquelme, D., et al. 2015, \aas, 579, A101

\begin{figure*}[hbt!]

\begin{subfigure}{0.5\textwidth}
\includegraphics[width=0.8\linewidth, height=5cm]{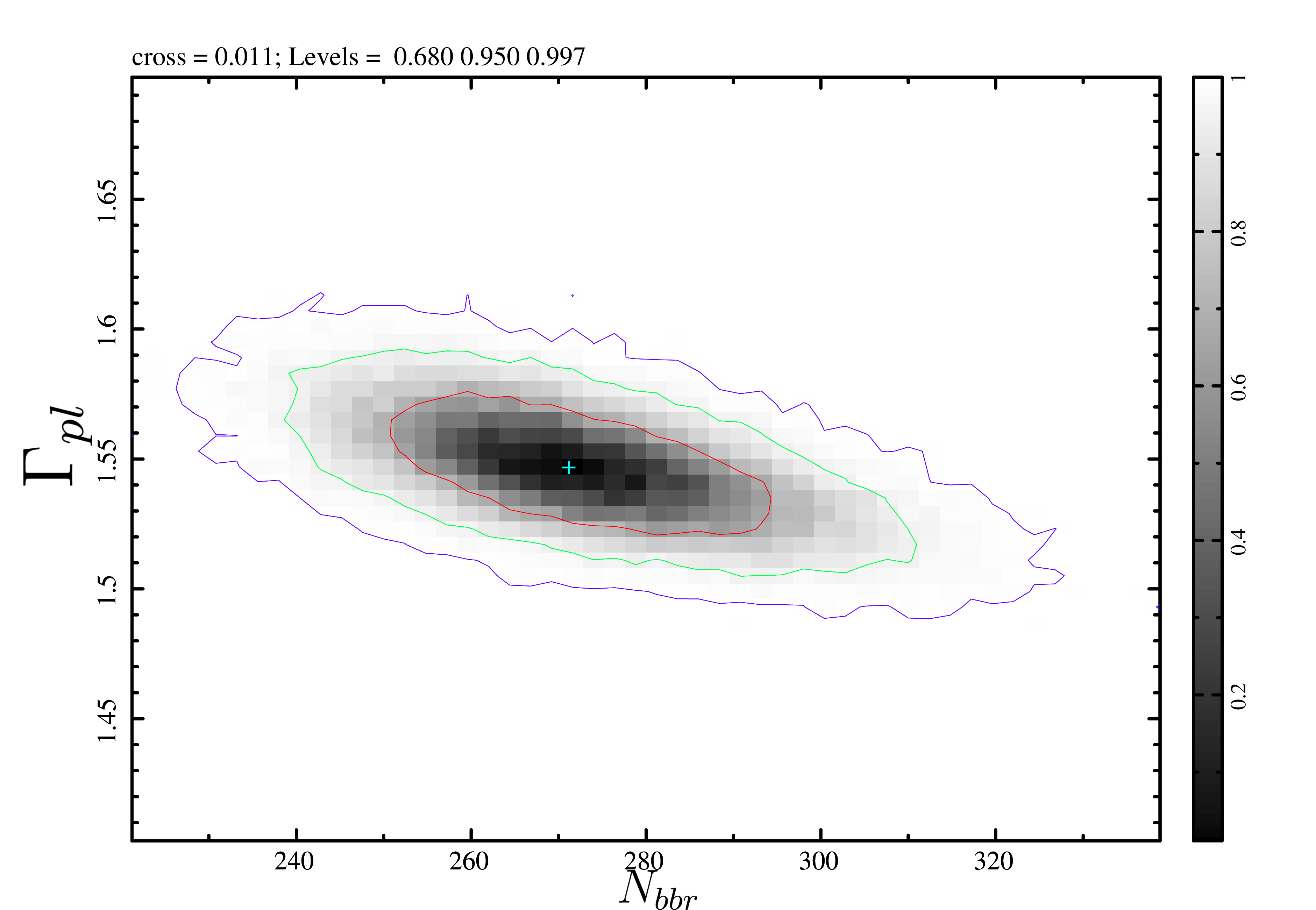} 
\caption{Sec A First 500s}
\label{fig:MCMC3_1}
\end{subfigure}
\begin{subfigure}{0.5\textwidth}
\includegraphics[width=0.8\linewidth, height=5cm]{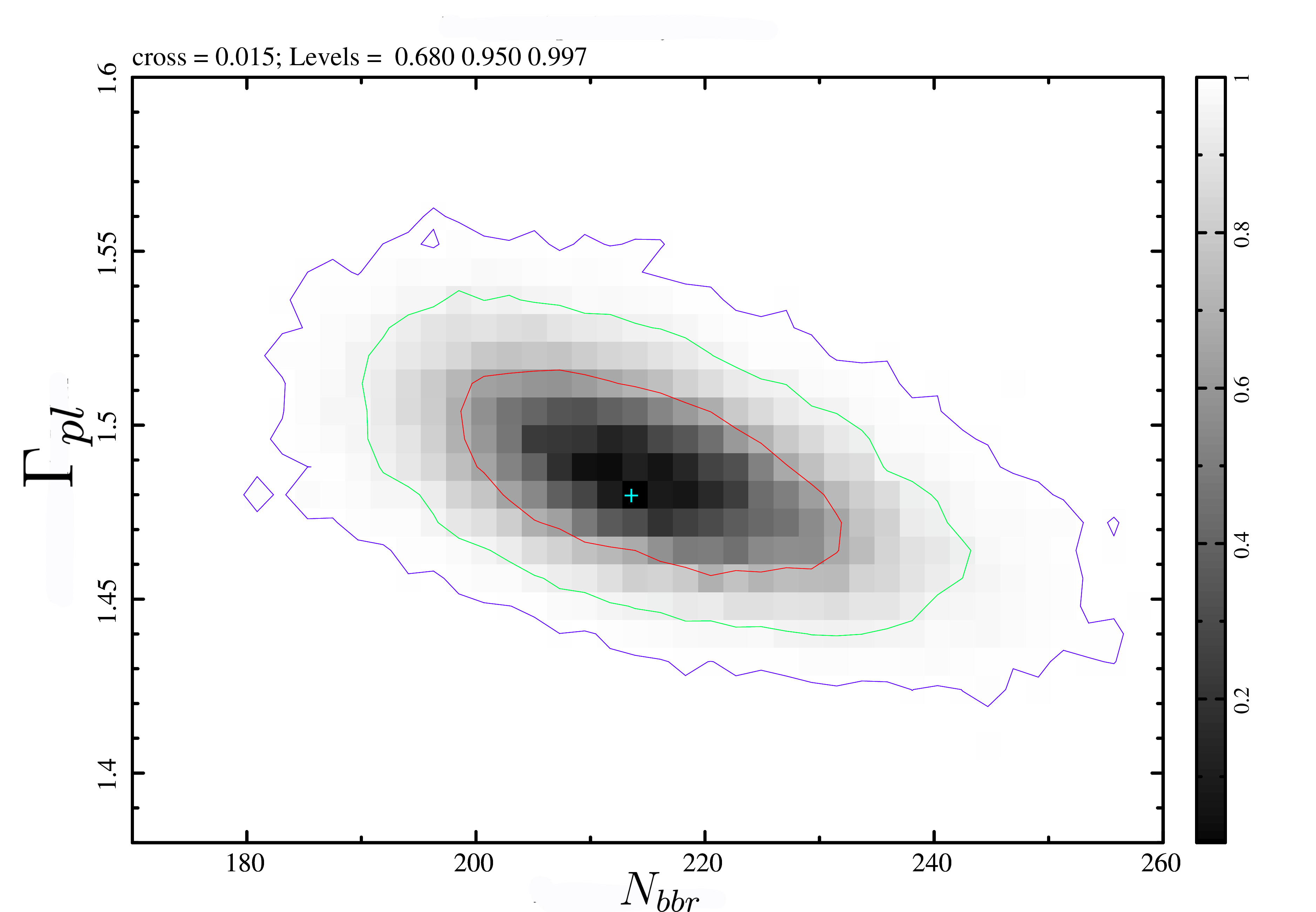}
\caption{Sec A Last 500s}
\label{fig:MCMC3_2}
\end{subfigure}
\label{fig:image2}

\begin{subfigure}{0.5\textwidth}
\includegraphics[width=0.8\linewidth, height=5cm]{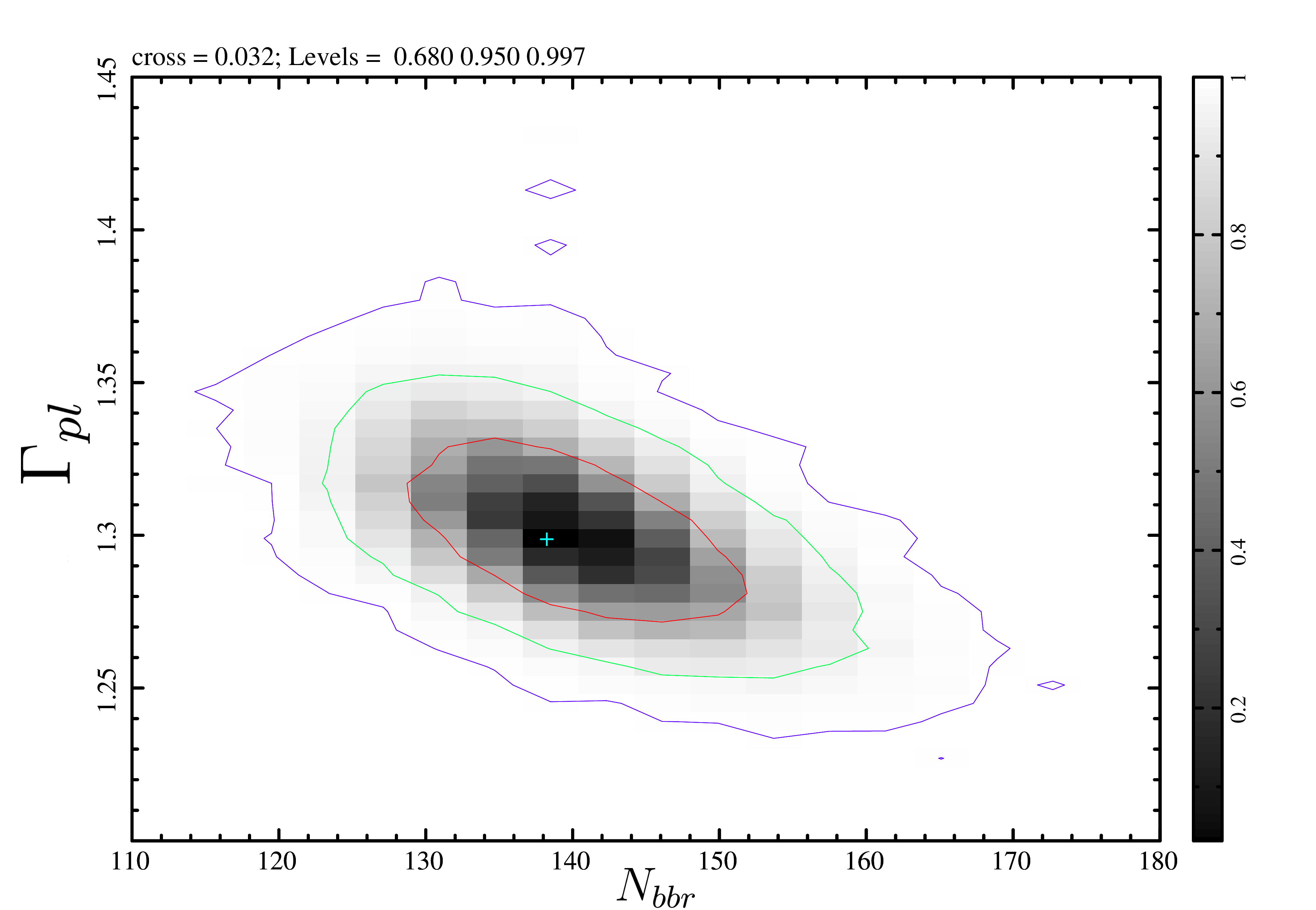} 
\caption{Sec B First 500s}
\label{fig:MCMC5_1}
\end{subfigure}
\begin{subfigure}{0.5\textwidth}
\includegraphics[width=0.8\linewidth, height=5cm]{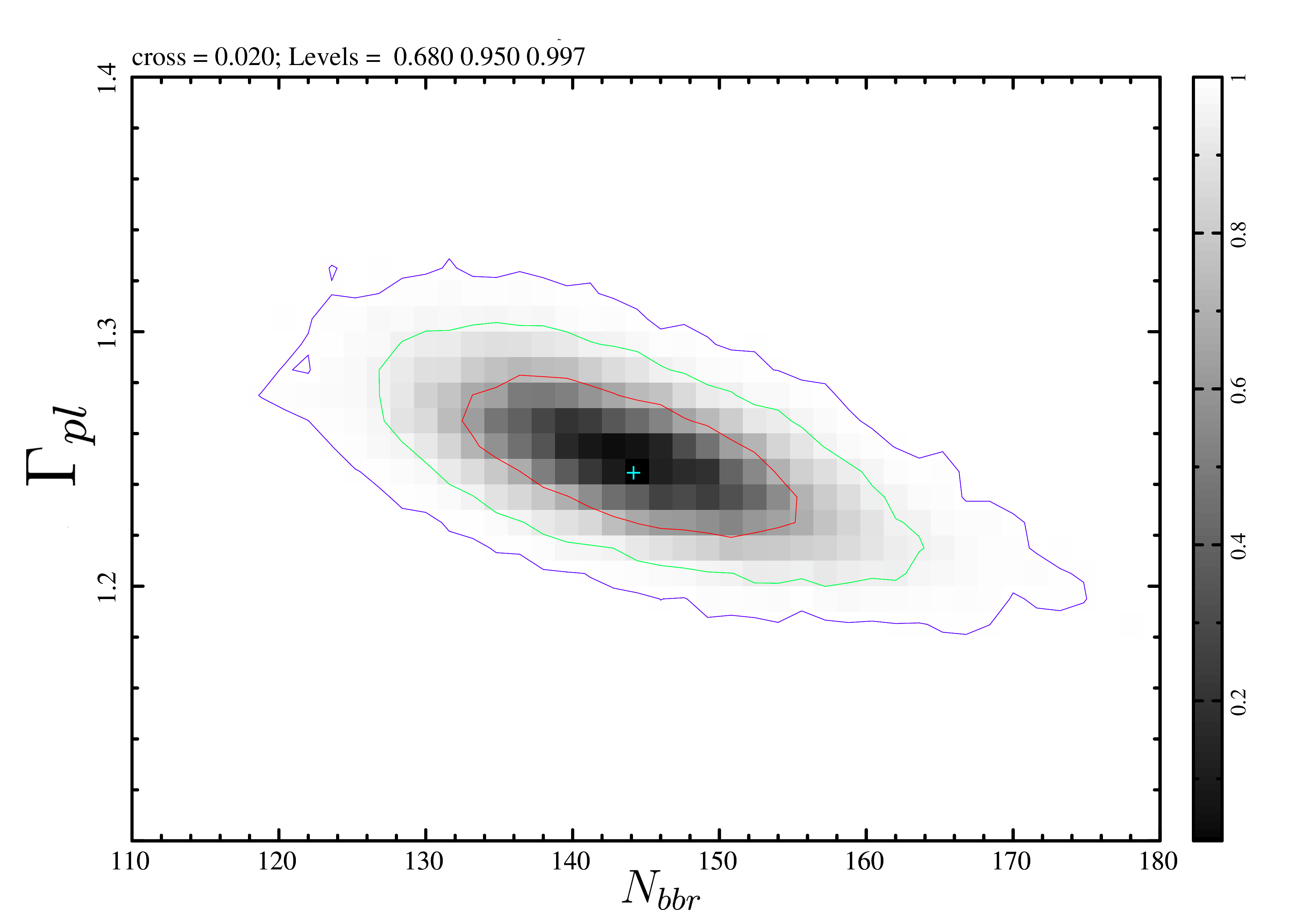}
\caption{Sec B Last 500s}
\label{fig:MCMC5_2}
\end{subfigure}

\label{fig:image2}

\begin{subfigure}{0.5\textwidth}
\includegraphics[width=0.8\linewidth, height=5cm]{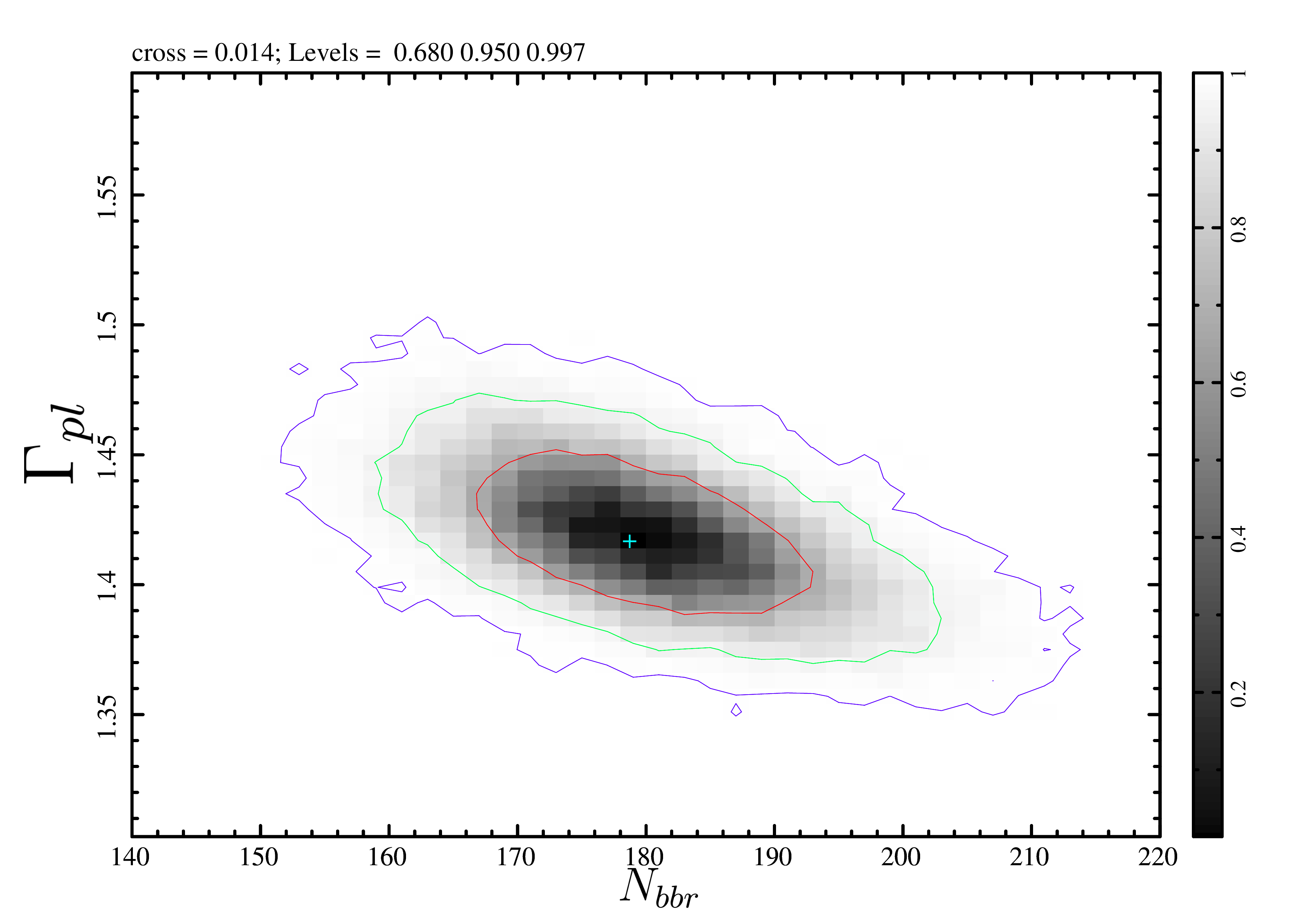} 
\caption{Sec C First 500s}
\label{fig:MCMC8_1}
\end{subfigure}
\begin{subfigure}{0.5\textwidth}
\includegraphics[width=0.8\linewidth, height=5cm]{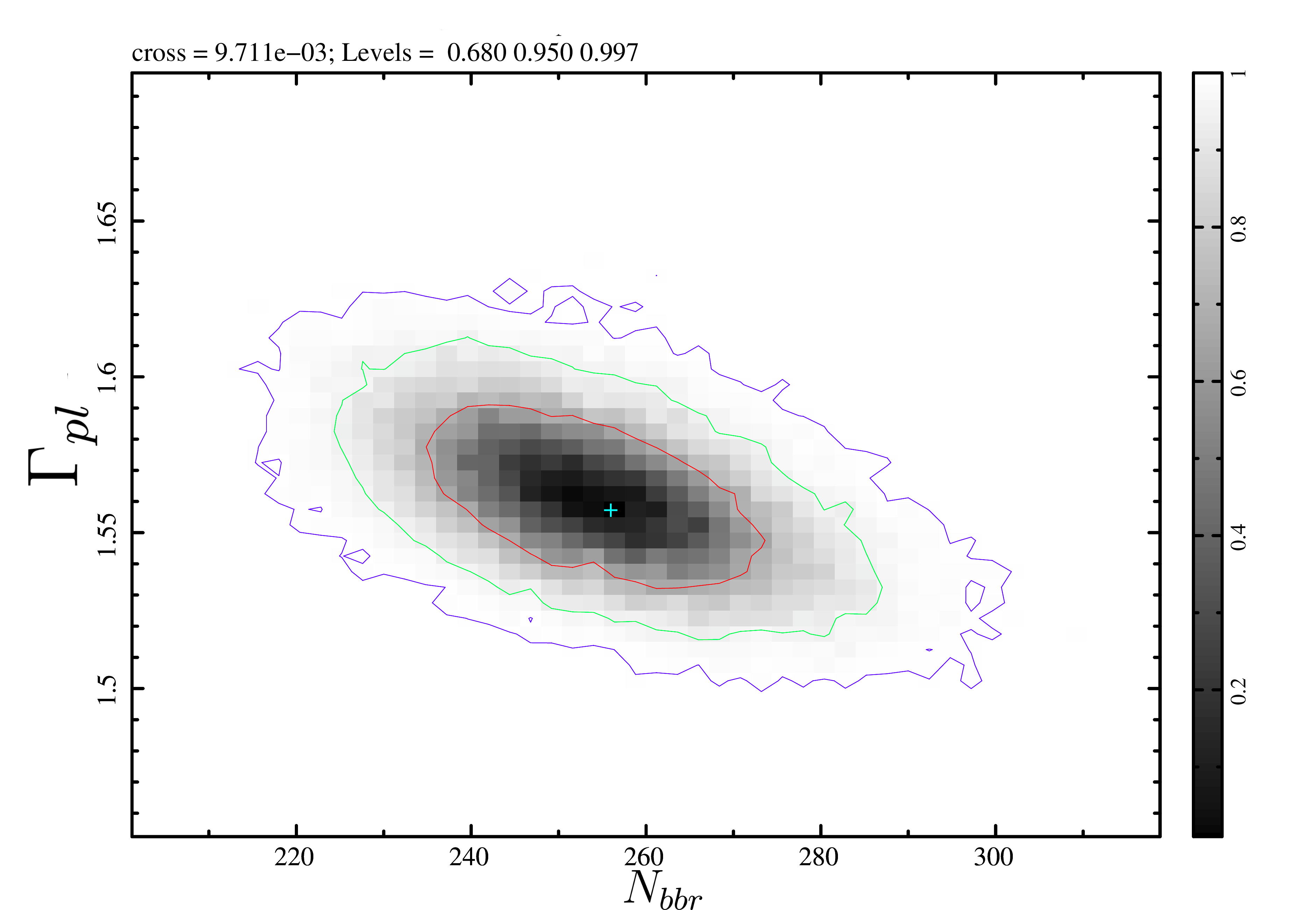}
\caption{Sec C Last 500s}
\label{fig:MCMC8_2}
\end{subfigure}

\begin{subfigure}{0.5\textwidth}
\includegraphics[width=0.8\linewidth, height=5cm]{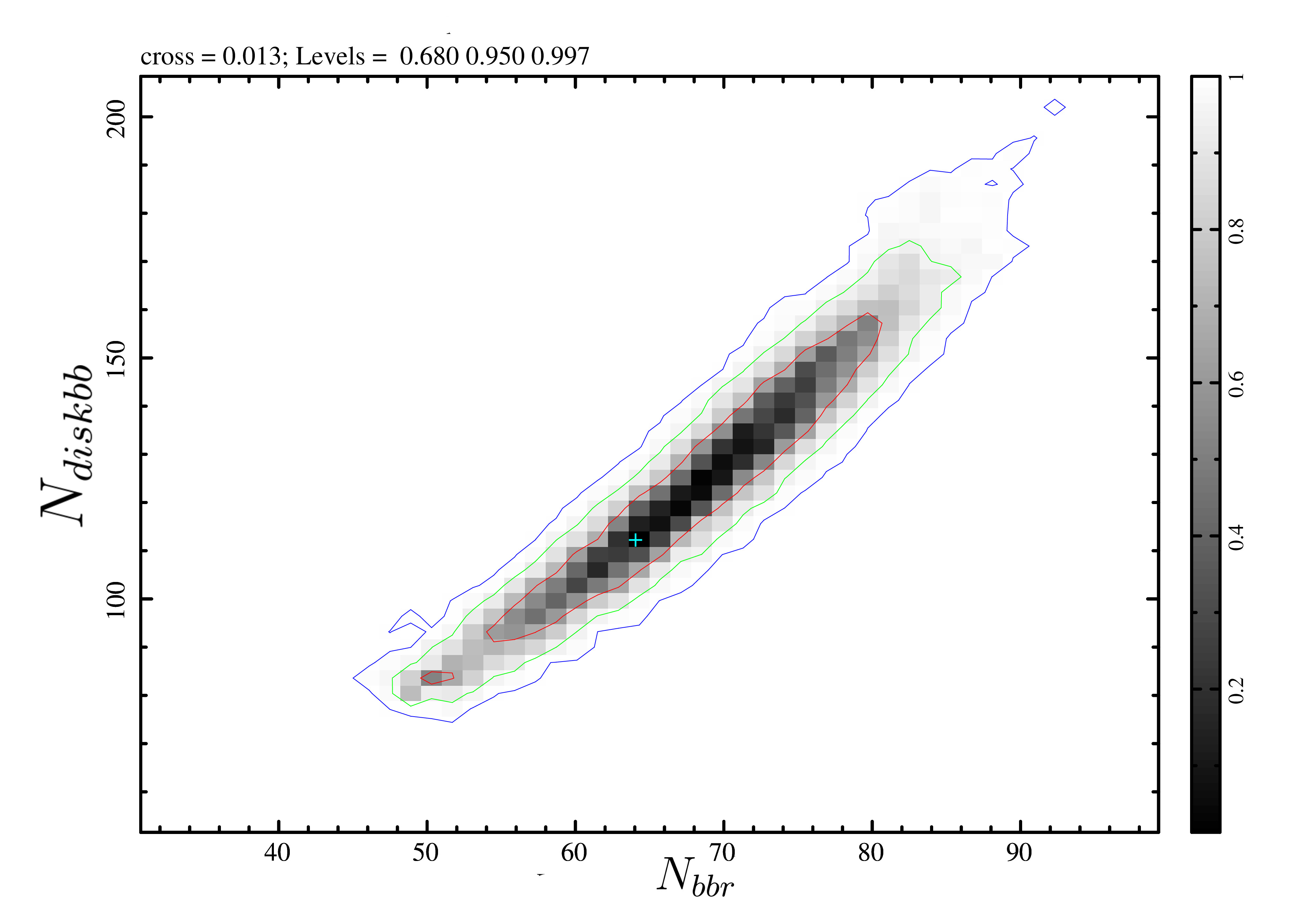} 
\caption{Sec B First 500s}
\label{fig:MCMC8_1}
\end{subfigure}
\begin{subfigure}{0.5\textwidth}
\includegraphics[width=0.8\linewidth, height=5cm]{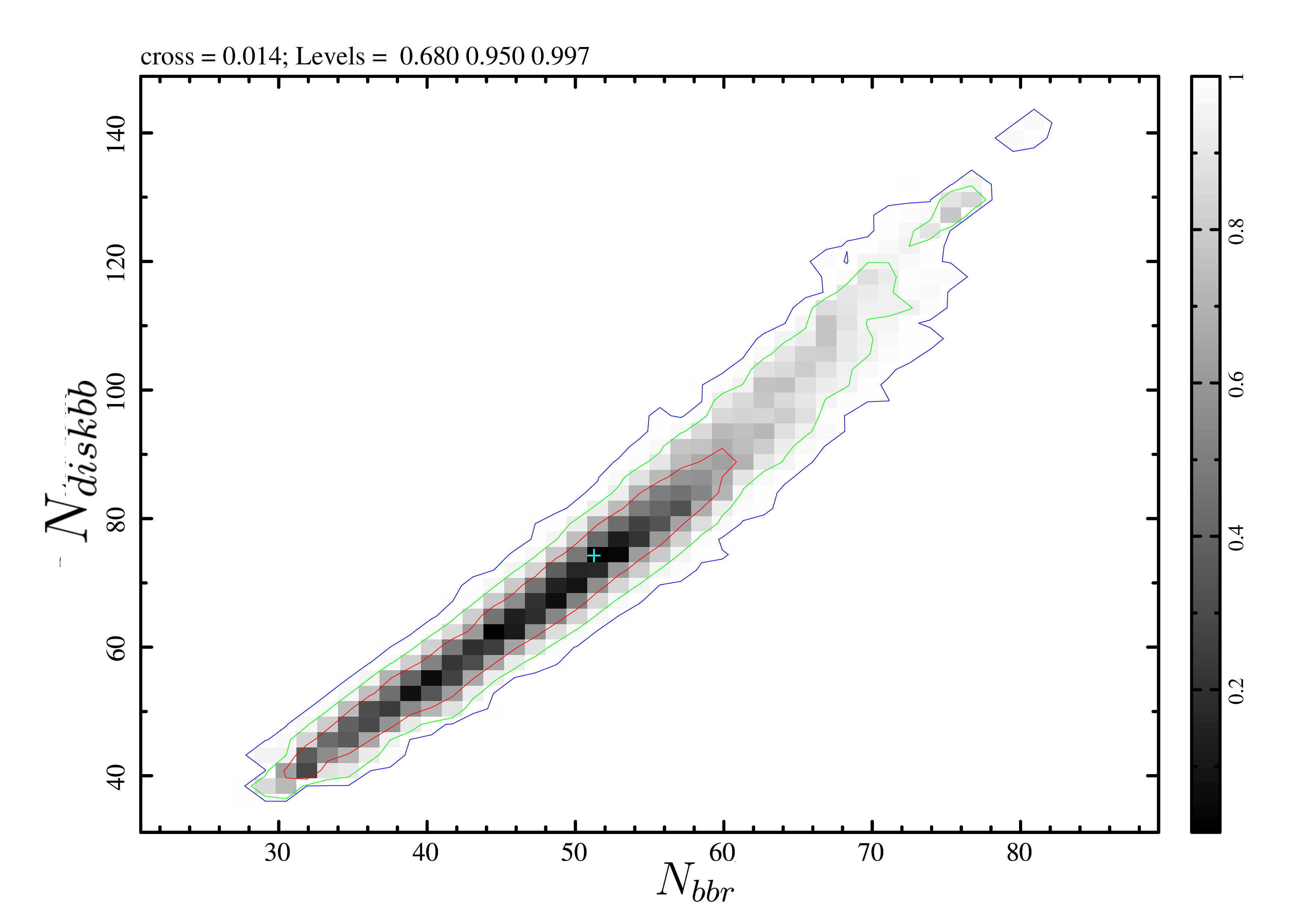}
\caption{Sec B Last 500s}
\label{fig:MCMC8_2}
\end{subfigure}

\caption{Section A, B, and C Contour plots of \textbf{Norm of bbodyrad} vs \textbf{$\Gamma$ of Powerlaw model}, from Model 1 ({\it Tbabs*(bbodyrad + powerlaw + Gaussian)}), where in each row, the left plot is of the First 500s of that section and the right plot is of the Last 500s of that section; the last row shows the Section B's Contour plot of \textbf{Norm of bbodyrad} vs \textbf{Norm of diskbb} for model 2 ({\textit{Tbabs*(bbodyrad+diskbb+Gaussian)}}).}
\label{fig:image2}
\end{figure*}

%%\end{thebibliography}

\end{document}